\documentclass[journal,review ,submit,pdftex,moreauthors]{Definitions/mdpi} 
\preto{\abstractkeywords}{\nolinenumbers} 
\usepackage{graphics}
\usepackage{epstopdf}
\usepackage{amsmath}
\usepackage{verbatim}
\usepackage{caption}
\usepackage{makecell}
\usepackage{mathrsfs}
\usepackage{url}    
\firstpage{1} 

\pubvolume{1}
\issuenum{1}
\articlenumber{0}
\pubyear{2024}
\copyrightyear{2024}
\datereceived{} 
\dateaccepted{} 
\datepublished{} 
\hreflink{https://doi.org/} 

\Title{Emergence and Causality in Complex Systems: A Survey of Causal Emergence and Related Quantitative Studies}

\Author{Bing Yuan 
$^{1}$ 
$^{,\dagger}$, {Jiang Zhang} $^{1,2,}$*$^{,\dagger}$, Aobo Lyu  $^{3}$, Jiayun Wu $^{4}$, Zhipeng Wang $^{2}$, Mingzhe Yang $^{2}$, Kaiwei Liu $^{2}$, Muyun~Mou~$^{2}$ and Peng Cui $^{4}$}

\AuthorNames{Bing Yuan, Zhang Jiang, Aobo Lyu, Jiayun Wu, Zhipeng Wang, Mingzhe Yang, Kaiwei Liu, Muyun Mou, Peng Cui}

\AuthorCitation{{Yuan, B.;} Zhang, J.; Lyu, A.; Wu, J.; Wang, Z.; Yang, M.; Liu, K.; Mou, M.; Cui, P.}

\address{%
$^{1}$ \quad Swarma Research, {Beijing 100085,} China; {yuanbing\_cn@hotmail.com} 
\\
$^{2}$ \quad School of Systems Sciences, Beijing Normal University, {Beijing 100875,} China; {wangzhipeng@mail.bnu.edu.cn (Z.W.); ymp66393866@163.com (M.Y.); 202221250005@mail.bnu.edu.cn (K.L.); moumuyun@mail.bnu.edu.cn (M.M.)}\\
$^{3}$ \quad Department of Electrical and Systems Engineering, Washington University, St. Louis, {MO 63130}, USA; {aobo.lyu@wustl.edu}\\
$^{4}$ \quad Department of Computer Science and Technology, Tsinghua University, {Beijing 100084}, China; {jiayun.wu.work@gmail.com (J.W.); cuip@tsinghua.edu.cn (P.C.)}}

\corres{Correspondence: zhangjiang@bnu.edu.cn}
\firstnote{These authors contributed equally to this work.} 

\abstract{Emergence and causality are two fundamental concepts for understanding complex systems. They are interconnected. On one hand, emergence refers to the phenomenon where macroscopic properties cannot be solely attributed to the cause of individual properties.~On the other hand, causality can exhibit emergence, meaning that new causal laws may arise as we increase the level of abstraction. Causal emergence (CE) theory aims to bridge these two concepts and even employs measures of causality to quantify emergence. This paper provides a comprehensive review of recent advancements in quantitative theories and applications of CE. It focuses on two primary challenges: quantifying CE and identifying it from data. The latter task requires the integration of machine learning and neural network techniques, establishing a significant link between causal emergence and machine learning. We highlight two problem categories: CE with machine learning and CE for machine learning, both of which emphasize the crucial role of effective information (EI) as a measure of causal emergence. The final section of this review explores potential applications and provides insights into future perspectives.}


\begin{document}

\section{Introduction}

Economic growth, environmental protection, sustainable development, the global climate crisis, social inequality, and many other issues are all intertwined with complex systems~\cite{ledford2015team,mensah2018systems}. Therefore, gaining a deep understanding of how complex systems operate, evolve, grow, stabilize, and collapse is of paramount importance. However, this task is exceptionally challenging due to the fact that complex systems consist of diverse and heterogeneous agents that interact through complex nonlinear relationships~\cite{bar2002general}. Moreover, they all exhibit emergent phenomena, which are highly common in complex systems but carry a sense of mystery~\cite{holland2000emergence}.

How did the first living cell emerge from the collisions between various molecules in the Earth's early environment~\cite{artime2022origin}?~How does the cognitive concept of ``I'' emerge from the intricate interactions among countless neurons in our brain~\cite{lagercrantz2009emergence}?~How do large neural language models suddenly exhibit emergent abilities~\cite{wei2022emergent}?~These fundamental questions revolve around the concept of emergence in complex living, cognitive, and artificial systems. Emergence refers to the phenomenon where macroscopic properties and phenomena cannot be solely attributed to or explained by the properties of individual components~\mbox{\cite{anderson1972more,meehl1956concept,bedau1997weak,holland2000emergence}}. This presents a formidable challenge to the traditional reductionist perspective while also shedding light on the underlying reasons for the enigmatic nature of emergent phenomena.

However, as elucidated by Bedau's theory of weak emergence~\cite{bedau1997weak}, many emergent phenomena can be comprehended through the interactions among the individuals within the system~\cite{holland2000emergence}. Complex systems, in fact, consist of extensive networks of interacting components~\cite{albert2002statistical}. Within these networks, even a minor cause, such as a perturbation of a single unit, can propagate through the interconnected network, resulting in a collective effect. The phenomenon described is commonly known as the butterfly effect~\cite{wiki:butterflyeffect}, which provides an explanation for the occurrence of emergence. On the other hand, emergent properties, such as  homeostasis~\cite{wiki:homeostasis}, can stabilize the system itself, thereby preserving the locality of causal effects and preventing the observation of macro-level effects. These phenomena demonstrate that complex systems achieve interactions through causal laws, where numerous local causal laws form interconnected causal networks as a whole. And this whole possesses unique causal characteristics.

Causality, or causation, refers to the connection between a cause and its resulting effect~\cite{granger1969investigating,pearl2000models,pearl2009causality}. It describes the phenomenon in which an event, known as the cause, leads to another event, known as the effect. Traditional studies of causality have typically focused on the causal relationship between two or a few variables. However, the unique characteristics of causality in complex systems present new challenges to classical causal science due to the vast number of variables involved and the presence of emergent phenomena. In complex systems, it is possible for one cause to have multiple effects, and conversely, one effect may be influenced by a multitude of causes. Furthermore, in complex systems, causality often exhibits cross-level properties, which are closely associated with emergence.

Emergence and causation interconnect with each other. On one hand, emergence is the causal effect of complex and nonlinear interactions between components in complex systems~\cite{anderson1972more,meehl1956concept}. On the other hand, emergent properties may have causal effects on individuals in complex systems~\cite{kim1992downward,holland2000emergence}. For example, the price of fossil fuel is the emerging result of the interactions between buyers and sellers in the market. At the same time, the price may also provide feedback to the market: it can affect the decision making of each individual.

Furthermore, we can understand emergence through the perspective of causation. What emergence means is that some phenomena and properties on the macroscopic level can not be attributed to the microscopic properties~\cite{fromm2005types}. Thus, emergent properties or phenomena lose their direct explanations as usual but may be attributed to the causes on the macro-level, as pointed out by~\cite{hoel2013quantifying}. Therefore, new causalities can be observed on larger scales.

In conclusion, gaining a deep understanding of emergence is crucial in the field of complex system studies. Specifically, the development of a quantitative theory of emergence is on the verge of emerging.~Such a theory holds the potential to address significant challenges, including the origins of life~\cite{ellis2023efficient}, the emergence of novel capabilities in large neural network models~\cite{wei2022emergent}, and the potential for intelligence, consciousness, and free will to arise in artificial systems~\cite{oizumi2014phenomenology}. Causality not only exhibits a profound connection with emergence but is also considered by many researchers as one of the most crucial perspectives for quantitatively comprehending emergence~\cite{hoel2013quantifying,crutchfield1989inferring,fromm2005types,seth2008measuring}. 

Two primary challenges take precedence in understanding emergence from a causal perspective. The first is establishing a quantitative definition of emergence, whereas the second involves identifying emergent behaviors or phenomena through data analysis.

To address the first challenge, two prominent quantitative theories of emergence have emerged in the past decade. The first is Erik Hoel et al.'s theory of causal emergence~\cite{hoel2013quantifying}, whereas the second is Fernando E. Rosas et al.'s theory of emergence based on partial information decomposition~\cite{rosas2020reconciling}. 

Hoel et al.'s theory of causal emergence specifically addresses complex systems that are modeled using Markov chains. It employs the concept of effective information (EI) to quantify the extent of causal influence within Markov chains and enables comparisons of EI values across different scales~\cite{hoel2013quantifying,hoel2017when}. Causal emergence is defined by the difference in the EI values between the macro-level and micro-level. Several examples of discrete Markov chains have been shown to exhibit causal emergence if their EI values of macro-level dynamics are larger than those of micro-level dynamics. Hoel and other researchers further extended the measures of effective information and causal emergence for dynamical systems with continuous variables~\cite{chvykov2020causal} and complex networks~\cite{klein2020emergence}. Other measures are also possible for quantifying causal effects and, consequently, causal emergence. In~\cite{comolatti2022causal}, Comolatti and Hoel systematically compared several causal effect measures and concluded that causal emergence is independent of the selection of the measure. 

However, in Hoel's theory of causal emergence, it is essential to establish a coarse-graining strategy beforehand. Alternatively, the strategy can be derived by maximizing the effective information (EI)~\cite{hoel2013quantifying}. However, this task becomes challenging for large-scale systems due to the computational complexity involved. To address these problems, Rosas et al. introduced a new quantitative definition of causal emergence~\cite{rosas2020reconciling} that does not depend on coarse-graining methods, drawing from partial information decomposition (PID)-related theory. PID is an approach developed by Williams et al., which seeks to decompose the mutual information between a target and source variables into non-overlapping information atoms: unique, redundant, and synergistic information~\cite{williams2010nonnegative}. Based on this groundwork, Rosas further developed the concept and introduced a theory called $\phi$ID to decompose the mutual information between multiple target and source variables~\cite{mediano2019beyond}. This framework provides a quantitative definition of causal emergence by measuring the positive synergy information between the source and target variables based on the inherent characteristics of the system.

The second challenge pertains to the identification of emergence from data. In an effort to address this issue, Rosas et al. derived a numerical method~\cite{rosas2020reconciling}. However, it is important to acknowledge that this method offers only a sufficient condition for emergence and is an approximate approach. Another limitation is that a coarse-grained macro-state variable should be given beforehand to apply this method. Hence, there is a need for the development of new methods. 

Recently, artificial intelligence, propelled by the rapid advancements in machine learning and deep neural network technology, has witnessed significant progress. In the context of causal emergence, there are two key aspects to consider. Firstly, machine learning and neural network technology can be employed to address the challenge of identifying causal emergence. By leveraging these tools, we can develop approaches to effectively detect and analyze causal emergence phenomena. Secondly, the concepts and techniques from causal emergence can be introduced into machine learning to enhance the generalization capabilities of models. This integration can potentially improve the ability of machine learning algorithms to generalize well beyond the training data, leading to more robust and adaptable systems. 

In a recent study by Zhang et al.~\cite{zhang2022neural}, a machine learning framework named the Neural Information Squeezer (NIS) was introduced to address the challenge of identifying causal emergence using Hoel et al.'s framework. Remarkably, the NIS neural network, functioning as a ``machine observer'' equipped with an internal model, exhibits a remarkable ability to identify causal emergence across various types of data. In the latest updated version of this work, the Neural Information Squeezer Plus (NIS+) has been developed to directly maximize the critical measure of causal emergence theory, namely effective information (EI)~\cite{yang2023finding}. Through extensive experiments conducted on both simulated data and real brain data, the NIS+ has demonstrated its ability to automatically find emergent macro-variables and macro-dynamics. Consequently, the NIS+ enables quantifying causal emergence in data with the learned macro-dynamics. The results of these experiments highlight the effectiveness and potential of the NIS+ in capturing and analyzing causal emergence phenomena.

Furthermore, the NIS+ showcases superior performance in terms of generalization ability by directly maximizing effective information (EI). This brings forth a second question: can we leverage the measure of causation, EI, in the context of causal emergence, to enhance the generalization capability of neural networks for out-of-distribution scenarios? This concept is referred to as causal emergence for machine learning. By exploring this idea, we aim to bridge the gap between causal emergence and machine learning, potentially unlocking new avenues for improving the generalization abilities of machine learning.

Finally, in Section 5, we address several important and related issues. Firstly, we explore the similarities and differences between two emerging fields: causal emergence and causal representation learning~\cite{scholkopf2021toward}.~This comparison sheds light on the interplay between these two domains. Secondly, we delve into a philosophical problem concerning ontological or epistemological causality and emergence, providing insights into the underlying philosophical implications. Lastly, we discuss the potential applications of causal emergence in complex systems and how it contributes to our understanding of complex systems from a causal emergence perspective. These discussions broaden the scope of this paper and offer intriguing avenues for future~research.

This paper aims to provide a comprehensive review of the latest research on the quantitative theory and applications of causal emergence and related works. It also explores the connections between causal emergence, machine learning, and complex systems. The subsequent section delves into the background of causal emergence, with a particular focus on the interplay between causation and emergence in complex systems. In Section 3, various quantitative theoretical frameworks are introduced, including \mbox{Crutchfield et al.'s} computational mechanics~\cite{crutchfield1989inferring}, Seth et al.'s Granger causal emergence~\cite{seth2008measuring}, Hoel et al.'s causal emergence theory, and Rosas et al.'s theory of emergence based on information decomposition. Additionally, related concepts such as the coarse-graining strategy, measures of effective information, and partial information decomposition are discussed, and a comparative analysis of these theories is presented. Section 4 addresses the connection between causal emergence theory and machine learning. It explores the use of machine learning and neural network techniques for identifying causal emergence and extends the measure of effective information (EI) to machine learning problems. Finally, this paper delves into other important topics and potential applications in the fields of machine learning and complex systems.

\section{Background: From Causality to Emergence}

\subsection{Causality}
\label{sec:causal_hierarchy}
Causality, which is a fundamental concept in many fields, including philosophy, natural science, and social science, refers to the relationship between a pair of events, where the first event (cause) can influence the second one (effect)~\cite{granger1969investigating,pearl2000models,pearl2009causality}. The relationship between cause and effect can exhibit either deterministic or probabilistic characteristics. In deterministic causality, the cause will always produce the same effect, whereas in probabilistic causality, the cause will only produce the effect with a certain probability.

In most physical systems governed by differential equations or Markovian dynamics, whether deterministic or probabilistic, causality is inherent. This arises from the fact that manipulating a variable within the system can result in observable changes in other variables, albeit with probabilistic outcomes. As pointed out by Y. Iwasaki and H. A. Simon~\cite{iwasaki1994causality}, one physical mechanism may correspond to several possible causal structures because causal graphs describe local causal interactions, whereas mechanisms are the global constraints for all variables. We further discuss the relationship between Markovian dynamics and causal models in detail below.

\begin{figure}[H]

\includegraphics[scale=0.7]{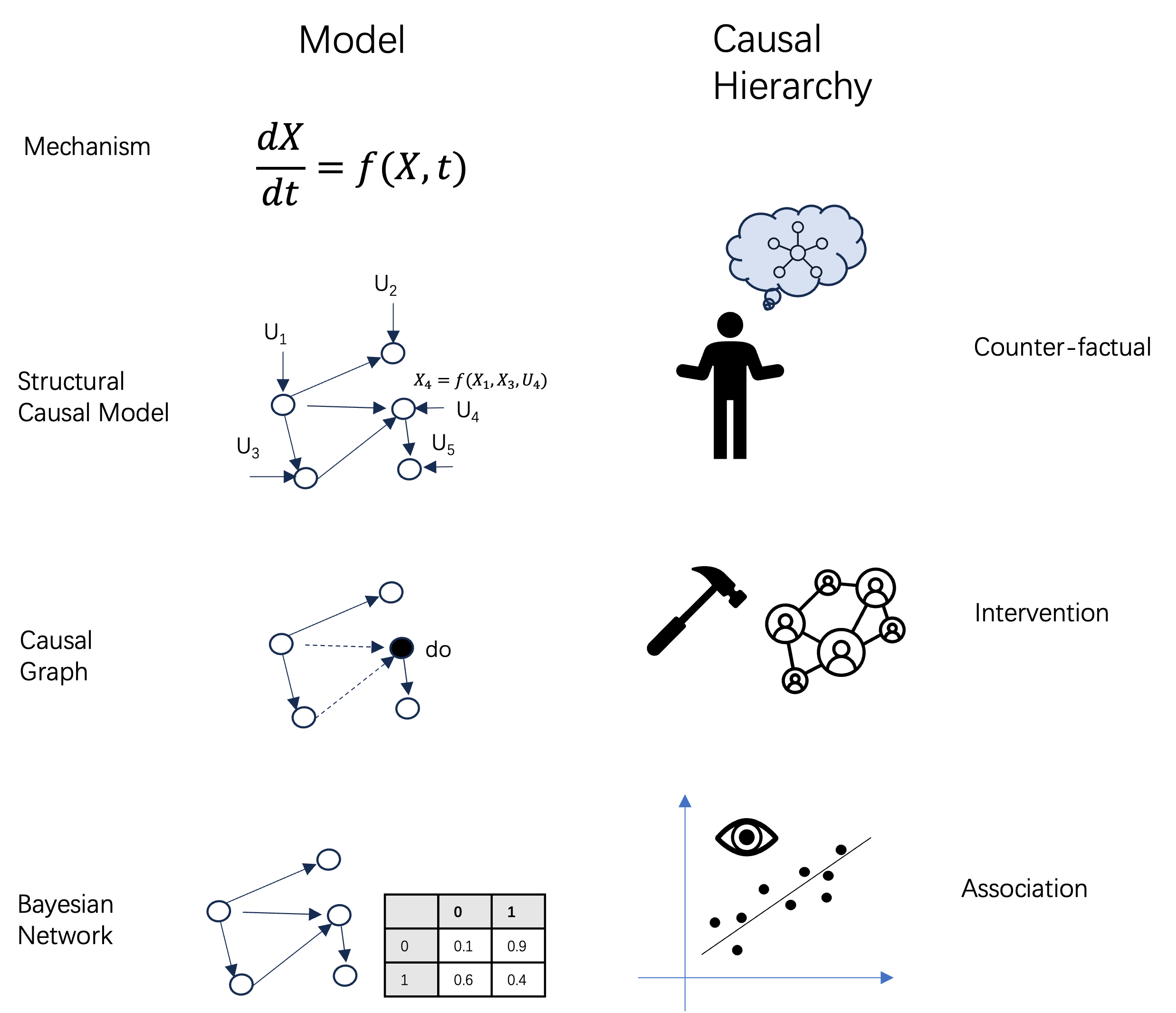}
\caption{The causal models (\textbf{left}) column and the corresponding causal hierarchy (\textbf{right}) column. 
}
\label{fig:causal_hierarchy}
\end{figure}

If we do not know the physical mechanism of a system, how can we model the causality behind the system? There are several parallel theoretical frameworks that can be used to do this. For example, Judea Pearl utilized probabilistic graphic models (e.g., Bayesian network, causal graph, and structural causal model) to characterize causal interactions~\cite{pearl2000models,pearl2009causality}. In these models, nodes represent random variables and acyclic links represent causal interactions. Pearl distinguished and quantified three levels of causality in different models.~The first level is association, i.e., the correlation between variables, which can be modeled using Bayesian networks~\cite{sucar2015probabilistic}.~The second level is intervention, which can be characterized using causal graph models~\cite{pearl2009causality}. The difference between a causal graph and a Bayesian network is that the former enables defining a special operator, called a ``do'' operator, to simulate the interventions performed by the experimenter. The last level is counterfactual, which can be quantified using structural causal models~\cite{pearl2000models}. The distinguishing characteristic of a structural causal model is that all interactions within a causal graph can be described by a set of deterministic functions that involve the input variables and unknown components, thereby introducing uncertainty. Counterfactual causal inferences, that is, inferences under an imaginary scenario, can be conducted only on structural causal models. The different types of causal models and the corresponding causal hierarchy in Pearl's theory are shown in Figure \ref{fig:causal_hierarchy}.

\vspace{-9pt}

{Another quantitative framework for causal inference is Donald B Rubin's ``potential outcome'' theory, which operates at the counterfactual level~\cite{rubin2005causal}.} 
According to this theory, each individual has two potential outcomes: the outcome that would be observed if the individual receives the treatment, and the outcome that would be observed if the individual does not receive the treatment. Rubin's theory suggests that the causal effect of a treatment can be estimated by comparing the difference between the observed outcome and the unobserved potential outcome for each individual.

Researchers have also developed a bundle of methods, named causal discovery algorithms~\cite{malinsky2018causal}, to automatically discover the causal relations between variables merely from data. For example, constraint-based algorithms~\cite{spirtes2000causation} and score-based algorithms~\cite{chickering2002learning} for causal discovery are two typical representations. Causal discovery has emerged as an interdisciplinary area at the intersection of machine learning, causal inference, and statistics. 

Due to the uncertainty and ambiguity behind the discovered causal relations, measuring the degree of causal effect between two variables is another important problem. Numerous independent historical studies have addressed the issue of causation measurement. These include Hume's concept of constant conjunction~\cite{hume2016enquiry}, Eells' and Suppes' measures of causation as probability raising~\cite{eells1991probabilistic,suppes1973probabilistic}, and Judea Pearl's measures of causation~\cite{pearl2009causality}. Effective information (EI) in the framework for causal emergence is also an indicator for measuring causal effect, and we explain this point in Section \ref{sec:EI}. All these measures can be roughly decomposed into two causal primitives, which characterize the necessary causation and sufficient causation, respectively~\cite{hoel2020emergence}. 

\textls[-15]{It is crucial to differentiate between Markovian dynamics or dynamical mechanism models and the causal models introduced above.~In current causal emergence theories~\mbox{\cite{hoel2013quantifying,rosas2020reconciling}},} Markovian dynamics, rather than causal models, including causal graphs and structural causal models, play a central role.~There are several distinctions between Markovian dynamics and these causal models. 

Firstly, in Markovian dynamics, the temporal evolution of one or multiple variables is typically described, with causal relationships between these variables extending across different points in time. Consequently, all variables are dependent on time, and a variable at one time step can causally impact another variable at the subsequent time step. Conversely, the aforementioned causal models do not explicitly incorporate the element of time. In these models, neither the variables nor the causal relationships are explicitly defined as functions of time.

Secondly, it is worth noting that circular interaction structures, including self-looped interactions, may exist in Markovian dynamics. However, such circular structures are typically not permitted in the causal models we introduced. This distinction arises from the fact that both causal graph models and structural causal models are constructed based on directed acyclic graphs (DAGs)~\cite{pearl2009causality}. However, this issue can be resolved by converting Markovian dynamics into DAGs. The initial step involves introducing static causal variables at each time step. Consequently, the same variable in Markovian dynamics can be transformed into a set of different variables at different time steps. This expansion allows for causal relationships across time steps while eliminating circular structures in causal graphs~\cite{koller2009probabilistic}.

Given the limitations of conventional causal models based on directed acyclic graphs (DAGs) \cite{textor2016robust}, there is a need for new representation methods. For instance, Richardson \cite{richardson2013discovery} expanded the causal modeling approach to include DAGs. Spirtes~\cite{spirtes2013directed} discussed DAG representations of feedback models and applied them to economic processes. Lacerda et al. also developed a method utilizing independent components analysis to discover cyclic causal models \cite{lacerda2012discovering}. Furthermore, Forr{\'e} and Mooij introduced a novel causal discovery method for nonlinear structural causal models with cycles and latent confounders \cite{forre2018constraint}. George T.H. Ellison proposed a methodology that leverages domain knowledge to extract DAGs from temporal relationships among multiple variables \cite{ellison2023using}. Causal discovery methods can also be extended to temporal data by segregating various temporal data analysis tasks, such as classification, clustering, and prediction. Recently, Gong et al. provided a comprehensive review of these techniques in \cite{gong2023causal}. These advancements have the potential to broaden the range of available causal models.

In summary, causality is a fascinating interdisciplinary topic spanning several subjects. Numerous quantitative frameworks and measures have been developed in recent decades.
\subsection{Emergence}

Emergence, known as ``the whole is greater than the sum of parts''~\cite{pepper1926emergence}, is a central concept in many philosophical and theoretical discussions about the nature of complexity and the relationship between micro- and macro-levels of organization.  Therefore, the scale of a system (micro or macro) and the cross-level interactions must be considered when we discuss emergence. This may increase the complexity of the study of emergence~\cite{winning2019complexity,hendry2019substance,Huxley1947-HUXEAE-3,mill_2011,gibb2019routledge,ross1908works,holland1996hidden}.

Although many examples of emergence have been pointed out in various fields~\cite{holland1996hidden,holland2000emergence}, including the herding behaviors of birds~\cite{reynolds1987flocks}, collective behaviors of simple computer programs~\cite{gardner1970fantastic}, and emerging ability of large language models~\cite{wei2022emergent}, a unifying understanding of this phenomenon does not exist right now. People try to understand emergence from different aspects, including self-organization~\cite{luhmann1995social}, order out of disorder~\cite{crutchfield1994calculi}, and causality. In this review, we mainly focus on the last point of view.

Emergence has deep connections with causation~\cite{fromm2005types,bedau1997weak,bedau2008emergence,bedau2002downward}. First, emergence emphasizes the new properties arising from the interactions between the components in the system, rather than the components themselves. Consequently, all emergent properties and processes at the macro-level can, in principle, be comprehended as the causal effects resulting from interactions between individuals at the micro-level. This is what the philosophical notion of ``supervenience'' tries to describe. 

However, this does not mean that all emergent phenomena at the macro-level can be easily attributed to micro-level individuals and their interactions. Two major reasons can account for this. The first reason is due to the principle of computational irreducibility proposed by Wolfram~\cite{wolfram2002new}, which claims that although we know emergent properties arise from a set of simple rules, there is no shortcut, in principle, to predict the results other than implementing the rules. For example, the complex behaviors (e.g., herding and aligning) of the Boid model are the results of simple rules, but these behaviors cannot be reduced to individual behaviors simply without simulating them~\cite{reynolds1987flocks}. 

The second reason is that emergent behaviors may be determined by other emergent properties at the macro-level~\cite{hoel2013quantifying,rosas2020reconciling}. For example, although the price of rice in the market is determined by all the interactions (bargaining) between buyers and sellers, it is also the causal effect of the price of fossil oil because severe shortages of fossil oil can cause inflation. Therefore, the cause of one emergent property (rice price) is another emergent macro-level coarse-grained variable. This means new causation can be observed between macro-level coarse-grained variables, as pointed out by the theory of causal emergence. Interestingly, the causation at different levels must be consistent to explain the micro-level event of fluctuating rice prices.~Otherwise, a double causation~\cite{merton1973sociology} fallacy may appear. This consistency between causality at different scales is called the causal equivalence principle in Yurchenko's latest article~\cite{yurchenko2023can}.

In order to better understand emergence, Bedau et al.~\cite{bedau1997weak,bedau2002downward} classified emergence into three categories according to the causal interactions between the micro- and macro-levels: nominal emergence~\cite{harre1985philosophies, baas1994emergence}, weak emergence~\cite{bedau1997weak,newman1996emergence}, and strong emergence~\cite{o1994emergent, kim1992downward}.  

Of these categories, the notion of nominal emergence is the least controversial. It can be described as a kind of property that can be possessed by macro-level patterns or processes but not by their micro-level components~\cite{harre1985philosophies, baas1994emergence}. For example, pixel patterns on a screen are nominal emergent properties. We can consider such nominal emergent patterns as ``supervenient'' because all macro-level properties derive from individuals. 

Weak emergence refers to macro-level properties or processes that derive through interactions between individual components in a complicated way such that they cannot be easily reduced to micro-level properties due to the principle of computational irreducibility~\cite{bedau1997weak,newman1996emergence}. For weak emergent patterns or processes, the causes may come from both the micro- and macro-levels; therefore, emergent causation may coexist with micro-level causation. 

In the discourse surrounding emergence, weak emergence is generally more widely accepted than strong emergence. However, there are some concepts and descriptions that remain somewhat ambiguous \cite{hedercriticism}. For instance, in Bedeau's original definition, a property is considered weakly emergent if and only if it can be derived from micro-dynamics through simulation. Nevertheless, the term ``simulation'' itself can be interpreted in different ways—whether it refers to digital simulations or analog simulations is not explicitly implied. Bedeau further elaborated on this concept by stating that if a macro-level property can, in principle, be simulated without requiring an actual simulation, then it can be considered weakly emergent \cite{bedau1997weak}. However, even this explanation is not entirely clear-cut.

There are more debates on strong emergence, which refers to properties or processes at the macro-level that cannot, in principle, be reduced to micro-level properties, including the interactions of individuals~\cite{o1994emergent, kim1992downward}. Thus, Boid's collective behaviors are not strongly emergent because they are the result of interactions within the Boid model. This notion is controversial because it rejects any mechanistic explanations for strong emergent properties, which cannot be explained by micro-level variables. Furthermore, it raises some long-standing philosophical debates about ``causal fundamentalism''~\cite{jackson1992defense} and ``supervenience''~\cite{kim1999making}, and the existence of strong emergence remains an open problem due to the scarcity of concrete~examples.

Jochen Fromm further explained strong emergence as the causal effect of downward causation~\cite{fromm2005types}. Consider a system that contains three different scales: micro, meso, and macro. Downward causation refers to the causal power from the macro-level to the meso-level or from the meso-level to the micro-level. Consequently, although the strong emergent properties or processes at the meso-level are supervenient to micro-level properties and interactions, their causes derive from the macro-level and, therefore, second-order patterns or processes. However, there are many debates on the notion of downward causation itself, e.g.,~\cite{yurchenko2023can,bedau2008emergence}.

\begin{figure}[H]

\includegraphics[scale=0.7]{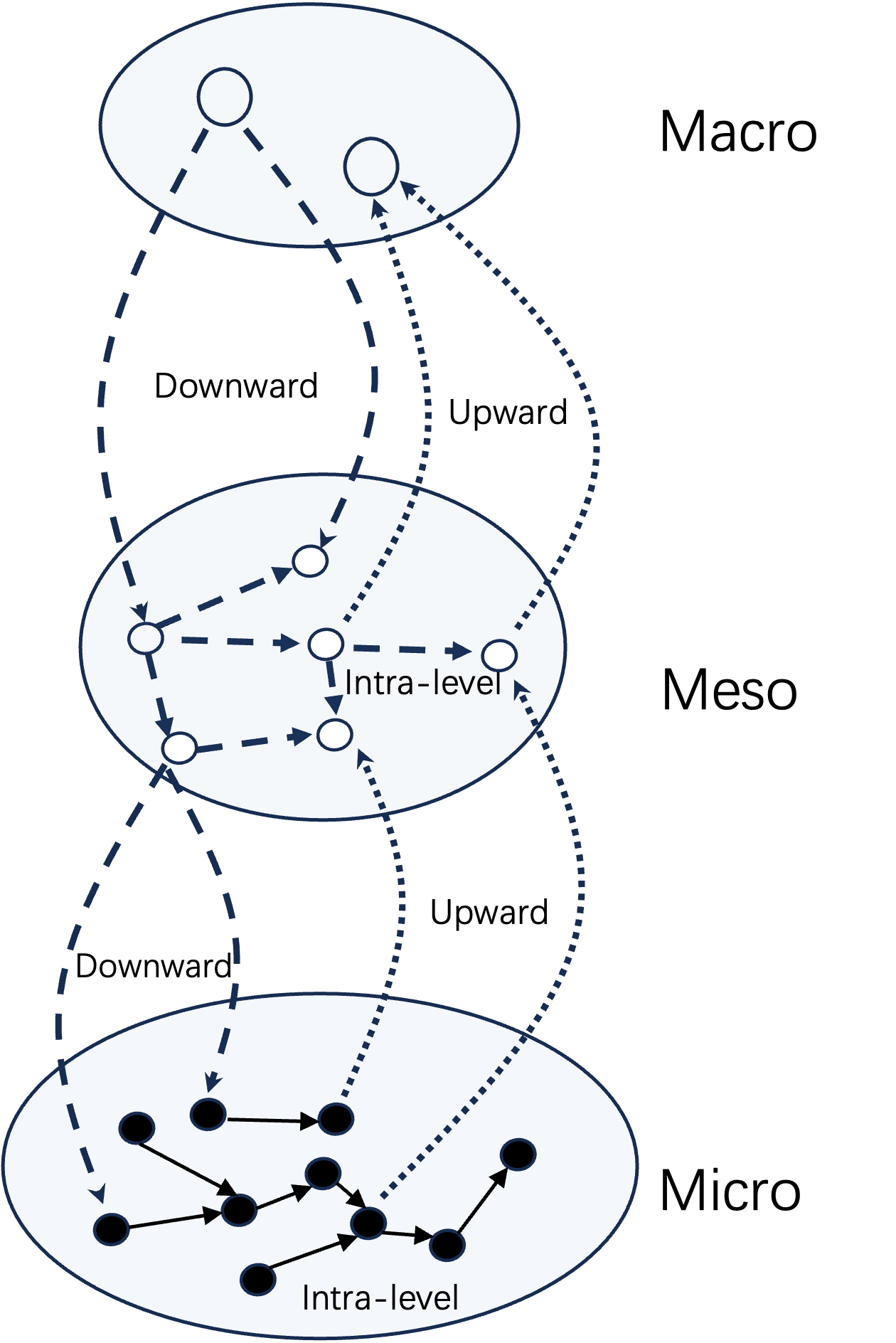}
\caption{The various types of causation that exist within multi-scale complex systems. In this figure, solid arrows represent common causation, which is generally accepted without much controversy. Dotted arrowed lines indicate a form of ``causation'' driven by supervenience, whereas dashed arrowed lines represent emergent causation, occurring either at an intra-level or downward.}
\label{fig:crosslevel_causation}
\end{figure}

In summary, there are three types of causation based on their cross-level properties, which are associated with emergence, as depicted in Figure \ref{fig:crosslevel_causation}: 

\begin{enumerate}[leftmargin=2.15em,labelsep=0.55mm]
\item {Upward} causation: This involves the supervenience relation, where macro-level effects can be attributed to micro-level variables.

\item Intra-level causation: This refers to causal effects occurring within the same scale or~level.

\item \textls[+15]{Downward causation: Here, macro-level properties influence micro-level~properties.}
\end{enumerate}

\vspace{-6pt}

In Figure \ref{fig:crosslevel_causation}, the black solid lines represent causation without controversy. However, according to a recent paper by Yurchenko~\cite{yurchenko2023can}, the causation represented by the dashed and dotted lines (except the intra-level ones in the meso-level) may not actually be causality but rather reasons or explanations. Yurchenko introduced the term ``reason'' to distinguish it from the conventional understanding of causality, as cross-level causal effects may not possess causal powers. Instead, they serve as reasons or explanations from the perspective of an observer. {Yurchenko accepted intra-level causation at the meso- or macro-level as ``real'', similar to causation at the micro-level. He even coined the term ``causal equivalent principle'' to represent the idea that all intra-level causations, except those that are cross-level, should be accepted and are equivalent to each other.}


However, other scholars think that compared to the causal relationship at the micro-level, the intra-level causation at the meso- or macro-level is problematic because the coarse-graining strategy is relevant~\cite{bonabeau2011detection,dewhurst2021causal,eberhardt2022causal}. As demonstrated in the numeric example in Section \ref{sec:exampleCE}, the causation measured by EI is dependent on the coarse-graining method. Hoel et al. tried to avoid this problem by introducing EI maximization. Indeed, EI maximization serves as an objective standard for selecting coarse-graining and macro-dynamics; however, a unique solution is required. We discuss this problem further in Section \ref{sec:ontological}.

\section{Quantifying Emergence by Causality}

Since causality and emergence have a strong connection and causality has multiple quantitative frameworks and measures, it becomes natural to use causality to quantify emergence. In this section, we review several  frameworks used to quantify multi-scale causality and emergence.

\subsection{Early Related Works}
Before the theory of causal emergence was proposed by Hoel et al., some works introduced very similar ideas to causal emergence theory. For example, Crutchfield et al.'s computational mechanics theory considered causal states, which are the partitions of the state space and may have good predictions. Furthermore, Seth et al. proposed G-causality to quantify emergence using Granger's causality. We discuss them in detail below.

\subsection{Computational Mechanics}
The theory of computational mechanics proposed by Crutchfield, Shalizi, and Feldman et al. tried to formulate this kind of emergent causation in a quantitative framework~\cite{crutchfield1989inferring}. In some sense, computational mechanics can be understood as the inverse of statistical mechanics. This is because statistical mechanics derives macro-level consequences from micro-dynamics, whereas the inverse process is performed by computational mechanics, which constructs a minimal causal model from the observations of a stochastic process that can generate the observed time series.


Let us assume that the stochastic process under consideration can be represented as $\overleftrightarrow{s}$. We can divide it into two segments beginning from time step $t$: the history before $t$, denoted as $\overleftarrow{s_t}$, and the future after $t$, denoted as $\overrightarrow{s_t}$. If the process is stationary, we can remove the denotation of $t$. Thus, all possible histories $\overleftarrow{s}$ form a set, denoted as $\overleftarrow{S}$, and all futures $\overrightarrow{s}$ form a set, denoted as $\overrightarrow{S}$. 

We aim to establish a model that can reconstruct and predict observed random sequences, with a higher accuracy being desirable. However, the randomness of the sequences prevents us from achieving a perfect reconstruction unless we record every randomly occurring character. This would make the model excessively long. To preserve useful information as concisely as possible, we need a coarse-grained mapping that captures the ordered structure in the random sequences, known as patterns~\cite{crutchfield1994calculi}. We can partition $\overleftarrow{S}$ into mutually exclusive and jointly comprehensive subsets that form a set $\mathcal{R}$. Any subset $R\in\mathcal{R}$ is called a "state". 
  We define a function from histories to states as $\eta:\overleftarrow{\mathcal{S}}\rightarrow \mathcal{R}$. Thus, $\eta$ is a method that can partition the history into mutually exclusive and jointly comprehensive subsets.

For a set of states $\mathcal{R}$, we can measure its simplicity using a complexity metric. Intuitively, the larger the cardinality of $\mathcal{R}$, the more complex it is. Additionally, we need to consider its distribution. For example, if one state occurs frequently while others occur rarely, it is less complex compared to a situation with a uniform distribution. Therefore, we can define the \textbf{{statistical complexity}} $C_\mu$ of a set of states using Shannon entropy~\cite{shalizi2001causal}:
\begin{align}
\label{eq:statistical complexity}
C_\mu(\mathcal{R})\triangleq -\sum_{\rho\in \mathcal{R}} P(\mathcal{R}=\rho)\log_2 P(\mathcal{R}=\rho).
\end{align}
When constructing predictive models using a set of states, the statistical complexity refers to the size of the model.

What kind of state set can achieve the best balance between predictiveness and parsimony? We can introduce an equivalent relationship called \textbf{{causal equivalence}}~\cite{crutchfield1994calculi}. Concretely, we say $\overleftarrow{s}$ and $\overleftarrow{s}'$ are causally equivalent if and only if:

$$P(\overrightarrow{s}|\overleftarrow{s})=P(\overrightarrow{s}|\overleftarrow{s}')$$

This equivalent relationship can partition all the histories into equivalent classes, and they are defined as \textbf{{causal states}}. We denote all the causal states of a history $\overleftarrow{s}$ as $\epsilon(\overleftarrow{s})$, where $\epsilon: \overleftarrow{\mathcal{S}}\rightarrow 2^{\overleftarrow{\mathcal{S}}}$ is a function that can map a history $\overleftarrow{s}$ to the causal state $\epsilon(\overleftarrow{s})\in 2^{\overleftarrow{\mathcal{S}}},$ which is a subset of the histories.

For any two causal states $\mathcal{S}_i$ and $\mathcal{S}_j$, we can define \textbf{{causal transitions}} as a set of labeled probabilities $T^{(s)}_{ij}$ that represent the transition from the causal state $S_i$ to the causal state $S_j$ while emitting the symbol $s\in \mathcal{A}$~\cite{crutchfield1994calculi}:

$$T^{(s)}_{ij} \equiv\text{P}(\mathcal{S}'=\mathcal{S}_j,\overrightarrow{\mathcal{S'}}^{L=1}=s|\mathcal{S}=\mathcal{S}_i),$$ 
where $\mathcal{S}$ is the current causal state, $\mathcal{S}'$ is its successor, and $\overrightarrow{\mathcal{S'}}^{L=1}=s$ denotes all the sequences of length $L=1$, in which the first symbol emitted is $s$ (we use 
 slightly different symbols from those in~\cite{shalizi2001causal}).~Hence, by combining: $\mathcal{S'}=\mathcal{S}_j$ and $\overrightarrow{\mathcal{S'}}^{L=1}=s$, we ensure that all subsequent causal states will have the identical initial emitted symbol $s$. We denote the set $\{T^{(s)}_{ij}:s\in \mathcal{A}\}$ by $T$. The definition of causal transitions leads to a direct conclusion: $T^{(s)}_{ij}=\text{P}(\overleftarrow{s}s\in \mathcal{S}_j|\overleftarrow{s}\in \mathcal{S}_i)=\frac{\text{P}(\overleftarrow{s}\in\mathcal{S}_i,\overleftarrow{s}s\in \mathcal{S}_j)}{\text{P}(\overleftarrow{s}\in\mathcal{S}_i)}$, where $\overleftarrow{s}s$ is read as the semi-infinite sequence obtained by concatenating the history $\overleftarrow{s}$ and the symbol~$s$~\cite{shalizi2001causal}.

The {\textbf{$\epsilon$-machine}} of a process is defined as the ordered pair $\left\{ \epsilon, T \right\}$, where $\epsilon$ is the causal state function (which can map a state $s$ to the partition $\epsilon(s)$), and $T$ is the set of the transition matrices~(the dynamics) for the states defined by $\epsilon$.

Up until now, we have defined $\epsilon$-machine, which is a pattern discovery machine where the patterns are unraveled from the set of histories. The remaining task is to show that it is, in some sense, optimal. It has the important properties of being minimally predictive, maximally statistically complex, and minimally stochastic~\cite{shalizi2001causal}.

We compare the causal state $\mathcal{S}$ to any state $R\in 2^{\overleftarrow{\mathcal{S}}}$ to show that for the purpose of predicting the future, the causal states do a better job, i.e., they provide more information. This property can be formulated as a mathematical theorem, called \textbf{{the maximal predictability theorem}}. It is stated as follows: if $\mathcal{S}$ is the causal state given by $\epsilon$, then for any other state $R\in 2^{\overleftarrow{\mathcal{S}}}$ and all $L\in\Bbb{Z}^+$, we have:
$$H[{\overrightarrow{\mathcal{S}}}^L|R]\ge H[{\overrightarrow{\mathcal{S}}}^L|\mathcal{S}].$$ 
where $H(\cdot|\cdot)$ is the conditional entropy, and $\overrightarrow{\mathcal{S}}^L$ is the $L$-length sequences for future. Thus, the inequality shows that the uncertainty of $\overrightarrow{\mathcal{S}}^L$ given the causal state $\mathcal{S}$ is less than all other~states.

After achieving optimal predictiveness, the causal state set remains the one with minimal statistical complexity. We first introduce the notion of \textbf{{prescient rivals}} denoted as $\hat{\mathcal{R}}$, which are the states that are as predictive as the causal states; viz., for all $L\in\Bbb{Z}^+$, $H[{\overrightarrow{\mathcal{S}}}^L|\hat{\mathcal{R}}]=H[{\overrightarrow{\mathcal{S}}}^L|\mathcal{S}]$.

\textls[-15]{Next, we present the \textbf{{minimum statistical complexity theorem}}: for all prescient rivals $\hat{\mathcal{R}}$,}
\begin{align}
    C_\mu(\hat{\mathcal{R}})\ge C_\mu(\mathcal{S})
\end{align}
Next, we show that the causal states are \textbf{{minimally stochastic}}. That is to say, compared with other competitors with the same ability to predict the future, the causal states and their transition dynamics have the least uncertainty. Then, we have the \textbf{{minimal stochasticity theorem}}, which is expressed as follows: for all prescient rivals $\hat{\mathcal{R}}$,
$$H[\hat{\mathcal{R}'}|\hat{\mathcal{R}}] \ge H[\mathcal{S}'|{\mathcal{S}}].$$
where $\mathcal{S}'$ and $\hat{\mathcal{R}}'$ are the next causal state of the process and the next state, respectively. This means that the causal state and $\epsilon$-machine provide the best intrinsic determinism.

Since the causal state set is considered the best, how can we compute the causal states and $\epsilon$-machine from the observed data? The authors of \cite{crutchfield1994calculi} introduced a \textbf{{hierarchical machine reconstruction algorithm}}; {however, the details are not reiterated here.}

Although this algorithm may not be applicable to all operational scenarios, the authors presented numerical computational results and corresponding machine reconstruction pathways for chaotic dynamics, hidden Markov models, and cellular automata as examples~\cite{crutchfield1994calculi,shalizi2001causal}. 

It is interesting to compare the theory of computational mechanics with causal emergence. Indeed, we can understand that all the histories $\overleftarrow{s}$ are micro-level states, and all the states $R\in\mathcal{R}$ are macro-states. Thus the function $\eta$ that can map a history $\overleftarrow{s}$ to a state $R$ is a possible coarse-graining strategy.

It is worth pointing out that the causal state $\epsilon(\overleftarrow{s})$ is the special state that can have at least the same predictive power as the micro-state $\overleftarrow{s}$, i.e., the full history. Therefore, $\epsilon$ is similar to the notion in the effective coarse-graining strategy in~\cite{zhang2022neural} (see Section~\ref{sec:causalemergence_identification}), and the causal transitions $T$ represent the corresponding effective macro-level dynamics. The feature of minimal stochasticity characterizes the deterministic property of the macro-dynamic. This property is characterized by effective information (EI) in causal emergence theory.

Although a clear definition and quantitative theory of emergence were not provided, the \textls[-15]{authors discussed the relationship between computational mechanics and emergence~\mbox{\cite{crutchfield1994calculi, shalizi2003macrostate}}}. In \cite{crutchfield1994calculi}, the authors explained that emergence can be conceptualized as a dynamic process in which a pattern acquires the ability to predictably adapt to different environments, as observed by an external observer. Additionally, they differentiated intrinsic emergence from emergence itself, as intrinsic emergence goes beyond the mere production of patterns and encompasses the formation of an embedded observer within the system through these~patterns. 

\subsection{G-Emergence Theory}

G-emergence theory, proposed by Seth in 2008~\cite{seth2008measuring}, is one of the earliest works on a quantitative measure of emergence. His basic idea is to use nonlinear Granger causality to measure weak emergence in complex systems.

Granger causality (G-causality) is formally defined as follows: Given two time series $A$ and $B$, if the past values of $B$ can help predict the future values of $A$, beyond what can be forecasted using the past values of $A$,  then $B$ is said to be ``Granger-cause'' of $A$. This implies the existence of Granger causality between $A$ and $B$. 

\begin{figure}[H]

\includegraphics[width=10cm,height=4cm]{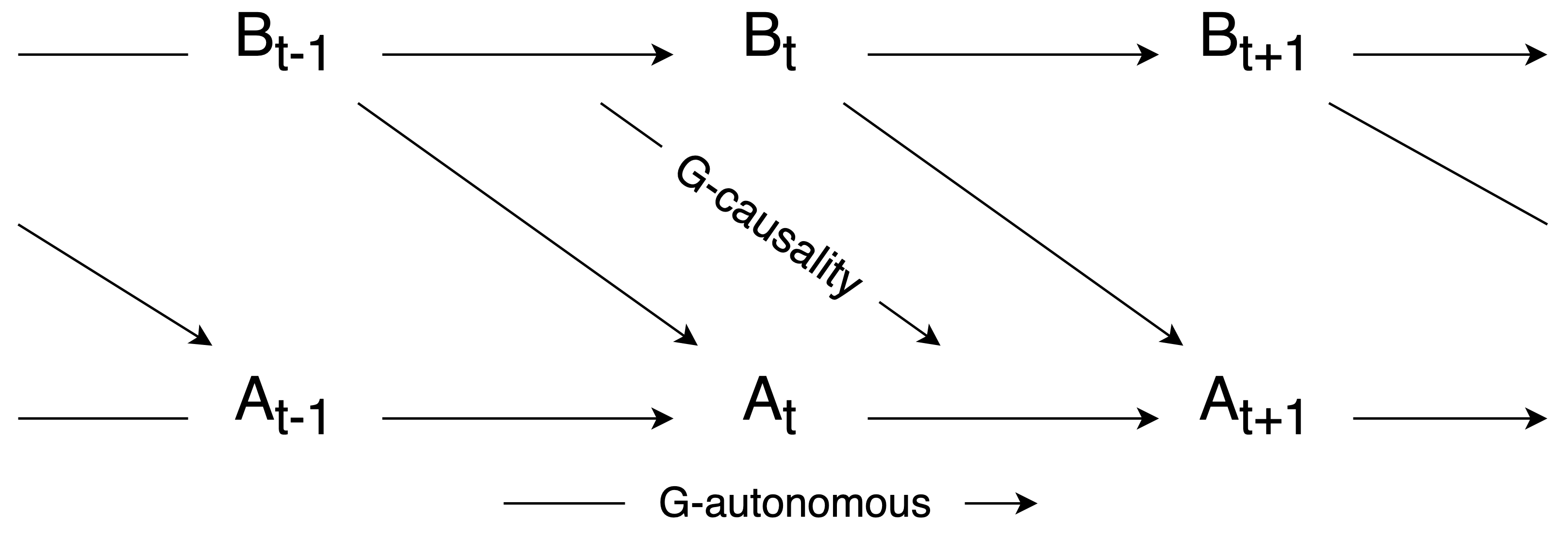}
\caption{G-causality and G-autonomous.}
\label{fig.gemergence}
\end{figure}

When applying a bivariate autoregressive model for predictions, residual terms are included in the equations of the two variables. Then, the residuals can be utilized to quantify the extent of the causal effect in G-causality. The degree of $B$ being the G-cause of $A$ is quantified by the logarithmic of the ratio of the two variances of residuals. One is the residual of the autoregression model of $A$ if all the terms of $B$ are omitted, and the other is the residual of the full prediction model, as shown in Figure \ref{fig.gemergence}.  In addition, in~\cite{seth2008measuring}, the author also defined G-autonomous as the ability of past values in a time series to predict its own future values. And the degree of G-autonomous can be measured in a similar way as G-causality.

\vspace{-2pt}

\begin{figure}[H]

\includegraphics[width=10cm,height=3cm]{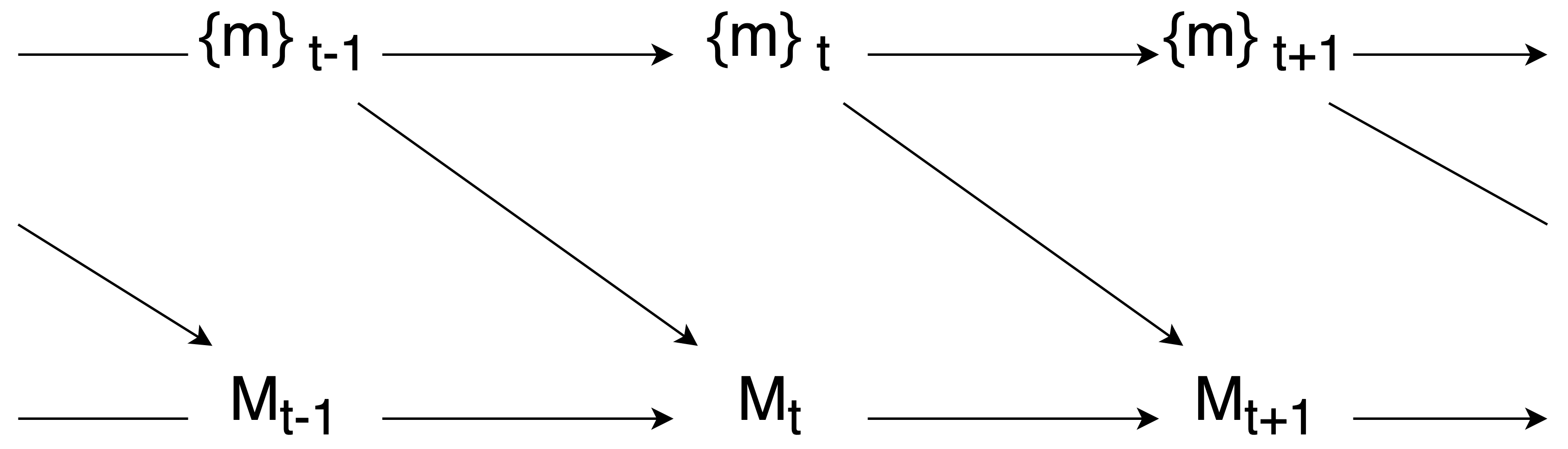}
\caption{G-autonomous.}
\label{fig.gautonomous}
\end{figure}

With these two basic notions in G-causality, as shown in Figure \ref{fig.gautonomous}, the author defined G-emergence on macro-variables as follows: a macro-variable $M$ is G-emergent from a set of micro-variables $\{m\}$ if and only if (i)  $M$ is G-autonomous with respect to $\{m\}$, and  (ii) $\{m\}$ is G-cause of $M$. The degree of G-emergence can also be quantitatively measured by multiplying the degree of G-autonomous of $M$ and the average of G-causes of $\{m\}$ on $M$.

\vspace{-2pt}

The author tested the G-emergence theory on the Boid model. This model is a famous artificial life model that simulates the flocking behavior of birds using three simple rules: cohesion, alignment, and separation~\cite{reynolds1987flocks}. These basic rules are realized by the virtual forces exerted on each bird, and the strength of each force can be controlled. The author found that by increasing the strength of cohesion, the G-emergence of the entire flock also increased. Here, the macro-variable is selected as the center of the flock (center of mass), and each bird is a micro-variable. The author also discovered a downward causality phenomenon in this simple model: the center of mass can be used to predict each individual bird. However, the author did not distinguish downward causation from other common causality in their work because Granger's causality is not a real causal relationship.


Seth's G-emergence theory was the first attempt to quantify the emergence phenomenon via a causality measure. However, the causality measure that the author used was Granger causality, which is not a strict causal measure, and it also depends on the regression method to be used. Furthermore, the measure is defined on variables but not dynamics, which means the result depends on the selection of variables.

\subsection{Other Quantitative Theories of Emergence}
There are alternative means of measuring emergence that do not require relying on causality. Two different methods have been discussed by the scholars. One is to understand emergence as a process from disorder to order, and the other is to understand emergence from the perspective that ``the whole is greater than the sum of parts''. 

For example, Moez Mnif and Christian Müller-Schloer used Shannon entropy~\cite{mnif2011quantitative} to measure order and disorder. In a self-organized process, emergence occurs when there is an increase in order. This increase can be quantified by measuring the Shannon entropy difference between the initial state and the final state, denoted as $H_{start}-H_{end}$. However, this definition has two limitations. Firstly, the measurement of entropy, denoted as $H$, is dependent on the abstract level of observation. Therefore, it is necessary to account for the entropy increase resulting from a change in the observer's abstract level. Secondly, the choice of the initial condition of the system is arbitrary. To address this limitation, one approach is to measure the relative level of Shannon entropy compared to the maximum entropy distribution. Finally, emergence can be quantified by:
\begin{equation}
\label{eq.emergence_shannon}
    Emergence = H_{max}-H-\Delta H_{view},
\end{equation}
where $H_{max}$ is the maximized entropy of the system. If no prior information is available, the maximum entropy, denoted as $H_{max}$, corresponds to the Shannon entropy of an equal probability distribution. On the other hand, $H$ represents the Shannon entropy of the system at the final moment of a self-organization process. Additionally, $\Delta H_{view}$ represents the entropy increase during this process resulting from a change in the observer's abstraction level. It should be noted that if the observer does not alter their abstraction level, this term would be zero. We can also normalize this quantity by dividing $H_{max}$ such that different features can be compared by the normalized quality. For a multivariate system, the authors suggested employing a radar plot to visualize the emergence fingerprint of the dynamical process, which showcases the normalized emergence measure across various variables. Then, the authors applied this method to a simulated system of chickens. M. Tang and X. Mao applied this indicator to artificial society models~\cite{tang2014information}.

Inspired by Moez Mnif and Christian Müller-Schloer's work, ~\cite{fisch2010quantitative} suggested using the divergence measure between two probability distributions to better quantify emergence. They understood emergence as being \textit{an unexpected or unpredictable change of the distribution underlying the observed samples 
}. However, this method suffers from computational complexity and estimation accuracy. To address these problems, \cite{fisch2012techniques} further proposed an approximating method using the Gaussian mixture model to estimate the density and introduced Mahalanobis distance to characterize the divergence between data and Gaussian components, leading to better results. In~\cite{fisch2011divergence}, the authors systematically compared the three aforementioned methods and applied them to a simple test example. Another Shannon entropy-based emergence measure was proposed by Holzer and de Meer et al.~\cite{holzer2008autonomy, holzer2011methods}. They considered a complex system as a self-organization process in which different individuals interact with communications. Emergence can then be measured based on a ratio between the Shannon entropy measure on all communications between agents and the total summation of the Shannon entropy for each communication as a separate source. 

Unlike the aforementioned methods,~\cite{teo2013formalization, szabo2015formalization} proposed a method to quantify emergence based on the idea that ``the whole is greater than the sum of its parts'',  defining emergence from the interaction rules and states of agents instead of the overall statistical measure of the whole system. Specifically, this measure consists of two terms that are subtracted from each other. The first term characterizes the collective states of the entire system, whereas the second term represents the summation of the individual states of all its constituent parts. This measure emphasizes the emergence that arises from the interactions and collective behavior of the system. This method was then tested on the example of bird flock simulation.

\subsection{Erik Hoel's Causal Emergence Theory}

\subsubsection{Basic Idea}

The first quantitative emergence theory based on Markov dynamics and causality measures through intervention was Erik Hoel's causal emergence theory. 

\begin{figure}[H]

\includegraphics[width=10cm]{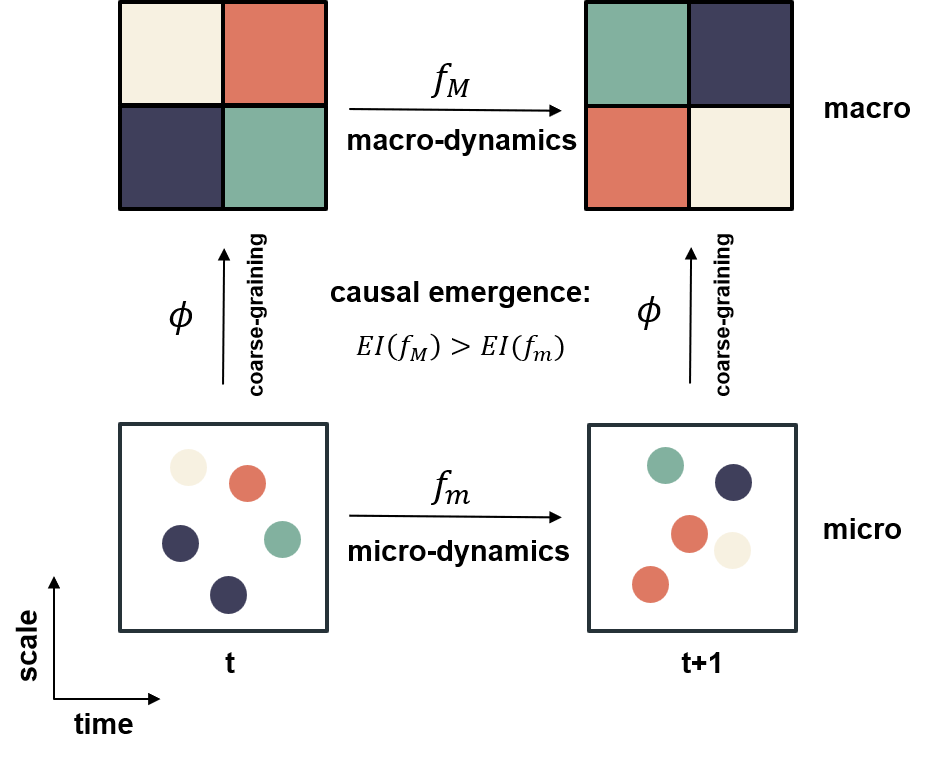}
\caption{The 
 basic idea of Erik Hoel's causal emergence theory. The colored circles represent micro-states, whereas the colored squares represent macro-states.} 
\label{fig:causal_emergence_illustration}
\end{figure}

In this framework, system properties can be characterized at various levels, ranging from micro to macro. If a system exhibits stronger causality at the macro-level than at the micro-level, it demonstrates causal emergence. Causality is reflected between successive states during the system's evolution. The strength of a system's causality reveals the extent to which its future state is influenced by its current state. The basic idea of causal emergence is illustrated in Figure \ref{fig:causal_emergence_illustration}.

For example, statistical mechanics is a typical theory of causal emergence. At the micro-level, a huge number of molecules collide and exhibit random behaviors such that probabilistic language must be used to describe them. However, if we coarse-grain the whole system into several thermodynamic physical variables like pressure, temperature, etc., we can use very concise and exact thermodynamic equations to describe their behaviors. Therefore, thermodynamic laws have a stronger causal effect than micro-level molecular dynamics. 

Formal tools are used to describe the elements in Hoel's causal emergence theory. Typically, it employs discrete Markov models to describe the micro-dynamics of systems, and the corresponding macro-level systems with different Markov dynamics can be derived by coarse-graining the micro-systems. Additionally, the inherent strength of the causal effect of the Markov model can be measured with effective information (EI), indicating how effectively a particular state influences the future state of a system.

EI is an intrinsic property of a system's dynamics and can be quantified by the transitional probability matrix (TPM). A coarse-graining strategy is a function that maps a set of micro-states into a particular macro-state, allowing for the derivation of a new dynamical model described by TPM from the micro-level TPM. The effective information of the coarse-grained model can also be computed. The phenomenon of causal emergence implies that as we coarse-grain microscopic states, the amount of effective information transmitted from the current state to the next state can possibly increase. At a certain macroscopic scale of coarse-graining, the effective information reaches a maximum; this scale represents the point at which the system state has the maximum causal power to predict future states in the most reliable and effective way.

To illustrate causal emergence, let us look at a particular Markov chain model with four possible states, whose transition probability matrix (TPM) is:
\begin{equation}
    S_m=
\begin{bmatrix}
1/3 & 1/3 & 1/3 & 0 \\
1/3 & 1/3 & 1/3 & 0 \\
1/3 & 1/3 & 1/3 & 0 \\
0 & 0 & 0 & 1
\end{bmatrix}
\end{equation}

In this model, if we take the first three states as group 1 and the last state as group 2, anyone in group 1 can transition to any of the three states in the same group with equal probability at the next moment. However, because there is only one state in group 2, so the fourth state will always stay in its position. Intuitively, we can conclude that the future states are not fully determined by the current states, and uncertainty mainly arises from group 1. 

However, if we merge the first three states in group 1 into one new state and keep the fourth state in group 2 as it is, we obtain a coarse-grained Markov model with two states, represented by the following~TPM:
\begin{equation}
    S_M = 
\begin{bmatrix}
1 & 0 \\
0 & 1
\end{bmatrix}
\end{equation}

Now, the future states of the new system can be fully determined by the current states. This shows that we can eliminate the uncertainty of a nondeterministic system by performing coarse-graining over the system states. 

\subsubsection{About Coarse-Graining}

The example above illustrates how a system's determinism can increase as it is coarse-grained. Coarse-graining is a process that simplifies the description of a system by grouping its components into larger, more slowly varying units. It is often used to identify the essential features of a system that determine its macroscopic behavior, without being burdened by the details of micro-scale interactions. Unlike dimension reduction techniques, such as PCA and SVD, coarse-graining takes into account the system's features at different spatial and temporal scales. Coarse-graining also differs from renormalization. In physics, the renormalization group method was invented to eliminate infinities in integrals. It is also used to coarse-grain a system such that the Hamiltonian or partition functions are similar between micro- and macro-levels~\cite{christensen2005complexity,mccomb2003renormalization}. However, the coarse-graining method does not have this requirement in general. Although both techniques are designed to describe a system from a coarse-grained level, coarse-graining focuses more on the states of the system, whereas renormalization cares about dynamics, system rules, partition functions, etc. Renormalization is frequently used in physics and the study of phase transitions to explore critical phenomena and symmetry breaking. These are aspects that coarse-graining typically does not consider.

Given a transitional probability matrix (TPM), there are many ways to partition the state space for coarse-graining the system. What kind of coarse-graining strategies are more reasonable? In~\cite{hoel2013quantifying,hoel2017when}, the authors suggested selecting an appropriate coarse-graining method by maximizing the effective information of the coarse TPM. In the literature, there is another criterion for selecting a coarse-graining strategy called lumpability, where the lumpability of a TPM refers to the coarse-grained TPM exhibiting similar dynamics and observation statistics~\cite{kemeny1983finite}. Commutativity is another requirement that requires the operations of coarse-graining and state transitions to be commutative~\cite{pfante2014comparison}. There are also other discussions or reasonable model reductions of Markov or hidden Markov models~\mbox{\cite{kotsalis2008balanced,white2000lumpable,wolpert2014optimal}}.

\subsubsection{About Effective Information}
\label{sec:EI}
In order to quantify emergence, a widely used method is to calculate the mutual information between the past states and future states of a system. This mutual information defines the upper limit of the information being transferred from the past to the future. If the mutual information is high, it suggests that a significant amount of information about the past is retained in the future. 

\begin{equation}
I(X_t; X_{t+1)} = \sum P(X_t, X_{t+1})\log\frac{P(X_t, X_{t+1})} { P(X_t)P(X_{t+1})}
\end{equation}

One limitation of utilizing mutual information is that the resulting value may fluctuate in response to changes in the joint probability of data. This can make it difficult to obtain a consistent outcome if the system is exposed to different inputs.

\subsubsection*{Effective 
 Information (EI)}
 
To address this limitation, Hoel introduced the concept of effective information to measure the causal effect of the current state on the future state of a system. Effective information is a scoring metric based on mutual information and can be calculated from the system's transition probability matrix (TPM), which is invariant to the input data.


By ``intervening'' in the current state distribution to follow the uniform distribution (the distribution of maximum entropy), denoted as $I_D$, the TPM enables the prediction of the future state distribution $E_D$ at the next moment. The effective information ($EI$) of the system is defined as the mutual information between $I_D$ and $E_D$, which can be expressed as

\begin{equation}
\label{eq:do_def_ei}
    EI \equiv I(I_D; E_D) = I(X_t; X_{t+1})|_{do(X_t) \sim U}
\end{equation}

The probability distribution of the random variable $X_t$, representing the initial state, is denoted as $do(X_t) \sim U$, indicating that it follows a uniform distribution. Hoel adopted the notion of the ``do'' operator from Judea Pearl's causal analysis framework~\cite{hoel2017when}.  It is worth noting that in Pearl's context, the do operator is typically used to assign a specific value to the intervened variable rather than applying a distribution~\cite{pearl2009causality}. 

Another point that needs clarification is that the ``do'' operator used here is purely a mathematical construct that specifies the distribution of $X_t$. It does not imply the need for actual intervention in the system. However, the imaginary ``do'' operation is equivalent to a real intervention in the context of this scenario, given that the dynamical mechanism (TPM) is provided. Consequently, we can perform any desired intervention on the system, just as in computer simulations.

The second issue pertains to why we apply the ``do'' operator to a uniform distribution. In \cite{hoel2013quantifying}, Hoel et al. claimed that the distribution should be the maximum entropy distribution, which is the most reasonable selection for the input variable $X_t$ if we have no prior information about the input variable~\cite{jaynes1957information}. As we know, uniform distribution can be derived by maximizing entropy if there is no constraint.  

The authors of this review believe that applying the ``do'' operator to a uniform distribution ensures that the objective measured by EI solely reflects the dynamical mechanism itself, i.e., the TPM, and is independent of any input data.

This point can be clearer if we re-express EI as a function of the TPM of the system (see Appendix \ref{sec:appendix_1} for the detailed derivation):

\begin{equation}
    EI = \frac{1}{N} \sum_{i,j} TPM(i,j)\log_2\Big(\frac{N \times TPM(i,j)}{\sum_k TPM(k, j)}\Big),
\label{eq:EI}
\end{equation}
where $TPM(i,j)$ represents the transitional probability of the system from the state $i$ to the state $j$. 

Therefore, effective information offers a solution to the aforementioned limitation of mutual information. Given that the TPM can capture the inherent nature of a system, EI is likewise an intrinsic property of the system. 

However, if we apply the ``do'' operator to the system using different distributions, EI will depend on the chosen distribution, and certain transition probabilities of specific rows may carry greater weight in the average calculation. Consequently, deriving a simple expression solely based on the transitional probability matrix (TPM) may not be possible. Furthermore, if there is no ``do'' operator, the effects of the dynamical mechanism and the input distribution in EI would not be distinguishable or separated, as the mutual information is a function of the input distribution $p(X_t)$.


\subsubsection*{The Derived Measures of EI}
Furthermore, we can calculate the EIs for the TPMs of both macro- and micro-dynamics, and their difference is defined as the measure of causal emergence, that is,

\begin{equation}
    \label{eq.causal_emergence_dfn}
    CE=EI(TPM_M)-EI(TPM_m),
\end{equation}
where $TPM_M$ represents the TPM of macro-level dynamics, and $TPM_m$ represents the TPM of micro-level dynamics. $CE$ measures the degree of causal emergence. If $CE>0$, then causal emergence occurs; otherwise, it does not occur.

However, there is a limitation for EI and CE when comparing two dynamics that differ significantly in size. This is because the value of effective information relies on the number of possible states within the system, denoted as $N$, with an upper bound of $\log_2(N)$. To facilitate comparisons between different coarse-graining strategies and scales, effective information is often normalized, resulting in a metric known as the effect coefficient $Eff$.
\begin{equation}
\label{eq:eff}
    Eff = \frac{EI}{\log_2(N)}
\end{equation}

The value of the effect coefficient is always between $0$ and $1$, representing the proportion of effective information being transferred from current states to future states. If the information is fully transferred, the effect coefficient is $1$. 

In addition to characterizing causal emergence, the effect coefficient can be further broken down into two meaningful components: ``$Determinism$
'', which represents the certainty of a current state evolving into a certain state or diverging into multiple states in the next moment, and ``$Degeneracy$'', which represents the possibility of multiple current states converging into one state in the next moment.
\begin{equation}
\label{eq:det_deg}
    Eff = Determinism - Degeneracy 
\end{equation}
where both ``$Determinism$'' and ``$Degeneracy$'' can be defined in terms of the TPM (refer to Equations (\ref{eq:EI}) and (\ref{eq:eff})):
\begin{equation}
\label{eq:det_deg_def}
\begin{aligned}
    Determinism &= \frac{1}{N\log_2(N)} {\sum_{i,j}{TPM(i,j)\log_2(N\times TPM(i,j))}} \\
    Degeneracy &= \frac{1}{N\log_2(N)} {\sum_{i,j}TPM(i,j)\log_2(\sum_k{TPM(k,j)})}
\end{aligned}
\end{equation}

It should be noted that the determinism of a system is always greater than its degeneracy, as the lower bound of the effect coefficient is $0$. The following examples illustrate what determinism and degeneracy look like in systems with varying TPMs.

\begin{figure}[H]
\centering
\includegraphics[width=1\linewidth]{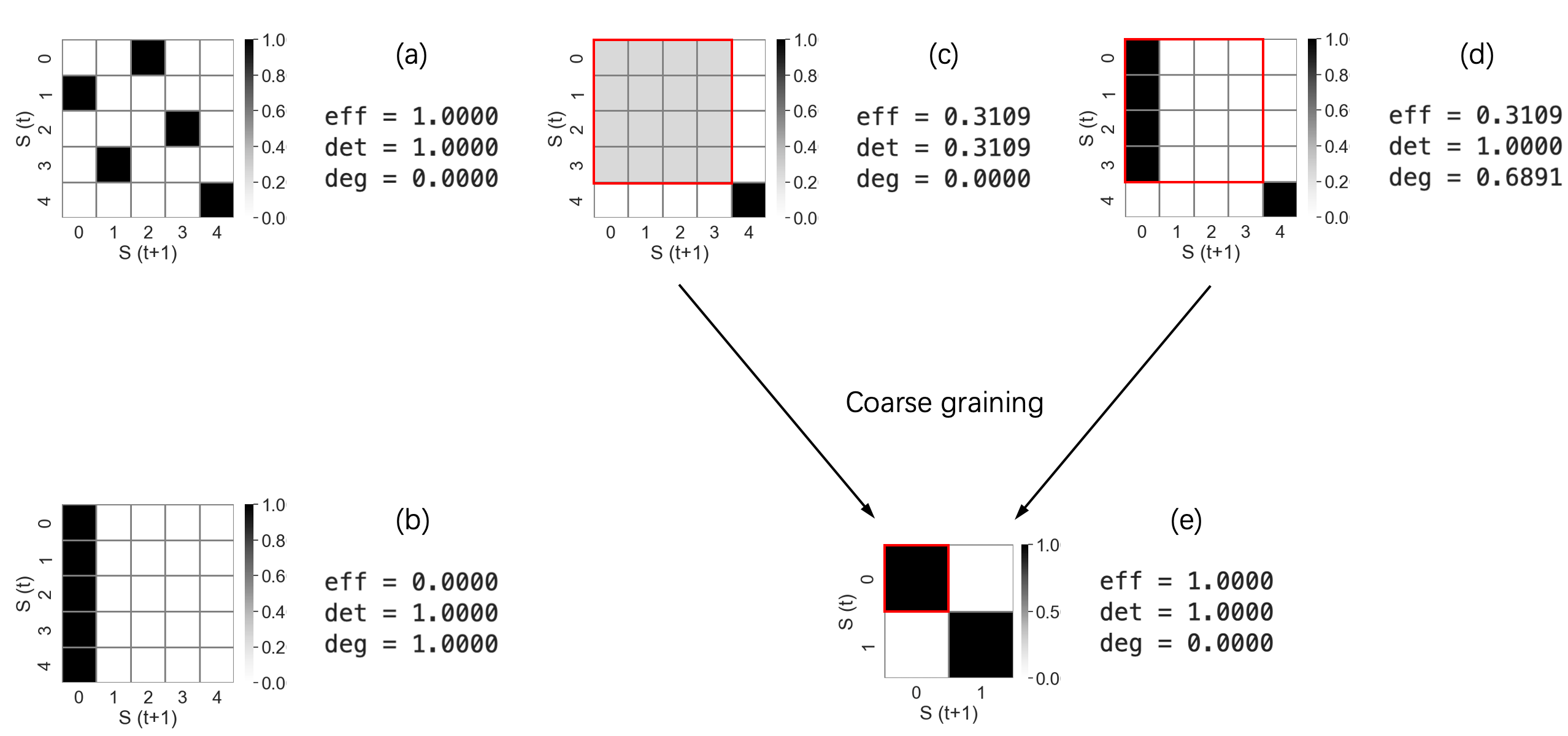}
\caption{Examples of effect coefficient, determinism, degeneration, and how coarse-graining can change them. (\textbf{a}) Deterministic system with zero degeneracy; (\textbf{b}) Deterministic system with zero effect coefficient due to high degeneracy; (\textbf{c}) Partially deterministic system; (\textbf{d}) Deterministic system with low effect coefficient; (\textbf{e}) Coarse-grained system from (\textbf{c}) or (\textbf{d}), resulting in a fully deterministic system with a high effect coefficient.}
\label{fig:ei_decomposition}
\end{figure}

In Figure \ref{fig:ei_decomposition}, the square cells represent the elements of the TPM, and the grayscale areas represent the values of the TPM elements. Example (a) is a bijective system, meaning that all information from current states can be transferred to future states without loss. It is fully deterministic with zero degeneracy. Example (b) is an extreme case where all current states lead to only one future state, illustrating that high determinism does not necessarily imply a high effect coefficient. Systems (c) and (d) differ but have the same effect coefficients. Finally, system (e) is a coarse-grained version of either system (c) or (d), demonstrating two important points: first, different microscopic systems can be coarse-grained to the same macroscopic system, and second, causal emergence can be quantitatively captured from coarse-graining by the increased effect coefficient.

One limitation of EI is its global nature, as highlighted by the fact that the summation of additional terms encompasses the entire state space, as defined in Equation (\ref{eq:EI}). Therefore, \cite{varley2022flickering} proposed the concept of flickering emergence, which decomposes EI into each term added in Equation (\ref{eq:EI}). This new feature can characterize the local properties of Markov dynamics.

\vspace{-9pt}
\subsubsection*{Comparison with Other Measures of Causation}

\label{sec:measure_comparisons}
Hoel's theory of causal emergence is based on effective information, however, is the selection of EI necessary for measuring causation? Is causal emergence a phenomenon that depends on the selection of the measure of causation? To address this problem, in Comolatti and Hoel's work~\cite{comolatti2022causal}, a systematic comparison was conducted between effective information (EI) and other measures of causation that are widely applied in various fields, ranging from philosophy to genetics.

The findings of this comparison revealed that causal emergence is not merely a peculiar phenomenon limited to a specific measure. Instead, it exhibits commonalities and shared characteristics across different measures and fields of study. EI is not the sole measure for capturing causal emergence; there are other measures of causation that can also reveal the phenomenon of causal emergence. This suggests that the concept of causal emergence has broader applicability and relevance. We introduce this work in detail below.

Firstly, Hoel highlighted that causation is not merely a singular relationship between a cause and an effect. Instead, it encompasses two fundamental dimensions known as \textit{causal primitives}: sufficiency and necessity.

The sufficiency aspect of causation refers to the scenario where the occurrence of the cause guarantees the occurrence of the effect. In other words, whenever the cause happens, the effect is also observed. This sufficiency dimension, denoted as $suff$, can be formally defined as the probability of the effect $e$ occurring given the condition that the cause $c$ has~occurred:
\begin{equation}
    suff(e, c)=P(e|c).
\end{equation}

In addition, a necessary ($nec$) relation in causation refers to the absence of the effect implying the absence of the cause. In other words, when the effect does not occur, it indicates that the cause also does not occur. This can be understood as a kind of causal effect measure for the counterfactual. This necessary dimension, denoted as $nec$, can be quantitatively defined as the probability that the effect $e$ does not occur when the condition that the cause $c$ has not occurred is given by:
\begin{equation}
\label{eq:nec}
    nec(e, c)=1-P(e|C/c),
\end{equation}
where $C/c$ represents the set of all possible causes in $C$ but with the particular cause $c$ being excluded. Therefore, $P(e|C/c)$ represents the probability that $e$ occurs if other causes except $c$ occur.

With these causal primitives defined, Hoel compared different measures of causation, including Hume's constant conjunction~\cite{hume2016enquiry}, Cheng's causal attribution~\cite{cheng1991causes}, Eells's measure of causation as probability raising~\cite{eells1991probabilistic}, Suppes's measure of causation as probability raising~\cite{suppes1973probabilistic}, and Judea Pearl's measures of causation~\cite{pearl2009causality}. The finding was that all of these measures can be expressed as the two causal primitives.

For example, Cheng's causal attribution can be expressed as:
\begin{equation}
    CS_{cheng}=\frac{suff(e,c)+nec(e,c)-1}{nec(e,c)}
\end{equation}

With this understanding, a natural question arises: Can effective information (EI) be expressed using causal primitives? The answer is affirmative. However, to clarify this point, two important distinctions need to be made. Firstly, in most measures of causation, the causal variables are binary, whereas EI is defined for variables with multiple values. Secondly, EI is an information-theoretic measure, whereas others are probabilistic measures. Despite these distinctions, EI can still be expressed using causal primitives.

To understand this, let us examine the equivalent measure of EI: normalized effective information ($Eff$) in Equation (\ref{eq:det_deg}).~This measure contains two terms. The first term is $determinism$, and the second one is $degeneracy$. These two terms can be expressed as $suff$ and $nec$:
\begin{equation}
    determinism = 1 - \frac{\sum_{c\in C}P(c)H(e|c)}{\log_2 N},
\end{equation}
where $H(e|c)\equiv -\sum_{e\in E}P(e|c)\log_2 P(e|c)$ is the Shannon entropy of the conditional probability $P(e|c)$, which is $suff(e,c)$, and $P(c), \forall c\in C$ is the distribution of all causes. In the definition of EI (Equation \ref{eq:do_def_ei}), this distribution is intervened as a uniform distribution such that equal weights are assigned to causes $c\in C$. It is not hard to see that $determinism$ is an information metric for $suff(e,c)$ and is averaged for all causes. Furthermore, as $suff(e,c)=P(e|c)$ increases and approaches one, indicating the emergence of causal effect, the value of $H$ decreases, whereas determinism concurrently increases. Thus, the determinism term in EI plays a similar role to that of $suff$ in causal primitives.

Another term is $degeneracy$, which can be re-written as:
\begin{equation}
    degeneracy = 1 - \frac{H(e|C)}{\log_2 N},
\end{equation}
where $H(e|C)\equiv -\sum_{e\in E}P(e|C)\log_2 P(e|C)$ is also the Shannon entropy of the conditional probability $P(e|C)$, which is calculated by averaging the causal effect for all elements in $C$ as $P(e|C) = \sum_{c\in C} P(c) P(e|c)$. The entropy $H(e|C)$ serves as a measure of the average causal effect for counterfactuals, as $P(e|C)$ can be interpreted as $P(e|C/c)$. This is due to the approximation $P(e|C)\approx P(e|C/c)$ when the number of elements in $C$ is significantly large, and $C$ certainly encompasses $C/c$. Consequently, $degeneracy$ in $EI$ acts as the counterpart to $nec$ in causal primitives.

Therefore, EI or $eff$ is a valuable measure of causation, particularly in cases where the cause and effect variables are not limited to binary values.

Comolatti and Hoel~\cite{comolatti2022causal} conducted additional experiments to investigate the phenomenon of causal emergence using various Markovian dynamics, employing different measures of causation, as discussed above. Their findings revealed the widespread occurrence of causal emergence, regardless of the specific measure of causation employed.

\subsubsection{Examples of Causal Emergence}
\label{sec:exampleCE}

\begin{figure}[H]

\begin{adjustwidth}{-\extralength}{0cm}
\centering 
\includegraphics[scale=0.37]{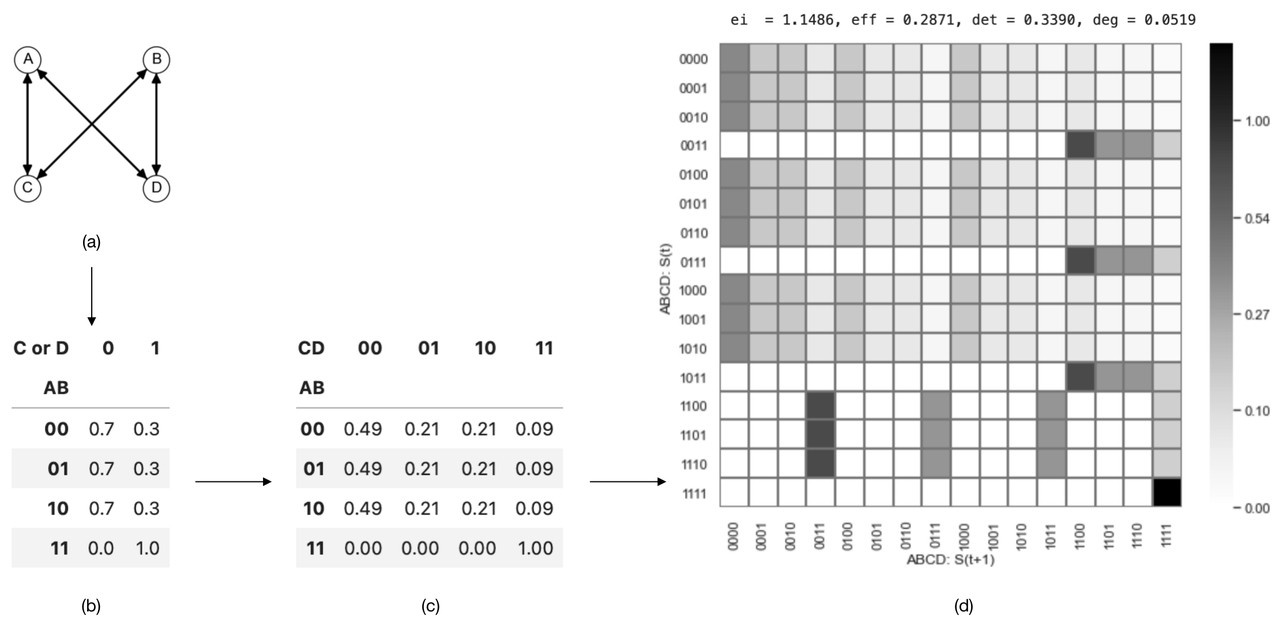}
\end{adjustwidth}
\caption{The mechanism and transition probability matrix of a boolean network. Each variable's next state is determined by the combination of the current states of two other variables, as shown in the boolean network (\textbf{a}). If we examine an atomic subsystem with two inputs and one output, such as AB \textrightarrow 
 C, the next state of C depends on the present states of A and B. If A = 0 and B = 0, the probability of C being 1 is 70\%, whereas the probability of it being 0 is 30\%, as indicated in the table (\textbf{b}). This same microscopic dynamics applies to other atomic subsystems such as AB \textrightarrow D, \mbox{CD \textrightarrow A}, and CD \textrightarrow B. When we expand to a subsystem of two inputs and two outputs, AB \textrightarrow CD and \mbox{CD \textrightarrow AB}, we obtain the transition probability matrix (TPM), as demonstrated in the table (\textbf{c}). Finally, we extend to the dynamics of the entire system ABCD \textrightarrow ABCD, whose TPM is represented as a heatmap~(\textbf{d}). The effective information and relative metrics of the entire system are as follows:
$EI_{mi} = 1.1486$, $Eff = 0.2871 $, $Determinism = 0.3390$, $Degeneracy = 0.0519$}
\label{fig:boolean}
\end{figure}

The example above demonstrates a coarse-graining strategy that is intuitive, as it involves aggregation at the state level. However, in reality, coarse-graining often occurs among variables, and the combination of these variables produces different states. Let us look at a more complicated example of a boolean network, given in~\cite{hoel2013quantifying}, and inspect the causal emergence when coarse-graining its state transition mechanism. A, B, C, and D are four binary variables whose state transition relations are shown in Figure~\ref{fig:boolean}a.

\begin{figure}[H]

\begin{adjustwidth}{-\extralength}{0cm}
\centering 
\includegraphics[scale=0.32]{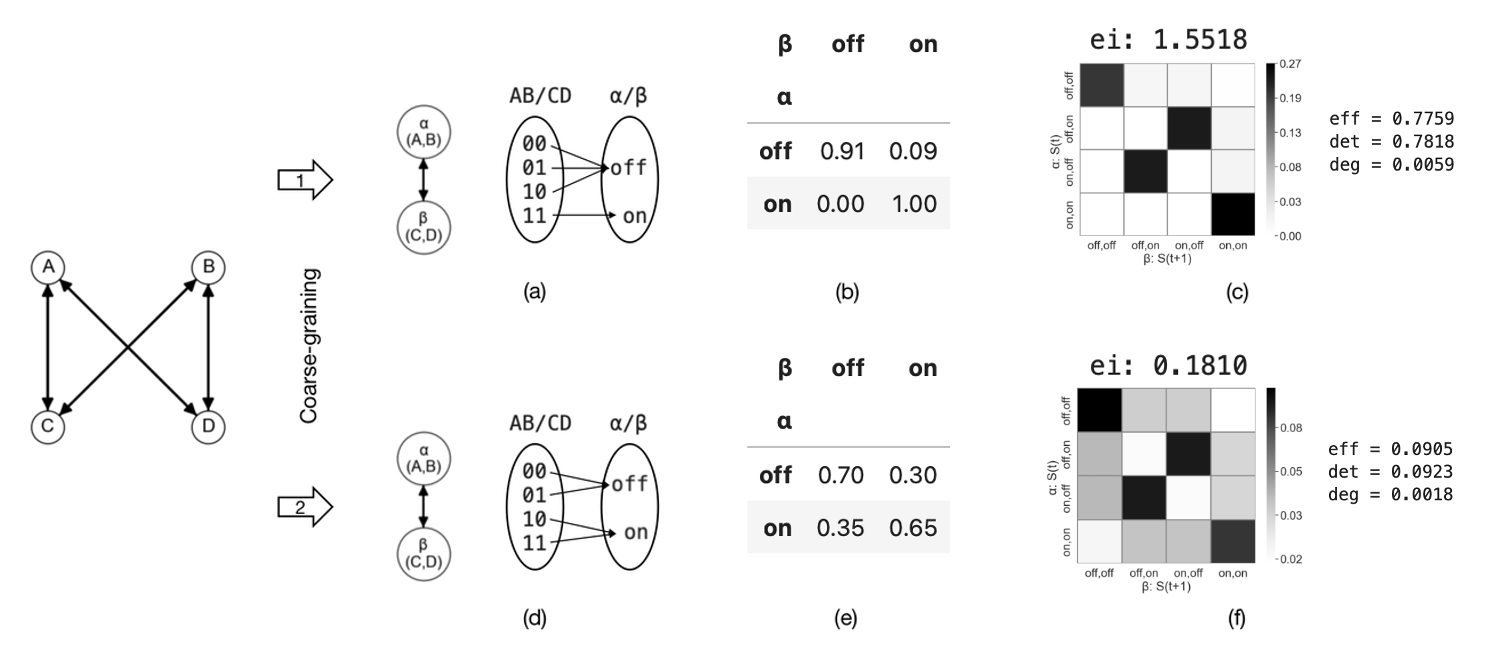}
\end{adjustwidth}
\caption{An example showing how different coarse-graining strategies applied to the same boolean network, shown in Figure \ref{fig:boolean}, can lead to different boolean dynamics. (\textbf{a}) Coarse-graining strategy 1. (\textbf{b}) Macro-dynamics resulting from strategy 1. (\textbf{c}) The corresponding heatmap of the TPM and the effective information metrics of the macroscopic system using method 1. (\textbf{d}) Coarse-graining strategy 2. (\textbf{e},\textbf{f}) The macro-dynamics and TPM corresponding to coarse-graining strategy 2 in (\textbf{d}), respectively.}
\label{fig:coarse_boolean}
\end{figure}

In order to demonstrate how different coarse-graining strategies can impact effective information, we can combine two micro-states into a single macro-state. For example, we can define $\alpha = (A,B)$ and $\beta = (C, D)$ as coarse-grained variables at the macro-level, where their values can be either `off' or `on'. Using various aggregation strategies, as illustrated in (a)/(b) and (d)/(e) in Figure \ref{fig:coarse_boolean}, we can construct the TPMs for the resulting coarse-grained systems. The TPM and the corresponding metrics are presented in Figure \ref{fig:coarse_boolean}c,f, respectively.

The effective information values for the two macroscopic systems are $EI_{ma1} = 1.55$ and $EI_{ma2} = 0.18$, respectively. It is evident that the effective information has increased in the first coarse-graining scenario ($EI_{ma1} (1.55) > EI_{mi} (1.15)$), whereas it has decreased in the second case ($EI_{ma2} (0.18) <EI_{mi} (1.15)$). Thus, causal emergence is observed in the first coarse-graining scenario. This implies that (1) proper coarse-graining can lead to causal emergence in a macroscopic system, and (2) not all coarse-graining approaches result in higher effective information. This example illustrates that coarse-graining can play an important role in emergence, and effective information is a well-defined quantification method. This gives us the opportunity to identify causal emergence in a given system. This raises the following question: What coarse-graining strategy can generate the maximum effective information? We address this question in Section \ref{sec:causalemergence_ai}. 

\vspace{-9pt}
\subsubsection{Extensions to Continuous Systems}

\label{sec:continuous_EI}



Effective information and its associated metrics have been shown to be useful in quantifying causal emergence. However, there are some limitations to be aware of. First, the method is only applicable to discrete-state systems. Second, knowledge of the system's transition probability matrix (TPM) is required to calculate the effective information metrics. Efforts have been made to apply or extend the concepts of effective information to continuous systems~\cite{chvykov2020causal,varley2020causal,zhang2022neural}.

\subsubsection*{Ordinal Partition Network}

One approach to incorporating effective information in continuous systems is to transform the continuous variables into discrete ones using the ordinal partition network (OPN) method~\cite{varley2020causal}. The basic idea is to transform the time-series data generated by a continuous dynamical system into a sub-time series of dimensionality $d$ with discrete values and then establish a transition probability matrix (TPM) between these discrete series. This approach involves three steps: firstly, sampling the original time series with a specified time interval $\tau$ to create sub-time series of dimensionality $d$; secondly, ranking these sub-series based on their values; and finally, utilizing the $d$-dimensional vectors of rank orders to represent the original sub-series. By employing rank orders, this method ensures that the values are restricted to integers within a certain range.

In \cite{varley2020causal}, the author applied the OPN method to the Rossler chaotic attractor system. By quantifying effective information, he noticed that effective information (such as determinism and degeneracy) was sensitive to the critical phase transition between period and chaos~\cite{varley2020causal}.

\subsubsection*{Causal Geometry}

Alternatively, Pavel Chvykov and Erik Hoel extended the causal emergence framework to continuous systems and proposed the concept of causal geometry~\cite{chvykov2020causal}. 

To understand the basic idea of this work, we suppose the continuous system considered is a continuous functional map and assume that the uncertainties are small disturbances added to this deterministic function, that is,

\begin{equation}
\label{eq:continuous_map}
y=f(x)+\varepsilon,
\end{equation}
where $x$ is the input variable defined on the interval $[-L/2,L/2]$, $L$ is a very large constant, $y$ is the output variable, and $\varepsilon$ is a Gaussian noise with a zero mean and standard deviation $\epsilon\ll L$. Then, an approximate form of the effective information for such a continuous system can be obtained using the Gaussian integral.

\begin{equation}
    \label{eq:ei_gaussian}
    EI \approx \ln(\frac{L}{\sqrt{2\pi e}})+\frac{1}{2L}\int_{-L/2}^{L/2}\ln \left(\frac{f'(x)}{\epsilon}\right)^2dx.   
\end{equation}
The reason we set the domain of $x$ to $[-L/2,L/2]$ is to guarantee that the Gaussian integral can be implemented and to simplify the definition of a uniform distribution on it.

This form can be easily generalized to a dynamical system with discrete time steps once we replace $x$ and $y$ with $x(t)$ and $x(t+1)$, respectively. The formula can also be extended to higher dimensions, with $\mathbf{x}$ and $\mathbf{y}$ represented in bold. Suppose $\mathbf{x}$ 
 $\in[-L/2,L/2]^n\subset\mathcal{R}^n$ and $\mathbf{y}$ $\in\mathcal{R}^m$, where $n$ and $m$ are positive integers. Equation ({\ref{eq:ei_gaussian}}) can be generalized to the following form:
\begin{equation}
    \label{eq:ei_gaussian_ndim}
    EI\approx \ln(\frac{L^n}{(2\pi e)^{m/2}})+\frac{1}{2}\mathbb{E}_{\mathbf{x}\sim U ([-\frac{L}{2},\frac{L}{2}]^n)}\ln\left|\det\left(\frac{\partial_\mathbf{x} f(\mathbf{x})}{\Sigma^{1/2}}\right)\right|^2,
\end{equation}
where $\Sigma$ is the covariance matrix of the Gaussian noise $\varepsilon$, $U([-L,L]^n)$ represents the uniform distribution on the hypercube $[-L,L]^n$, $|\cdot|$ is the absolute value operation, and $\det$ is the determinant.

Note that the conditional distribution of $\mathbf{y}$ given $\mathbf{x}$ is a Gaussian distribution, that is, $p(\mathbf{y}|\mathbf{x})=\mathcal{N}(f(\mathbf{x}),\Sigma)$. Thus, the term in the expectation in Equation (\ref{eq:ei_gaussian_ndim}) can be written as:

\begin{equation}
\label{eq:expectation}
\left|\det\left(\frac{\partial_\mathbf{x} f(\mathbf{x})}{\Sigma^{1/2}}\right)\right|^2=\left|\det\left(\mathbb{E}_{\mathbf{y}|\mathbf{x}}\left[\partial_{\mu}\partial_{\nu}\ln p(\mathbf{y}|\mathbf{x})\right]\right)\right|,
\end{equation}
and this corresponds to the determinant of the negative Fisher information metric for the distribution $p(\mathbf{y}|\mathbf{x})$:
\begin{equation}
    g_{\mu\nu}\equiv -\mathbb{E}_{\mathbf{y}|\mathbf{x}}\left[\partial_{\mu}\partial_{\nu}\ln p(\mathbf{y}|\mathbf{x})\right],
\end{equation}
which measures the sensitivity of the logarithmic $p(\mathbf{y}|\mathbf{x})$ to changes in $\mathbf{x}$, where $\partial_{\mu}\equiv\frac{\partial}{\partial \mathbf{x}_{\mu}}$ represents the partial derivative with respect to the $\mu$th component of $\mathbf{x}$. Therefore, we have defined a distance metric for the Riemann manifold $\mathcal{M}=\{p(\mathbf{y}|\mathbf{x})\}$ on the parameter space \mbox{$\mathbf{x}\in[-L/2,L/2]^n$}, encompassing all possible distributions of $p(\mathbf{y}|\mathbf{x})$. This is the origin of the term ``geometry'' in this framework.

Finally, we can obtain an expression for EI with the Fisher information metric:
\begin{equation}
    \label{eq:ei_fisher}
    EI\approx \ln\frac{L^n}{(2\pi e)^{m/2}}-\frac{1}{2}\mathbb{E}_{\mathbf{x}\sim U ([-\frac{L}{2},\frac{L}{2}]^n)}\ln|\det(g_{\mu\nu})|.
\end{equation}
This formula can be generalized to cases where $p(\mathbf{y}|\mathbf{x})$ is a non-Gaussian distribution. Once the distribution function is known, we can obtain its Fisher information metric, and then EI can be calculated using Equation (\ref{eq:ei_fisher}). The reason behind this is that the whole manifold $p(\mathbf{y}|\mathbf{x})$ for any $\mathbf{x}$ can be understood as a concatenation of local Gaussian distributions.

In \cite{chvykov2020causal}, the authors considered a more general case in which intervention noise is added to the input (intervention) variable $x$ in Equation (\ref{eq:ei_gaussian}). The intervention noise is denoted as $\xi\sim \mathcal{N}(0,\delta^2)$, where $\delta$ is the standard deviation of $\xi$. In contrast, $\varepsilon$ is called observational noise distinguishing it from $\xi$, and is added to the output (observational) variable $y$ according to Equation (\ref{eq:ei_gaussian}). Finally, if two kinds of noise are both considered and if $L=1$ and $\epsilon\ll 1$, Equation (\ref{eq:ei_gaussian}) becomes:
\begin{equation}
    \label{eq:full_ei_gaussian}
    EI\approx -\frac{1}{2}\int_{-1/2}^{1/2}\ln\left[\left(\frac{\epsilon}{f'(x)}\right)^2+\delta^2\right]dx.
\end{equation}
This is the formula for EI for the continuous mapping function (Equation (\ref{eq:continuous_map})) given by \cite{chvykov2020causal}. Using this equation and a logistic function for $f(x)$, the authors compared the EIs of a more continuous map (Equation (\ref{eq:continuous_map})) and a more discrete map with an adjustable parameter in the logistic function. They discovered that when the noise level is low, the continuous map can exhibit higher EI compared to the discrete map. However, as the noise level increases, discretizing the mapping function can lead to a model with higher EI. This phenomenon helps explain why digital circuits eventually outperform analog circuits in mitigating noise interference; the binarization and coarse-graining strategy of digital circuits suppresses the propagation and diffusion of noise.

To generalize the information geometry to the case with intervention noise and observational noise, let us introduce a new intermediate variable $\theta\in \Theta\subset\mathcal{R}^l$ with dimensionality $l$ such that we cannot control $\mathbf{y}$ by directly intervening $\mathbf{x}$. Instead, we can intervene $\mathbf{x}$ to influence $\theta$ and indirectly affect $\mathbf{y}$. Therefore, the three variables form a Markov chain: $\mathbf{x}\rightarrow \theta\rightarrow \mathbf{y}$. 

In this case, two manifolds can be obtained: the effect manifold $\mathcal{M}_E=\{p(\mathbf{y}|\theta)\}_{\theta\in \Theta}$ with metric $g_{\mu\nu}=-\mathbb{E}_{p(\mathbf{y}|\theta)}\partial_{\mu}\partial_{\nu}\ln p(\mathbf{y}|\theta)$ and the intervention manifold $\mathcal{M}_I=\{\Tilde{q}(\mathbf{x}|\theta)\}_{\theta\in \Theta}$ with metric $h_{\mu\nu}=-\mathbb{E}_{\Tilde{q}(\mathbf{x}|\theta)}\partial_{\mu}\partial_{\nu}\ln \Tilde{q}(\mathbf{x}|\theta)$, where $\Tilde{q}\equiv \frac{q(\theta|\mathbf{x})}{\int q(\theta|\mathbf{x})d\mathbf{x}}$ and $\partial_{\mu}=\partial/\partial \theta_{\mu}$. These two manifolds together are called causal geometry.

Finally, the EI calculation formula for causal geometry is:
\begin{equation}
    \label{eq:ei_causal_geometry}
    EI_g=\ln\frac{V_I}{(2\pi e)^{n/2}}-\frac{1}{2V_I}\int_\Theta\sqrt{|\det(h_{\mu\nu})|} \ln\left|\det\left( I_n+\frac{h_{\mu\nu}}{g_{\mu\nu}}\right)\right|d^l\theta,
\end{equation}
where we have set $L=1$ and $m=l=n$ to reduce the number of free parameters, and $I_n$ is the identity matrix with size $n$, $V_I=\int_\Theta \sqrt{|\det(h_{\mu\nu})|}d^l\theta $.

\subsubsection*{EI Calculation for Neural Networks}

One of the application areas for continuous EI is neural networks. In \cite{marrow2020examining}, the authors applied EI to analyze the causal effects for different layers of neural networks and compared the results with the information bottleneck theory. The method they used to calculate EI was to convert continuous variables into discrete ones by dividing the domains of the input and output variables into small regions. However, this method has high computational complexity and is challenging to extend to high dimensions.

In \cite{zhang2022neural}, the authors proposed a new numeric method for calculating EI on neural networks. The basic idea behind this method is to understand a well-trained feed-forward neural network as a stochastic mapping, following:
\begin{equation}
\mathbf{y}=f(\mathbf{x})+\varepsilon,
\end{equation}
where $\mathbf{x}\in \mathcal{R}^n$ is the input variable, $\mathbf{y}\in\mathcal{R}^m$ is the output variable, and $f:\mathcal{R}^n\rightarrow\mathcal{R}^m$ is the deterministic map of the neural network. Additionally, $\varepsilon$ is a Gaussian noise with a mean of zero and a covariance matrix $\Sigma=diag(\sigma_1^2,\sigma_2^2,\cdot\cdot\cdot,\sigma_m^2)$, where $\sigma_i$ is the mean square error on dimension $i$ across the entire training dataset. In this way, the neural network can be regarded as a Gaussian distribution $p(\mathbf{y}|\mathbf{x})=\mathcal{N}(f(\mathbf{x}),\Sigma)$.

Therefore, Equation (\ref{eq:ei_gaussian_ndim}) can be used on this neural network to calculate EI. In \cite{zhang2022neural}, the authors used the Monte Carlo integration method to estimate the expectation in Equation (\ref{eq:ei_gaussian_ndim}). This technique significantly reduces the computational complexity. 

However, when the authors compared the results calculated using Equation (\ref{eq:ei_gaussian_ndim}) for various neural networks with different dimensions, they obtained unreasonable results: EI increased with the number of dimensions. Thus, causal emergence was not always observed because micro-dynamics exhibited unreasonably larger EI. Another drawback of Equation (\ref{eq:ei_gaussian_ndim}) is that the parameter $L$ always dominates the value of the expression and should be removed from the equation. One way to solve this issue is to use Eff instead of EI (see Equation (\ref{eq:eff})). However, not all $L$s in Equation (\ref{eq:ei_gaussian_ndim}) can be eliminated. Therefore, the authors introduced the notion of dimension-averaged EI, defined as
\begin{equation}
    \label{eq:dimension_averaged_ei}
    dEI\equiv \frac{EI}{m}
\end{equation}
which simply divides EI by the dimension of the output variable $\mathbf{y}$. By using dimension-averaged EI, $L$ can be removed when calculating dimension-averaged causal emergence for the neural networks $f_M$ and $f_m$ representing macro- and micro-dynamics, respectively, given by

\begin{equation}
    \label{eq:dim_CE}
    dCE(f_M,f_m)\equiv \frac{EI(f_M)}{n_M}-\frac{EI(f_m)}{n_m},
\end{equation}
where dCE is the dimension-averaged causal emergence measure. If we incorporate Equation~(\ref{eq:ei_gaussian_ndim}) into Equation (\ref{eq:dim_CE}), we obtain:
\begin{equation}
\label{eq:dim_CE_gaussian}
    dCE\approx \frac{\mathbb{E}_{\mathbf{x}_M}\log|\det(\partial_{\mathbf{x}_M}f_M)|+\sum_{i=1}^{x_M}\log\sigma_{i,M}^2}{n_M}-\frac{\mathbb{E}_{\mathbf{x}_m}\log|\det(\partial_{\mathbf{x}_m}f_m)|+\sum_{i=1}^{x_m}\log\sigma_{i,m}^2}{n_m},
\end{equation}
in which $L$ is eliminated.

\subsubsection{Causal Emergence in Complex Networks}

Many complex systems can be represented by networks; therefore, applying the framework for causal emergence to networks is necessary. However, two problems should be solved in advance to apply Hoel's framework: one is to assign dynamics to the studied network because networks are static structures without any dynamical properties and the other is to coarse-grain the network. 

Klein and Hoel~\cite{klein2020emergence} addressed the first problem by considering random walks on complex networks. Subsequently, the TPM of the system was defined based on the transfer probabilities between nodes of a large number of random walkers. Due to the good mathematical properties of the random walk dynamics on graphs, the authors established an explicit expression for EI on networks, and the final form only relied on the weighted normalized adjacency matrix $w_{ij}$($\sum_jw_{ij}=1$) of the network. The expression is as follows:
\begin{equation}
    \label{eq:EI_networks}
    EI = H(\langle W^{out}_{i}\rangle) - \langle H(W^{out}_{i})\rangle.
\end{equation}
where $H(\langle W^{out}_{i}\rangle)$ represents the Shannon entropy calculated from the distribution of the averaged out-weights across all nodes~($W^{out}_i\equiv\{w_{ij}|j=1,2,\dots,N\}$), which characterizes the determinism of the random walk dynamics; $\langle H(W^{out}_{i})\rangle$ is the average Shannon entropy for all nodes, which characterizes the degeneracy of the dynamics; and $H(W^{out}_{i})$ is the uncertainty of node $i$.

To address the second problem, in \cite{klein2020emergence}, the greedy algorithm was used to group nodes to form a macro-network. It is worth noting that when merging micro-nodes into macro-nodes, the TPM of the resulting macro-network can be derived by merging the probabilities from the TPM of the micro-level. To ensure that the grouped macro-network maintains the same random walk dynamics as the original one, a dynamic consistency test is implemented.

In the experimental section, the aforementioned paper explored the application of a greedy search for identifying causal emergence in complex networks. Experiments were conducted on artificial networks as well as four types of real networks. In the case of ER random networks, the size of effective information was solely dependent on the connection probability $p$ and converged to $-\log_{2}p$ with the increase in network size. Additionally, a significant finding was the presence of a phase transition point, where the average degree of the network reached approximately $\log_2N$. Beyond this point, the random network structure did not contain additional information with increasing size. In the case of preferential attachment~(PA) networks, when $\alpha<1.0$ ($\alpha$ represents the degree of preferential attachment), the effective information of the network increased as the network size expanded. Conversely, when $\alpha>1.0$, the opposite was observed. The scale-free network corresponding to $\alpha=1.0$ represented the critical boundary of growth. Regarding real networks, the authors found that biological networks exhibited the lowest EI due to the presence of significant noise, which can be removed through effective coarse-graining, and causal emergence was the most significant compared to other types of networks. However, technical networks exhibited sparsity and non-degeneracy, resulting in higher average efficiency and more specific node relationships. Consequently, they exhibited the highest EIs among the studied networks. 

The network coarse-graining method mentioned above is the greedy algorithm. However, when the network is very large, the efficiency of this method remains considerably low. Following that, Griebenow et al.~\cite{griebenow2019finding} introduced a spectral clustering-based approach for identifying causal emergence within networks. More specifically, the method involved performing eigenvalue decomposition of the TPM, followed by constructing a similarity matrix using the eigenvectors of the nodes. The OPTICS algorithm was employed to cluster the nodes, and nodes belonging to the same cluster were aggregated into a macro-node. Subsequently, the maximum value of EI was selected by utilizing the linear search distance hyperparameter $\epsilon$. In their paper, the authors additionally proposed a gradient descent algorithm based on deep learning. This approach encompassed several steps, beginning with the random initialization of a grouping matrix. Then, this matrix was used to construct a macro-network by combining the micro-networks. Finally, the grouping strategies were automatically learned through the process of maximizing effective information within the macro-network. However, this method often falls into the local optimal solution.

\subsubsection{Other Applications}
Once the method for quantifying causal emergence in complex systems is developed, it can be applied across various fields that possess abundant network data. The first type of network studied was biological networks. 

As previously discussed, biological networks are full of noise, which poses challenges in comprehending their internal operating principles. On one hand, such noise arises from inherent fluctuations within the system itself, whereas on the other hand, it can be introduced through measurement or observation processes. Consequently, Klein et al.~\cite{klein2022exploring} further explored the relationship between noise, degeneracy, and certainty in biological networks and their specific meanings. For instance, in gene expression networks, highly deterministic relationships indicate that the expression of one gene almost invariably leads to the expression of another gene. Simultaneously, degeneracy is a prevalent phenomenon in the evolutionary processes of biological systems. Due to these two factors, it remains unclear at which scale biological systems should be analyzed to gain a deeper understanding of their functions. 

To address this, Klein et al.~\cite{klein2021evolution} conducted an analysis of protein interaction networks across more than 1800 species. They employed EI as a measure for assessing the levels of noise and uncertainty in protein interactions. The findings revealed that the macro-scale network exhibited lower levels of noise and degeneracy. Additionally, the nodes within the macro-scale interaction group demonstrated greater resilience compared to the nodes that were not part of the macro-scale interaction group. Through robust analysis, the authors demonstrated that eukaryotes exhibited a stronger degree of causal emergence compared to archaea. Additionally, to address the 'deterministic paradox,' the authors introduced the concepts of the neutral process and the selection process in biological evolution. The neutral process operates at the micro-scale, leveraging mutations to promote interactions and enhance species diversity. On the other hand, the selection process operates at the macro-scale, effectively eliminating noise that hampers system operation and efficiency. Therefore, in order to adapt to the demands of evolution, it becomes essential for evolved biological systems to function across multiple scales.

Hoel et al.~\cite{hoel2020emergence} conducted further research on the causal emergence within biological systems. The authors elaborated that macro- and micro-systems exist widely in biological systems. For example, the micro-scale of a group of cells can involve potential ion channel changes, whereas the macro-scale corresponds to the membrane potential changes of cells. Furthermore, the authors utilized EI to analyze gene regulatory networks, aiming to identify the most informative models for controlling mammalian heart development. By quantifying the causal emergence within the largest component of the gene network of Saccharomyces cerevisia, it was revealed that informative macro-scale structures were prevalent across biological systems. Additionally, the authors emphasized the importance of evolutionary systems operating at multiple scales due to the significant advantages they offer. Natural selection requires variation between populations, and degradation observed in biological systems serves as a crucial factor for evolution. Degradation provides the conditions for the evolutionary process. However, organisms also need to maintain predicted consistency in phenotype, behavior, and structure to ensure survival and reproduction. Consequently, evolved systems need to operate at multiple scales, and the function of multi-scale systems is malleable in changing environments.

Swain et al.~\cite{swain2022interactions} conducted an investigation into the impact of ant colony interaction history on task assignment and task switching. They employed effective information to examine how noise information propagates among ants and explored the relationship between EI and the proportion of ants assigned to different tasks within ant colonies. The study revealed that the extent of historical interaction between ant colonies influenced task assignment. Additionally, the specific type of ant colony involved in an interaction determined the level of noise present within that interaction. For example, the interactions between foragers displayed significantly higher levels of noise when contrasted with interactions between nurses or cleaners. Furthermore, even when ants switched functional groups, the cohesion within ant colonies ensured the stability of the overall colony, and different functional ant colonies also played different roles in maintaining group cohesion.

The EI indicator and causal emergence theoretic framework can also be applied to artificial systems. For example, Marrow et al.~\cite{marrow2020examining}  quantified and monitored the changes in the causal structure of neural networks during training, in which EI was employed to evaluate the degree of causal influence of nodes and edges on downstream tasks at each layer. By observing the changes in EI, including determinism (sensitivity) and degeneracy, throughout model training, the generalization ability of the model could be determined, thus helping better understand and explain the working principle of neural networks. 

Varley et al.~\cite{varley2020causal} attempted to apply the causal emergence framework to both discrete cellular automata and continuous Rossler systems. In the case of cellular automata systems, the authors select 88 unique rules corresponding to four types: static, periodic, chaotic, and complex. By considering each state as a node, where each state determines the subsequent state, a directed state transition graph is constructed. The analysis revealed that rules 1, 2, and 4 correspond to the strongest causal emergence of dynamics. Notably, the network constructed by these three rules exhibited a significant presence of star and spoke motifs. In addition, the authors also draw some quantitative conclusions, for example, among the 17 rules belonging to the third and fourth categories, 30\% exist causal emergence, 70\% show causal degradation, and the CE of cellular automata with the same rules remained relatively consistent across different sizes. 

Furthermore, the authors employed the OPN algorithm to transform the continuous system into a discrete-state transition graph for comparative analysis~(see Section~\ref{sec:continuous_EI}). They found that chaos dynamics demonstrated a correlation with low determinism, and the variations in the degeneracy and efficiency coefficients aligned with the changes observed in the determinism curve.

\subsubsection{Critiques}
\label{sec:critiques}
In the extensive literature on causality and emergence, Hoel's theory has attracted attention for linking emergence and causality through interventionism, introducing the concept of causal emergence in a quantitative manner. However, Dewhurst~\cite{dewhurst2021causal} provided a philosophical clarification of Hoel's theory, arguing that it was epistemological rather than metaphysical. This suggests that Hoel's macroscopic causality is merely a causal explanation based on information theory, rather than involving ``genuinely novel causal powers''. This also raises concerns about the assumption of a uniform distribution, as there is no empirical basis to favor it over other distributions.

The computation of Hoel's effective information relies on two premises: (1) knowledge of the system's microscopic dynamics, and (2) knowledge of the coarse-graining scheme. However, in practice, it is rare for both conditions to be simultaneously satisfied, especially in observational studies where both are unknown. This limitation hinders the practical applicability of Hoel's theory.

It has been pointed out that Hoel's theory neglects the constraints on coarse-graining methods, and certain coarse-graining methods can lead to ambiguity~\cite{eberhardt2022causal}. Additionally, the combination of some coarse-graining operations over states and coarse-graining operations over time does not exhibit commutativity. For instance, by assuming $A_{m \times n}$ is a coarse-graining operation over states (merging $n$ states into $m$ states), and $(\cdot) \times (\cdot)$ is a coarse-graining operation over time (combining two time steps into one), the equation $A_{m\times n}(TPM_{n \times n}) \times A_{m \times n}(TPM_{n \times n}) = A_{m \times n}(TPM_{n \times n} \times TPM_{n \times n})$ does not always hold. This indicates that certain coarse-graining operations can result in a discrepancy between the evolution of macroscopic states and the coarse-grained states of the evolved microscopic systems. It implies the need for consistent constraints on coarse-graining strategies.

This means that solely maximizing EI may raise some problems, and further constraints must be added to the framework. We discuss this problem in Section \ref{sec:nis}.

\subsection{Fernando E. Rosas's Quantification of Causal Emergence}
\subsubsection{Basic Idea}

In Hoel's framework, it is essential to find a coarse-graining strategy in order to determine the occurrence of causal emergence, and the outcome is influenced by the choice of the coarse-graining method. Although it has been suggested by Hoel~\cite{hoel2013quantifying,hoel2017when} that an optimal strategy can be identified by maximizing EI, certain issues have been raised by Dewhurst~\cite{dewhurst2021causal}. EI is a global measure because it requires the input variable $X$ to be a uniform distribution over the whole domain~\cite{rosas2020reconciling}. However, many regions are not observable from the data. Consequently, there is a pressing need for an alternative theoretical framework for causal emergence that does not rely on a coarse-graining method.

In response, Fernando E. Rosas took an approach that did not require a coarse-graining strategy as a prerequisite, attempting to break down excess entropy---the mutual information between a system's past and future states---into non-overlapping parts to identify the information components most relevant to causal emergence. To accomplish this, he relied on the partial information decomposition (PID) framework proposed by Williams and Beer, which provides a method for the non-overlapping decomposition of joint mutual information~\cite{williams2010nonnegative}. Below, we introduce Williams and Beer's theoretical framework. 

\subsubsection{Partial Information Decomposition}

The partial information decomposition (PID) framework investigates the general informational relationship between source variables and a target variable. To simplify the problem description without sacrificing generality, let us consider a system with two input variables ($X_1$, $X_2$) and one output variable ($Y$) as an example, as depicted in the Venn diagram below.

The mutual information between the target variable and individual source variables, $I(X_1;Y)$ and $I(X_2;Y)$, as well as the mutual information between the target variable and the joint source variable, $I(X_1,X_2;Y)$, exhibits a complex relationship. Intuitively, one cannot be converted to another. Nonetheless, it is an intuitive notion that the overlapped circular plates of $I(X_1;Y)$ and $I(X_2;Y)$ divide the oval region of $I(X_1,X_2;Y)$ into four adjacent non-overlapping regions, representing the three types of information components of $I(X_1,X_2;Y)$:
\begin{equation}
I(X_1,X_2;Y)=Red(X_1,X_2;Y)+Un(X_1;Y|X_2)+Un(X_2;Y|X_1)+Syn(X_1,X_2;Y)
\end{equation}

Specifically: 

\begin{itemize}
    \item $Red$: Redundant information refers to the information held by both sources;
    \item $Un$: Unique information refers to the information held by one source but not the other;
    \item $Syn$: Synergistic information is the information held by all sources together, but not any individual one.
\end{itemize}

If we can identify variable representations corresponding to these information components, the non-overlapping feature in the Venn diagram indicates their independence from one another.

For a more intuitive understanding, let us consider a few simple toy examples.

\begin{figure}[H]
\centering
\includegraphics[scale=0.30]{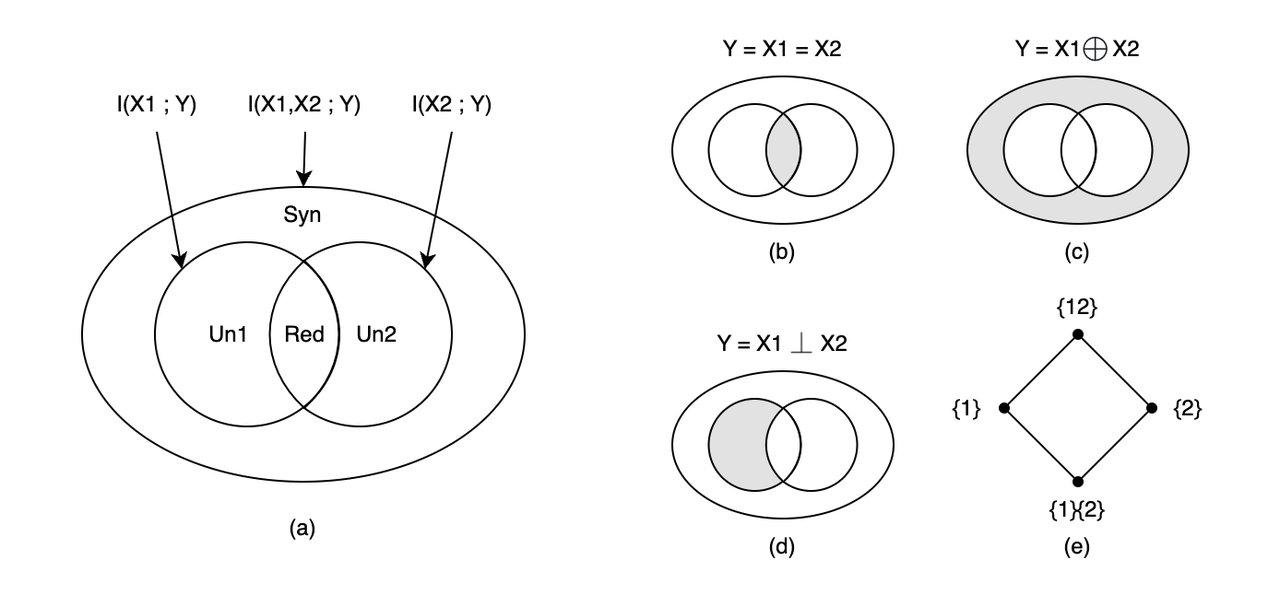}
\caption{Venn diagram of PID. (\textbf{a}) Relationship of mutual information between Y and X1, X2. (\textbf{b}--\textbf{d}) Examples showing that joint mutual information is solely contributed by redundant, synergistic, and unique information. (\textbf{e}) Redundancy lattice representation of (\textbf{a}).}
\label{figure8}
\end{figure}

Case 1, 
$Y=X_1=X_2$: This is a scenario in which a single source variable can predict the target variable, and the addition of another source variable does not enhance the prediction of the target variable. In this case, we have $I(X_1,X_2;Y)=Red(X_1,X_2;Y)$, as depicted in Figure~\ref{figure8}b. 

Case 2, $Y=X_1\oplus X_2,X_1\perp{X_2}$:  In another scenario, neither of the source variables can predict the target variable individually, but together, they can predict the target variable synergistically. In this case, $I(X_1,X_2;Y)=Syn(X_1,X_2;Y)$, as depicted in Figure~\ref{figure8}c.

Case 3, $Y=X_1, X_1\perp{X_2}$: In this scenario, the target variable can be predicted by one of the source variables but not the other, which implies $I(X_1,X_2;Y)=Un(X_1;Y|X_2)$, as depicted in Figure~\ref{figure8}d.

For general cases, Williams and Beer presented a method for calculating redundant information, defined as $Red_{min}$. This method reflects the concept of redundancy, identifying it as the shared information across all sources, which is equivalent to the minimum amount of information contributed by any single source:

\begin{equation}
Red_{min}(X_1,X_2;Y)\sum_{y \in \mathcal{Y}} P(Y=y)\min_{Xi}I(Y=y;X_i)
\end{equation}

Figure~\ref{figure8}e presents an alternative way of visualizing the PID framework, known as the redundancy lattice~\cite{williams2010nonnegative}. In this representation, \{12\} denotes synergistic information, $\left\{1\right\}$ and $\left\{2\right\}$ denote unique information, and $\left\{1\right\}$$\left\{2\right\}$ denotes redundant information. In the subsequent sections, we demonstrate that the $\phi$ID framework employs the notation of the redundancy lattice.

Although the PID framework remains compatible with scenarios involving more than two source variables, it should be noted that the corresponding Venn diagrams and redundancy lattices for these scenarios can become substantially more complicated and difficult to decipher, as discussed in~\cite{williams2010nonnegative}.

\subsubsection{Integrated Information Decomposition}
The PID framework provides a useful framework for analyzing the non-overlapping information composition in a multivariate system.~However, its application to causal analysis of dynamic systems is limited by the fact that it only allows for a single target variable. This limitation prevents the framework from fully capturing the transitions of multiple states across time steps. To address this challenge, Fernando E. Rosas developed Integrated Information Decomposition ($\phi$ID)~\cite{mediano2019beyond}, which takes its name from Integrated Information Theory (IIT)~\cite{oizumi2014phenomenology}. This extension of PID provides a more comprehensive method for analyzing dynamic systems.

To introduce Rosas's framework clearly, we consider a system with only two variables. All the definitions and calculations can be generalized to systems with more variables.

The objective of the $\phi$ID framework is to decompose excess entropy into non-overlapping information components. In a two-variable Markovian system, excess entropy is given by $E=I(X^1_t,X^2_t ; X^1_{t+1}, X^2_{t+1})$, where $X^1_t$ and $X^2_t$ represent the current states, and $X^1_{t+1}$, $ X^2_{t+1}$ represent the future states. From a causation perspective, $X^1_t$ and $X^2_t$ represent causes, whereas $X^1_{t+1}$ and $ X^2_{t+1}$ represent effects.

\begin{figure}[H]
\centering
\includegraphics[width=13cm,height=4cm]{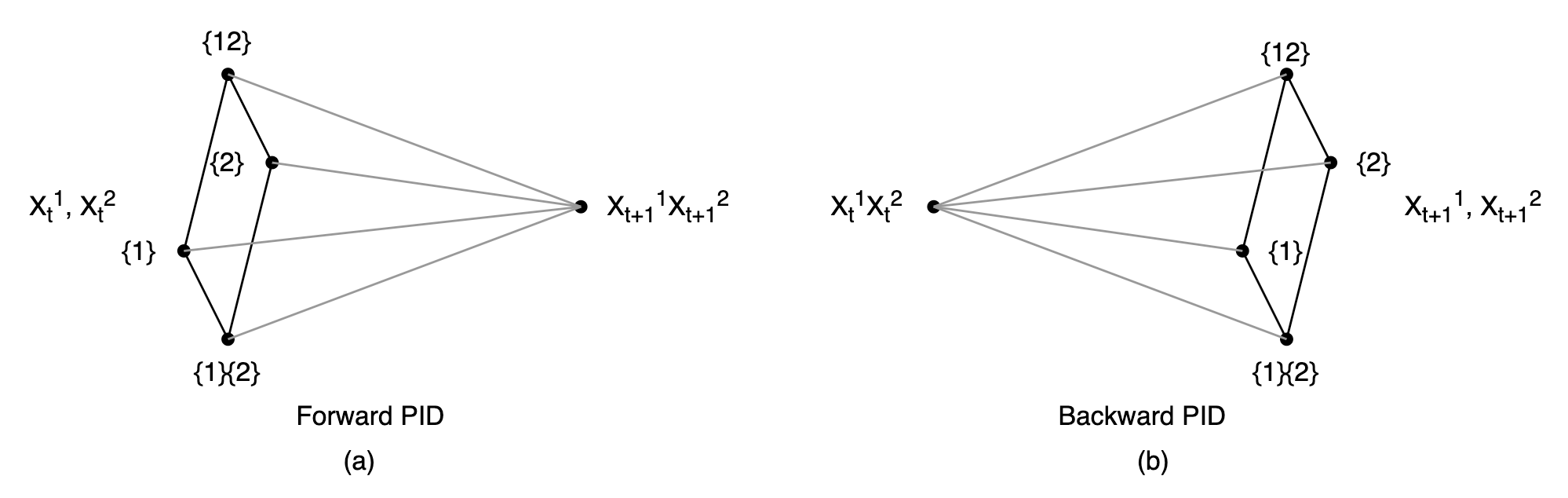}
\caption{(\textbf{a}) Redundancy lattice of ``forward PID'' and its target variable $X^1_{t+1}X^2_{t+1}$. (\textbf{b}) Redundancy lattice of ``backward PID'' and its target variable $X^1_tX^2_t$.}
\label{figure9}
\end{figure}

Rosas first noted that mathematically, the PID framework is symmetric with respect to the source and target variables. There are two viewpoints for analyzing the above Markovian system with PID. One viewpoint takes $X^1_{t+1}$ and $X^2_{t+1}$ together as the target variable (aiming to decompose the ``cause'' to elements), whereas the other viewpoint takes $X^1_t$ and $X^2_t$ together as the target variable (aiming to decompose the ``effect'' to elements). The redundancy lattices of these perspectives are illustrated in Figure~\ref{figure9}a,b, and respectively, are referred to as ``forward PID'' and `` PID''.

\vspace{-3pt}

\begin{figure}[H]
\centering
\includegraphics[width=13cm,height=6cm]{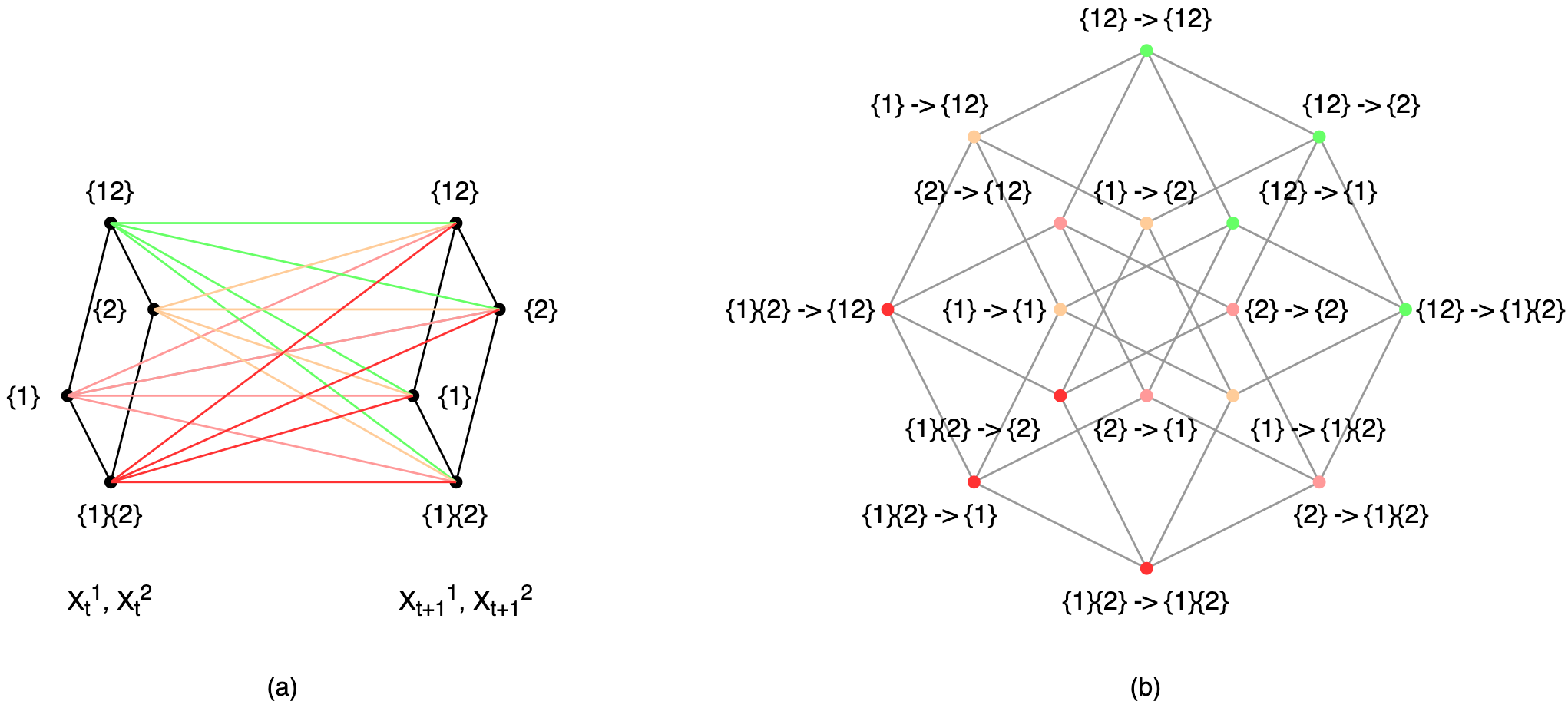}
\caption{(\textbf{a}) Relationships 
 between redundancy lattices of forward PID and backward PID. (\textbf{b})~Double-redundancy lattice, whose vertices correspond to the edges in (\textbf{a}). Distinct colors of dots and lines are employed to distinguish the source elements of $X_t$.}
\label{figure10}
\end{figure}

Next, Rosas introduced the $\phi$ID framework, consolidating forward PID and backward PID into a single framework. In this framework, the one-to-many relationship in PID was expanded to include many-to-many relationships. He built full connections between the elements of the redundancy lattices of forward PID and backward PID. This approach generates 16 relations between the source and target, as depicted by the colored lines in Figure~\ref{figure10}a (each color corresponds to a specific source element of $X_t$). These relations are referred to as ``$\phi$ID atoms'' and can be represented as vertices in a lattice, as shown in Figure~\ref{figure10}b. This more complex lattice is referred to as the double-redundancy lattice because it is a ``product'' of two redundancy lattices (of forward PID and backward~PID).

In the double-redundancy lattice, each vertex is defined as a ``$\phi$ID atom'', denoted as $I^{\alpha \rightarrow \beta}_\partial,\alpha,\beta\in\mathcal{A}$, where $\mathcal{A}=\{\{\{1\}\{2\}\}, \{1\}, \{2\}, \{12\}\}$. For instance, the $\phi$ID atom of the vertice ${\{12\}\rightarrow \{12\}}$ is denoted as $I^{\{12\}\rightarrow \{12\}}_\partial$. The $\phi$ID atom represents the broken-down information transmitted from the cause elements to the effect elements, and their cumulative sum amounts to the excess entropy of the entire system. Consequently, the formal definition of Integrated Information Decomposition ($\phi$ID) is:

\begin{equation}
    \bold{E}=I(\bold{X_t};\bold{X_{t+1}})=\sum_{\alpha,\beta \in \mathcal{A}}{I^{\alpha \rightarrow \beta}_\partial}
\end{equation}


The main advantage of $\phi$ID is that it can provide a more fine-grained analysis of complex systems compared to PID by introducing the temporal dynamic. By decomposing the mutual information of the system at different times, $\phi$ID provides another approach for the quantitative study of causal emergence. 

\subsubsection{Reconciling Emergences}
\label{sec:rosas}
Building upon the concept of synergistic information in the PID framework, Rosas introduced a quantitative definition of causal emergence using the $\phi$ID framework to tackle the challenge of identifying an appropriate coarse-graining strategy. The definition includes two aspects: firstly, determining whether the system has the capacity to generate causal emergence, and secondly, assessing the occurrence of causal emergence given a specific macroscopic feature. 

Regarding a system's capacity to exhibit causal emergence, this definition establishes a connection between causal emergence and synergistic relationships among variables across different time points. Consequently, a system denoted as $X_{t}$ is said to possess the capacity for causally emergent features if and only if:
\begin{equation}
    \label{sec:syn_emergence}
    Syn(X_{t};X_{t+1})>0
\end{equation}
In this context, causal emergence is understood as the synergistic effect between variables at preceding and subsequent moments within a Markovian dynamics system.

Then, Rosas further divided causal emergence into two parts in the $\phi$ID framework, downward causality and causal decoupling, based on the distinct characteristics of the information atoms. Among the sixteen $\phi$ID atoms obtained by decomposing the mutual information $I(X_{t};X_{t+1})$ using $\phi$ID, there are four information atoms corresponding to the synergistic effect, which is regarded as the composition of causal emergence. These atoms are denoted as $I^{\{12\} \rightarrow \alpha}_\partial(X_{t},X_{t+1})$, $\alpha \in \mathcal{A}=\{\{\{1\}\{2\}\}, \{1\}, \{2\}, \{12\}\}$. 

\begin{figure}[H]

\includegraphics[width=8 cm]{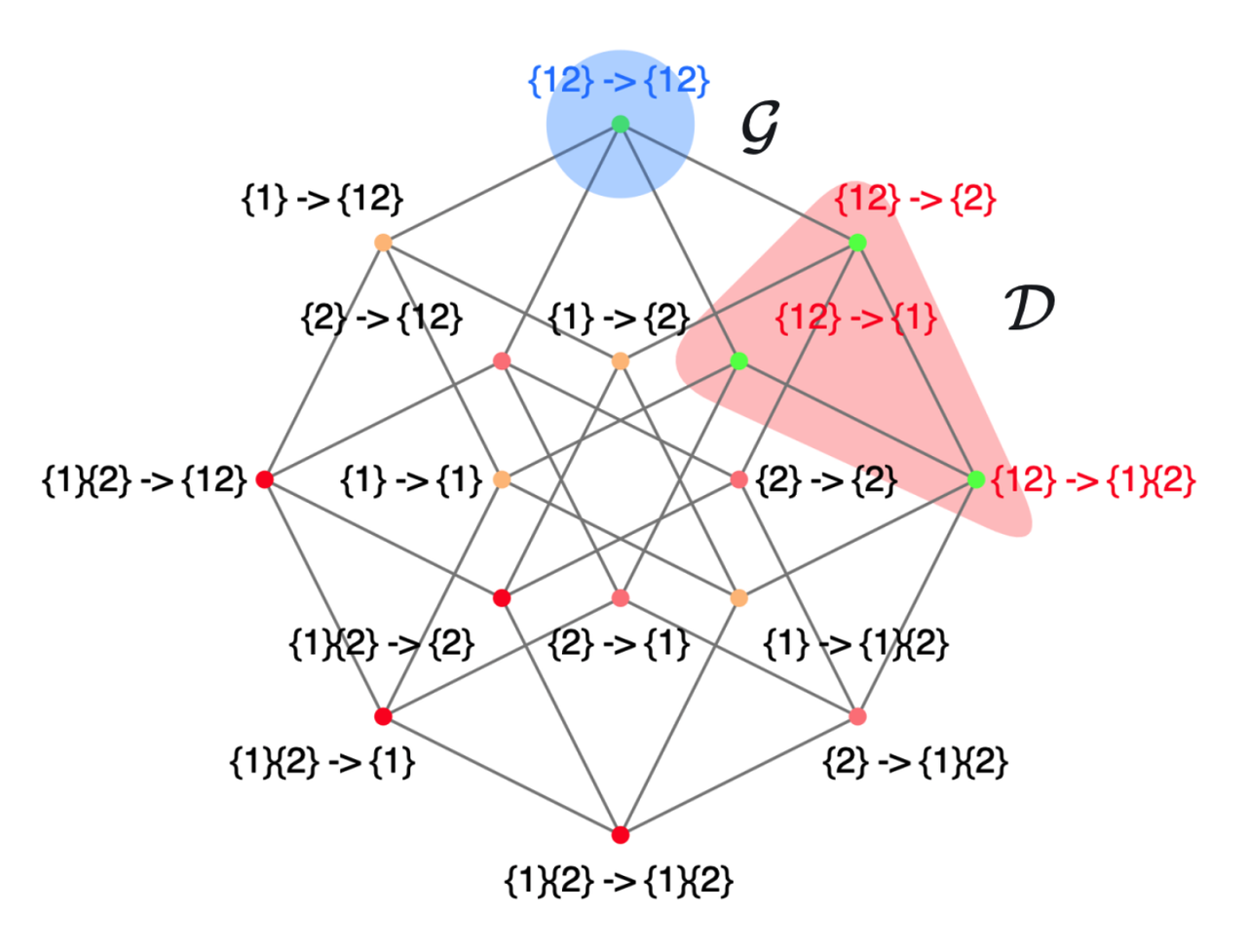}
\caption{\textls[-25]{Causal 
 decoupling ($\mathcal{G}$) and downward causation ($\mathcal{D}$) are represented by a double-redundancy~lattice. Distinct colors of dots are employed to distinguish the source elements of $X_t$, (colors are corresponding to Figure \ref{figure10}).}}
\label{figure11}
\end{figure}

\begin{figure}[H]

\includegraphics[width=7 cm]{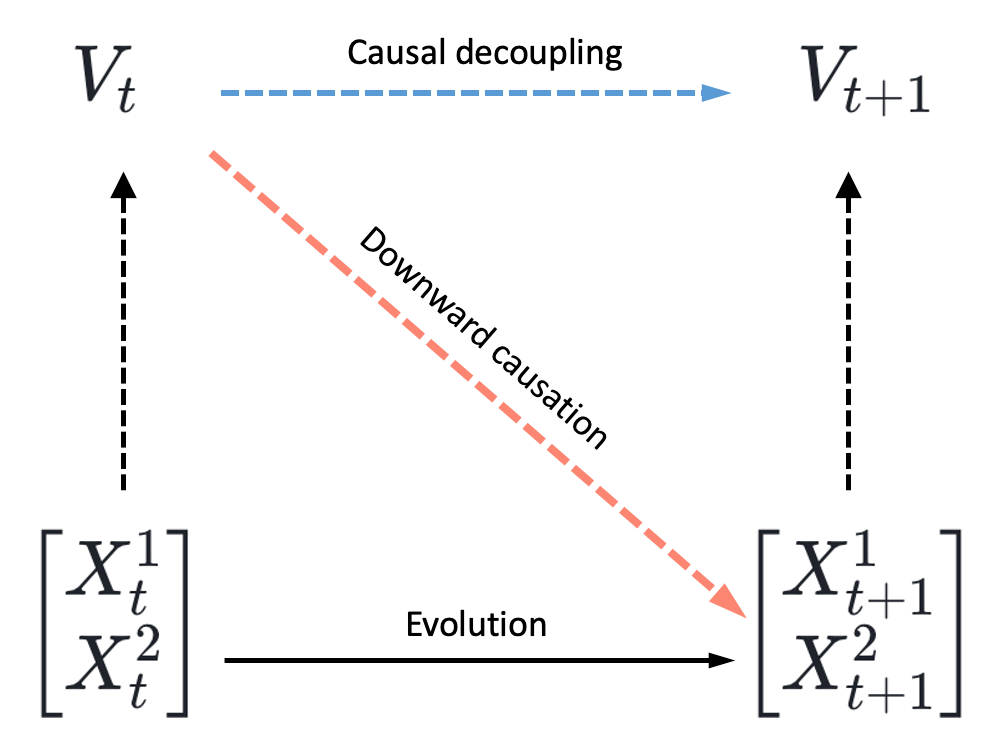}
\caption{Synergistic 
 causes can be further divided into causal decoupling and downward causation. }
\label{figure12}
\end{figure}

Downward causality is denoted as $\mathcal{D} (X_{t},X_{t+1})$, encapsulating the information atoms that exclusively manifest a synergistic effect in forward PID, as shown in Figure~\ref{figure11},~\ref{figure12}: 

\begin{equation}
\mathcal{D} (X_{t},X_{t+1}):=\sum_{\substack {\alpha \in A/\{12\}}}{I^{\{12\} \rightarrow \alpha}_\partial}(X_{t},X_{t+1})
\end{equation}

While causal decoupling is denoted as $\mathcal{G} (X_{t},X_{t+1})$, pinpointing the specific information atom where both forward and backward PID exhibit synergy:

\begin{equation}
\mathcal{G} (X_{t},X_{t+1}):={I^{\{12\} \rightarrow \{12\}}_\partial}(X_{t},X_{t+1})
\end{equation}

Since $\phi$ID is a non-overlapping decomposition of all information, this classification takes into account all cases of causal emergence, that is, 
\begin{equation}
    \label{eq:causal_decompose}
    Syn(X_{t};X_{t+1})=\mathcal{G} (X_{t},X_{t+1}) + \mathcal{D} (X_{t},X_{t+1}).
\end{equation}

\vspace{-20pt}

In addition to that, Rosas also provided an approach to quantify the causal emergence of a specific macro-variable, i.e., a coarse-graining strategy. If a system has the capacity to generate causal emergence, there can be some macroscopic features that exhibit causal emergence. A feature variable $V$ is said to be supervenient on the underlying system if it does not provide any predictive power for future states at times $t+1$ when the complete state $X$ of the system at time $t$ is known with perfect precision. This is equivalent to $V_{t}$ being statistically independent of $X_{t+1}$ given $X_{t}$. Then, for a system described by $X_{t}$, a supervenient feature $V_{t}$ is said to exhibit causal causation if:
\begin{equation}
    \label{eq:rosas_causal_uni}
    Un(V_{t};X_{t+1}|X_{t})>0.
\end{equation}
\textls[-35]{For this definition, the system's capacity of causal emergence is required, where \mbox{$Syn(X_{t};X_{t+1})$ > 0,}} since $Un(V_{t};X_{t+1}|X_{t}) \le Syn(X_{t};X_{t+1})$ holds for any supervenient feature $V_{t}$. Corresponding to the classification of the system's capacity, the downward causation (as indicated by the red dotted line in Figure~\ref{figure12}) 
of a feature variable $V$ exists when

\begin{equation}
\label{eq:downward_causation}
\begin{aligned}
Un(V_{t};&X_{t+1}^{1}|X_{t}) >0 \\
         & or \\
Un(V_{t};&X_{t+1}^{2}|X_{t}) >0  
\end{aligned}
\end{equation}

and causal decoupling (as indicated by the blue dotted line in Figure~\ref{figure12}) exists when

\begin{equation}
    \label{eq:causal_decoupling}
    Un (V_{t};V_{t+1}|X_{t},X_{t+1}) >0,
\end{equation}
which also depends on the capacity of the system. Furthermore, $V_{t}$ is said to have pure causal decoupling if $Un(V_{t};X_{t+1}^{\alpha}|X_{t}) =0$ and $Un(V_{t};X_{t+1}|X_{t})>0$.~If all the emergent features exhibit pure causal decoupling, the system is said to be perfectly~decoupled.

Although a rigorous quantitative definition for causal emergence was proposed, the mathematical formulations used in $\phi$ID can be complex and computationally demanding, making it difficult to apply the method to real-world systems. Also, In addition, the inconsistency in PID calculation causes the definition of causal emergence to depend on a specific PID calculation. To address these problems, Rosas relaxed the calculation of causal emergence and established a identification criteria based one the sufficient conditions of causal decoupling and downward causation. 

Specifically, to avoid exploring the specific quantitative methods of synergistic and redundant information, this criteria repeatedly subtracts redundant information, making the result a sufficient condition for causal emergence, which lose some of its generality but improves the reliability. The three indicators to be used are:

\begin{equation}
\label{eq:psi_identification}
    \Psi_{t,t+1}  (V):={I(V_{t};V_{t+1})} - \sum_{\substack {j}}I(X_{t}^{j};V_{t+1}),
\end{equation}
which measures the mutual information between macro-variables across two time steps minus the mutual information between each micro-state and macro-state.
\begin{equation}
\label{eq:rosas_emergence1}
   \Delta _{t,t+1}  (V):= \max_{\substack {j}}\left (   {I(V_{t};X_{t+1}^{j})} - \sum_{\substack {i}}I(X_{t}^{i};X_{t+1}^{j}))\right ),
\end{equation}
which is the maximum of the difference between mutual information of $V_t$ and $X^j_{t+1}$ and the summation of the mutual information of $X_t^i$ and $X_{t+1}^j$.

\begin{equation}
\label{eq:rosas_emergence2}
   \Gamma _{t,t+1}  (V):= \max_{\substack {j}}{I(V_{t};X_{t+1}^{j})},
\end{equation}
which is the maximum mutual information between $V_t$ and $X^j_{t+1}$. For the above indicators, $V$ is a predefined macro-variable, but the specific method for finding such a macro-variable is not discussed in \cite{rosas2020reconciling}.

The specific usage of the indicators are as follows: Firstly, $ \Psi_{t,t+1}  (V)> 0$ is a sufficient condition for the causal emergence of $V_{t}$. Secondly, $\Delta _{t,t+1}  (V)> 0$ is a sufficient condition for $V_{t}$ to show downward causation. Thirdly, $\Psi_{t,t+1}  (V)> 0$ and $\Gamma _{t,t+1}  (V) = 0$ together constitute a sufficient condition for causal decoupling. 

In summary, Rosas proposed an approach to quantitatively characterize and classified causal emergence based on $\phi$ID by establishing the relationship between causal emergence and the synergistic effect of variables at different time points and further categorized causal emergence. The definition not only provides an objective assessment  of a system's capacity for causal emergence but also enables the measurement the causal emergence associated with a specific macro-feature. His significant contributions include bridging the gap between the study of causal emergence and quantitative empirical research, classifying the different types of causal emergence, and complementing philosophical discussions on the topic.

The PID and $\phi$ID frameworks hold the potential for explaining data in various applications. For instance, Luppi et al. recently employed the $\phi$ID method to analyze brain BOLD signals~\cite{luppi2022synergistic}, aiming to identify the synergistic core of the brain. Their findings revealed that synergistic information facilitates the integration of different brain regions, whereas redundancy contributes to robustness. This study offers valuable insights into understanding the underlying mechanisms of the brain.

The advancement of partial information decomposition techniques allows for further analysis of the mutual information between variables, enabling a deeper understanding of system properties from multiple perspectives. In a study by Varley et al.~\cite{varley2022emergence}, the authors applied partial information decomposition to decompose the mutual information of a system. They calculated an indicator of synergy bias to assess how synergistic information is distributed across different levels of the system using the method proposed by Williams and Beer~\cite{williams2010nonnegative}. A higher synergy bias indicated a greater amount of partial information involved in synergistic relationships. Subsequently, the authors observed that in certain systems exhibiting causal emergence, when the system is simplified or reduced, the synergy bias increases.~This suggests that as we coarse-grain the system, partial information undergoes a transformation from redundancy to synergy. The overall conclusion drawn was that emergence can be understood as a form of information conversion.

\subsubsection{Causal Emergence Identification from Data}
\label{sec:causalemergence_identification}
In the previous sections, we introduced several works on quantifying emergence through causality and other information-theoretic concepts. All of these works tried to propose quantitative measures, conceptual frameworks, and numeric examples based on Markovian dynamics for causal emergence. However, in practical applications, for a theoretical framework to be implemented, we need to automatically identify causal emergence from real data, especially time-series data of dynamic systems, and provide explanations for the results. 

The first method for causal emergence identification was introduced in Rosas's paper on causal emergence \cite{rosas2020reconciling} and involves the three indicators defined by Equations (\ref{eq:psi_identification})--(\ref{eq:rosas_emergence2}), as mentioned in Section \ref{sec:rosas}. The identification criteria were exemplified in three case studies, leading to the following conclusions: particle collisions emerge as a distinctive feature within Conway's Game of Life, flock dynamics emerge as a characteristic feature in simulated bird behavior, and the representation of motor behavior in the cortex emerges from neural~activity.

Although these three indicators avoid the problem of redundant information calculation, it is important to note that they serve as sufficient conditions, rather than as definitive proof of emergence. In other words, indicators greater than 0 can suggest the presence of emergence, but indicators less than 0 do not  necessarily imply the absence of emergence. The construction of this indicator faces challenges in identifying emergence in systems with significant redundant information or a large number of variables, which is often the case in many real-world systems. Additionally, a limitation of this method is the requirement of a predetermined coarse-grained variable $V$, and different choices of this variable can significantly impact the results. As a result, the development of an automatic coarse-graining strategy based on data remains an unresolved issue.

As a result, the current theoretical frameworks for causal emergence lack a practical and effective identification algorithm. Although previous studies have proposed methods based on static network structures and approximations for information decomposition, there is still a need for a comprehensive approach that can be applied to general Markovian dynamic systems. One of the main challenges is the necessity to search for all possible functions of coarse-graining or decomposing subsets within the data to identify causal emergence in complex systems. However, conventional numerical methods are unable to handle the computational costs associated with such an extensive search in a vast functional space. Therefore, the development of new methods is imperative to address these issues.

In Section \ref{sec:causalemergence_ai}, we explore the application of machine learning techniques to address the challenge of identifying causal emergence within time-series data.

\subsubsection{Comparison with Hoel's Framework}
When comparing Hoel's framework to Rosas's quantification framework for causal emergence, several clear advantages can be observed in the latter. Firstly, Rosas's theory does not require a predetermined coarse-graining method, making it more mathematically rigorous and formal. Secondly, it offers a detailed decomposition of causal emergence, specifically downward causation and causal decoupling. Lastly, it effectively avoids cases of fake causal emergence, where the macro-variable depends solely on unique or redundant information from the micro-variables.

However, there are also some disadvantages to consider. Firstly, in order to obtain the full information lattice, a systematic iteration of all variable compositions is required. Additionally, despite the use of Formula (\ref{eq:rosas_causal_uni}), it is necessary to define a macro-variable. Unfortunately, the authors do not provide any method to identify such a variable. Secondly, all the mutual information and its decomposition are based on correlations rather than causality. It is crucial to discuss how to incorporate causal elements, such as intervention and counterfactuals, into the framework.

Finally, the issue of identifying whether causal emergence occurs in a system based on the given time-series data of its behaviors has not been addressed in the preceding discussion. To address this problem, the application of emerging technologies in machine learning and artificial intelligence is required. These technologies can provide valuable tools and techniques for detecting and analyzing causal emergence.

\section{Causal Emergence and Machine Learning}
\label{sec:causalemergence_ai}
Recently, newly emerging machine learning technologies have made significant breakthroughs in addressing a range of important and challenging problems~\cite{silver2016mastering,dabney2020distributional, senior2020improved,lecun2015deep,wang2018cosface,mikolov2013efficient,devlin2018bert,vaswani2017attention,mnih2015human}. Examples include defeating human champions in complex games like Go~\cite{silver2016mastering, dabney2020distributional}, predicting the intricate structures of protein folding~\cite{senior2020improved}, and generating human language using large language models~\cite{fan2023bibliometric}. These achievements have been made possible through the application of machine learning methods, which leverage well-designed neural network architectures and automatic differentiation techniques. 

However, there is a crucial limitation of conventional machine learning: it can only capture the information or associations within the data, without being able to uncover the underlying causal relationships. According to the theory of causal hierarchy by Judea Pearl~\cite{pearl2000models,pearl2009causality,cui2022stable}, causality is very different from association because the former always represents a more stable relationship that is invariant in different environments, whereas the latter may be more dependent on contexts and reflects the limitation of the data. Therefore, it is necessary to develop new machine learning frameworks to incorporate the consideration of causality\cite{scholkopf2021toward,kaddour2022causal,cui2022stable}. Numerous studies have provided evidence to show that incorporating aspects of causality can lead to improved performance in machine learning tasks, including out-of-distribution generalization, adaptation to diverse environments, and handling interventions~\cite{peters2016causal,cui2022stable,kuang2018stable}.

Machine learning and causal inference are also connected to causal emergence in two aspects, which we call ``causal emergence with machine learning'' and ``causal emergence for machine learning''. On one hand, machine learning can be employed to address the challenge of identifying causal emergence from data, referred to as ``causal emergence with machine learning''. On the other hand, causal emergence theory and effective information (EI) measures have potential applications in the field of machine learning, known as ``causal emergence for machine learning''. However, the extension depends on a deeper understanding of how causal emergence theory is related to~causality.

\subsection{Causal Emergence with Machine Learning}
\label{subsec:ei_with_ml}
In Section \ref{sec:causalemergence_identification}, we discussed the challenge of identifying causal emergence. Although Rosas et al. presented a method to address this issue by calculating an approximate measure of causal emergence based on their theoretical framework (Equation (\ref{eq:psi_identification})), the method requires the availability of a macro-variable, denoted as $V_t$, which needs to be defined in advance. However, defining this variable in a general sense when confronted with data remains challenging. Hence, the question arises of whether we can leverage machine learning to automatically learn the macro-variable. In this section, we explore two recent approaches that utilize machine learning and neural network techniques to address the problem of causal emergence identification from the perspective of Heol et al.'s theoretical framework for EI maximization. These methods aim to learn the coarse-graining strategy, macro-variable, and macro-dynamics directly from the data, eliminating the need for explicit definitions. By leveraging the power of machine learning, we can potentially overcome the limitations of manual variable definition and enable a more automated and data-driven approach to identifying causal emergence.


\subsubsection{Neural Information Squeezer (NIS)}
\label{sec:nis}
The first work we introduce has just been published in this special issue. In this paper, the authors formulated the problem of causal emergence identification in continuous Markov dynamics. Suppose the time-series data $\boldsymbol{x}_{1}, \boldsymbol{x}_{2}, \dots, \boldsymbol{x}_{T}$ are generated by a stochastic dynamical system continuously, which can be described by a differential equation,

\begin{equation}
\label{eq:dynamical_model}
    \frac{d\boldsymbol{x}}{dt}=g\left(\boldsymbol{x}(t),\xi\right),
\end{equation}
where $\xi$ is a Gaussian noise. The time series $\{\boldsymbol{x}_t\}$ represents the observations of the micro-states of the system. The problem we confront is identifying whether causal emergence occurs for the original system. We also want to know the appropriate coarse-graining strategies, the macro-dynamics corresponding to the chosen coarse-graining method, and the dimension of the phase space at the macro-level when the causal emergence occurs.

To address this problem, the authors proposed a mathematical framework by converting the original issue into an optimization problem for functions. Specifically, the goal is to find the appropriate coarse-graining strategy~$\phi_q$, defined on $\mathcal{R}^p\rightarrow \mathcal{R}^q$, where $q<p$ is the dimension of the macro-states. Additionally, the aim is to determine the macro-level Markov dynamics $f$, defined on $\mathcal{R}^q\rightarrow \mathcal{R}^q$ with the random noise $\zeta$, by maximizing $\mathcal{J}$---the dimension-averaged EI of the macro-dynamics (see Equation (\ref{eq:dimension_averaged_ei})). This is expressed as,

\begin{equation}
\label{eq:maximize_ei}
\max \mathcal{J}(\hat{f}_{\phi_q}),  \qquad 
\end{equation}

However, this problem has a trivial solution, e.g., $\phi$ is a constant map for all micro-states, and the macro-dynamic is an identity. Surprisingly, this solution has a large EI because the identity map has the largest determinism and lowest degeneracy among all $q$-dimensional functions. This is trivial because too much information has been abandoned by the coarse-graining strategy such that the macro-dynamic is useless.

To address the problem, the authors added a reasonable constraint to the original optimization framework, called the effectiveness requirement. A coarse-graining strategy and its corresponding macro-dynamics are considered effective if there exists another decoarse-graining function $\phi^{\dagger}$ such that the functions $\phi$, $f$, and $\phi^{\dagger}$ can predict the micro-state, instead of the macro-state, in the next time step based on the state in the previous step. The constraint can be written as,

\begin{equation}
\label{eq:constraint}
||\phi^{\dagger}(f(\phi(\boldsymbol{x}_t)))-\boldsymbol{x}_{t+1}||<\epsilon,
\end{equation}
for all time steps $t$, where $\epsilon$ is a given constant. This way, an effective coarse-graining strategy and macro-dynamics not only maximize the EI but also reproduce the original micro-dynamics as much as possible. 

This constraint is an important complement to the framework for EI maximization because the decoarse-graining function $\phi^{\dag}$ can map macro-states back to micro-states, addressing the problem of the ambiguity of macro-states. This constraint also ensures commutativity. To demonstrate this, we can apply $(\phi^{\dag})^{-1}\approx \phi$ to both sides of Equation (\ref{eq:constraint}), resulting in $||f(\phi(x_t))-\phi(x_{t+1})||<\phi(\epsilon)$. 

However, the problem is still hard to solve. In~\cite{zhang2022neural}, the authors proposed a two-stage method, which can minimize the prediction error under a given dimension $q$ of macro-states in the first step and maximize EI for different values of $q$ in the second step. The first step can be solved by training a neural network, and the second step converts the complex functional optimization problem into a line search in one-dimensional space.

\begin{figure}[H]

\includegraphics[width=10cm,height=8.9cm]{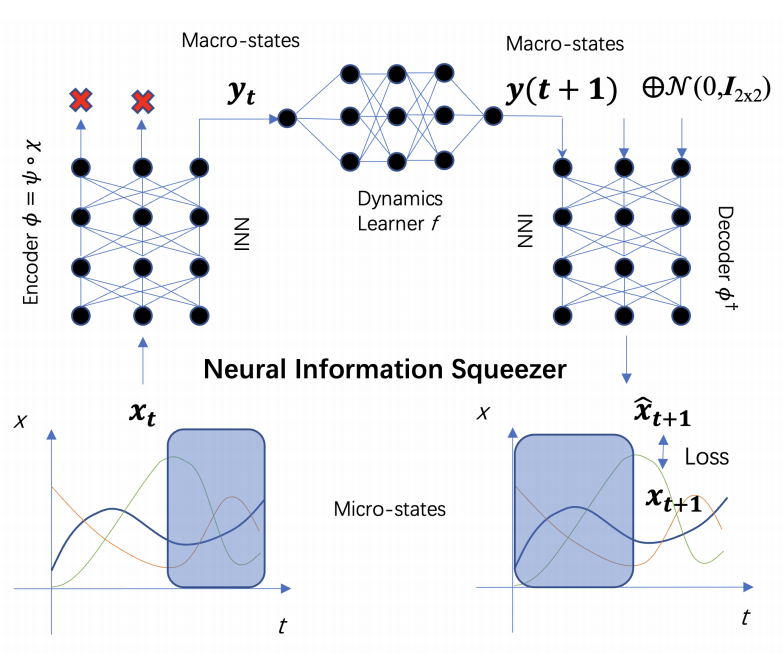}
\caption{The 
 architecture and workflow of the Neural Information Squeezer (NIS). The NIS consists of three main components: an encoder $\phi$, a dynamics learner $f$, and a decoder $\phi^{\dag}$. The encoder $\phi$ comprises an invertible neural network $\psi$ and a projection operator $\chi$, which simply discards fixed dimensions output by $\phi$. On the other hand, the decoder $\phi^{\dag}$ incorporates the same invertible neural network as the encoder but implements its inverse $\psi^{-1}$. In cases where the input is incomplete, it can be complemented using Gaussian random noise.}
\label{fig:nis}
\end{figure}

To realize 
the abstract mathematical framework, the authors proposed an encoder-dynamics \textls[-15]{learner-decoder framework for neural networks, named the Neural Information Squeezer} (NIS), vividly describing the basic working principle of the framework as depicted in 
Figure~\ref{fig:nis}. 

\vspace{-12pt}

In order to reduce the complexity of the encoder and decoder, corresponding to $\phi$ and $\phi^{\dagger}$, respectively, the authors used an invertible neural network to implement the encoder and inverted the input-output workflow of the same neural network to implement the decoder.  Concretely, the encoding process was decomposed into two computation steps: information conversion with the invertible neural network~$\psi: \mathcal{R}^p\rightarrow \mathcal{R}^q$ on a $p$-dimensional input space, and information discarding with a simple projection operation. This is expressed as:
\begin{equation}
    \phi(\boldsymbol{x})=\chi_q(\psi(\boldsymbol{x})),
\end{equation}
where $\chi_q$ is the projection operator that projects a $p>q$-dimensional vector into a $q$-dimensional vector, retaining dimensions from $1$ to $q$. The decoding process also consists of two stages: randomly sampling Gaussian noise to complete the details of macro-states, and information conversion but in an inverted way, that is,
\begin{equation}
    \phi^{\dag}(\boldsymbol{x})=\psi^{-1}(\boldsymbol{x}\bigoplus \zeta),
\end{equation}
where $\zeta\sim \mathcal{N}(0,I_{p-q})$ is a $p-q$-dimensional standard Gaussian random vector, and $\bigoplus$ is the operation of vector concatenation. 

The adoption of an invertible neural network in the framework serves two main purposes. Firstly, by inverting the workflow of the neural network $\psi$ to obtain the decoder $\phi^{\dag}$, a significant reduction in the number of parameters is achieved. This reduction can lead to improved efficiency and computational benefits. Secondly, the use of invertible functions in the neural network provides favorable mathematical properties that facilitate the analysis of the entire framework in a tractable manner. These properties enable researchers to gain insights and understand the characteristics of the framework more easily.

This architectural design with an invertible neural network not only reduces the complexity of computation but also separates the input information from the micro-states into noise and EI for prediction. The authors proved a mathematical theorem stating that, as the neural network converges through training, the mutual information between adjacent macro-states tends to approach the mutual information of two adjacent-step micro-states in the data. This means that all the information discarded by the encoder is almost irrelevant to the prediction. {All the useful information converges into the information bottleneck within the macro-level dynamics learner.} 
{Another theorem demonstrates that the degree of information compression for micro-states increases as we reduce the dimension of the macro-states.} 
{Thus, the macro-state dynamics learner, acting as a narrower information bottleneck, progressively discards more and more extraneous information from the original micro-state data.} 

To address the problem of causal emergence identification, we first optimize the neural networks, and the dimension-averaged EI, $\mathcal{J}_q$, can be obtained for different $q$. When $q=p$, the derived $\mathcal{J}_p$ is the dimension-averaged EI for the micro-dynamics. Finally, the calculation~of 
\begin{equation}
    \Delta \mathcal{J}=\mathcal{J}_q-\mathcal{J}_p
\end{equation}
is the measure of dimension-averaged causal emergence (see Equation (\ref{eq:dim_CE}). If $\Delta\mathcal{J}>0$, we can state that causal emergence occurs.

Numeric examples are provided to show the effectiveness of the whole framework. Interestingly, the Neural Information Squeezer can handle time-series data with discrete micro-states, even though the whole framework was initially designed for continuous dynamical systems. Particularly, the model determines the same coarse-graining strategy and macro-level dynamic as illustrated in the example of a boolean network with four nodes in Hoel's original paper, although no information about node grouping and state mapping was provided in the~framework.


It is also interesting to compare the NIS with the frameworks and models mentioned in previous sections. Compared to the theory of computational mechanics, the NIS can be treated as a kind of $\epsilon$-machine because any encoded macro-states can be understood as states. When the whole framework is well trained, such that it can precisely predict future micro-states, the encoded macro-states converge to effective states, which can be treated as causal states in computational mechanics.  However, the objective that maximizes the effective information (EI) in the NIS has no correspondence.    

The NIS also shares a similarity with G-emergence because it adopts the ideas of Granger causality: the effective macro-state is optimized by predicting the micro-state in the next time step. However, there are several obvious differences between the two frameworks. In the theory of G-emergence, the macro-state must initially be selected manually, although it is optimized automatically. In addition, the NIS uses neural networks to predict future states, whereas G-emergence uses auto-regressive techniques to fit the data. Due to the universal approximate theorem (see~\cite{hornik1991approximation,teshima2020coupling}), neural networks are more advantageous.

The NIS provides a solution for causal emergence identification from data, and it is the first solution for addressing the problem of the automatic construction of the coarse-graining strategy and macro-dynamics from data for causal emergence identification. 

Another important benefit of the NIS is that it essentially addresses a generalized causal emergence identification problem, extending beyond mere justification. This is because following training, the NIS not only identifies whether causal emergence occurs but also obtains a neural network for coarse-graining the data and another for simulating the macro-dynamics. This can provide us with richer information compared to a mere justification.

\subsubsection{Neural Information Squeezer Plus (NIS+)}
\label{sec:nis+}
One of the biggest problems of the NIS is that it does not directly address the EI maximization problem by optimizing the neural networks (encoder, decoder, and dynamics learner). Instead, it searches for the optimal dimension of the macro-state space $q$ such that EI can be maximized. Therefore, for a given $q$, the neural networks are optimal for reconstructing the micro-dynamics (minimizing Equation (\ref{eq:constraint})) but not for EI. In addition, the NIS cannot support complex computations such as coarse-graining functions involving multiple steps of information conversion and discarding. 

\begin{figure}[H]
	\centering
	\includegraphics[width=\textwidth]{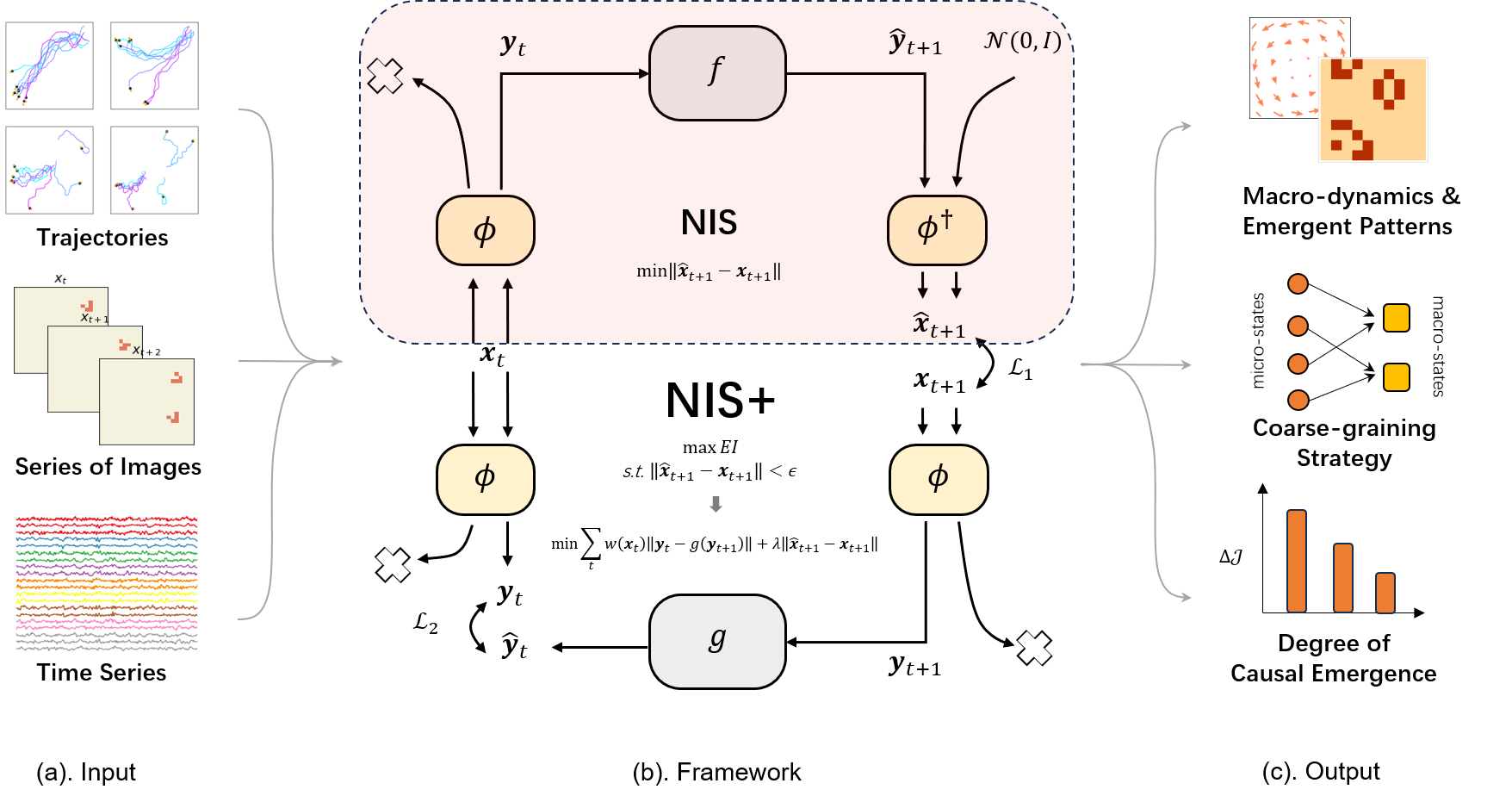}
	\caption{The workflow and architecture of our proposed framework, the Neural Information Squeezer Plus (NIS+), for discovering causal emergence within data. (\textbf{a}) Various forms of input data. (\textbf{b}) The framework for our proposed model, the NIS+, incorporates our previous model, the NIS. The boxes represent functions or neural networks, and an arrow pointing to a cross represents the operation of information discarding. $\boldsymbol{x}_t$ and $\boldsymbol{x}_{t+1}$ represent the observational data of micro-states, whereas $\hat{\boldsymbol{x}}_{t+1}$ represents the predicted micro-state. $\boldsymbol{y}_t=\phi(\boldsymbol{x}_{t})$ and $\boldsymbol{y}_{t+1}=\phi(\boldsymbol{x}_{t+1})$ represent the macro-states obtained by encoding the micro-states using the encoder. $\hat{\boldsymbol{y}}_t=\phi(\hat{\boldsymbol{x}}_{t})$ and $\hat{\boldsymbol{y}}_{t+1}=\phi(\hat{\boldsymbol{x}}_{t+1})$ represent the predicted macro-states obtained by encoding the predictions of micro-states. The mathematical problems that each framework aims to solve are also illustrated in the figure. (\textbf{c}) The various output forms of the NIS+, which include the learned macro-dynamics, captured emergent patterns, functions of coarse-graining, and the degree of causal emergence. }
	\label{fig:nis+}
\end{figure}

To address these problems, the NIS+ was proposed, as discussed in~\cite{yang2023finding}. Mathematically, the problem of EI maximization can be transformed into a machine learning problem based on the definition of EI and the variational inequality and probability reweighting technique. After this conversion, the minimization problem can be solved by training three neural networks $\psi_{\omega}, f_{\theta},$ and $g_{\theta'}$ with parameters $\omega,\theta$, and $\theta'$, respectively. Formally, the minimization problem without constraints can be written as:
\begin{gather}
\label{new optimization}
\min_{\omega,\theta,\theta'} \sum_{t=1}^{T-1} w(\boldsymbol{x}_t)||\boldsymbol{y}_t-g_{\theta'}(\boldsymbol{y}_{t+1})||+\lambda||\hat{\boldsymbol{x}}_{t+1}-\boldsymbol{x}_{t+1} ||,
\end{gather}
where $\boldsymbol{y}_t=\phi(\boldsymbol{x}_t)=Proj_q(\psi_{\omega}(\boldsymbol{x}_t))$ and $\boldsymbol{y}_{t+1}=\phi(\boldsymbol{x}_{t+1})=Proj_q(\psi_{\omega}(\boldsymbol{x}_{t+1}))$ are the macro-states.  $\lambda$ is a Lagrangian multiplier, which is regarded as a hyperparameter in experiments. $w(\boldsymbol{x}_t)$ is the inverse probability weight, which is defined as:

\begin{equation}
    \label{eq:probreweight}
    w(\boldsymbol{x}_t)=\frac{\Tilde{p}(\boldsymbol{y}_t)}{p(\boldsymbol{y}_t)}=\frac{\Tilde{p}(\phi(\boldsymbol{x}_t))}{p(\phi(\boldsymbol{x}_t))},
\end{equation}
where $\Tilde{p}$ is the new distribution of macro-states $\boldsymbol{y}_t$ after intervention for $do(\boldsymbol{y}_t\sim U_q)$, and $p$ is the natural distribution of the data. In practice, $p(\boldsymbol{y}_t)$ is estimated through kernel density estimation (KDE)~\cite{rosenblatt1956remarks}. The approximated distribution, $\Tilde{p}(\boldsymbol{y}_t)$, is assumed to be a uniform distribution, which is characterized by a constant value. Consequently, the weight $w$ is computed as the ratio of these two distributions. Theorems have been proven to ensure the correctness of this conversion. The overall framework for the NIS+ is depicted in Figure 
 \ref{fig:nis+}.

Additionally, to apply the NIS+ to data generated by complex systems like multi-agent systems and cellular automata, the structures of the encoder can be extended by defining two kinds of combinations: stacking the basic encoder and concatenating the basic encoder. Here, a basic encoder is the composition of an invertible neural network and a projection operation. By inverting the directions of all the basic encoders and supplementing empty inputs with Gaussian noise vectors, we can obtain the corresponding decoder.

Serious numerical experiments have demonstrated that the NIS+ can automatically learn emergent dynamics and the coarse-graining strategy from data and can integrate these results to identify causal emergence. 

For instance, in experiments utilizing training data generated by introducing Gaussian noise to SIR dynamics, the NIS+ demonstrated superior performance compared to the NIS in recovering the vector field of derivatives that aligned with the actual macro-level SIR dynamics. To assess the ability of the NIS+ to learn emergent dynamics, the authors generated data based on macro-states using the SIR dynamics as the ground truth. However, the NIS+ cannot directly utilize this data; only micro-level noisy data can be employed. These micro-level data were generated by introducing Gaussian noise to the macro-states.

In the experiments conducted with the Boid model, training data were generated using the renowned Boid model, which simulates the herding behavior of birds. Both the NIS+ and NIS demonstrated the ability to automatically learn macro-level states and dynamics. In this simulation, the authors divided all artificial birds (boids) into two distinct groups. Interestingly, the learned macro-states could also be categorized into two groups. Within each group, the predicted trajectories generated by the learned macro-dynamics of this group accurately followed the center of the boid groups.

The ability to capture emergent patterns was verified through experiments conducted in the ``Game of Life''. The NIS+ demonstrated the capability to automatically discover both static and dynamic patterns, such as the ``glider'', within the learned latent space. Additionally, when clear emergent patterns appeared in the data, larger effective information (EI) was observed for the learned emergent macro-dynamics.

Across all the training data generated through simulations, the NIS+ outperformed comparative models, including the NIS, variational autoencoders, and feed-forward neural networks, in experiments testing the model on data distributions that differed from the training data.

The paper also presented results on real data. Two sets of real fMRI time-series data from human subjects were used. The first dataset, AOMIC ID1000~\cite{snoek2021amsterdam}, contained fMRI scanning data collected while subjects watched the same movie clip. As a comparison, another fMRI dataset, AOMIC PIOP2~\cite{snoek2021amsterdam}, containing resting-state data from 50 subjects, was also utilized. The primary training objective was to predict the fMRI signal at the next time step using the learned macro-dynamics and decoder while simultaneously maximizing the effective information (EI) of the macro-level dynamics. Ultimately, a one-dimensional macro-state was found to represent all 100-dimensional micro-state data. Through attribution analysis, the authors found that micro-level signals with larger weights contributing to the one-dimensional macro-state were located in areas associated with visual tasks. However, in the comparison dataset with resting-state data, seven-dimensional macro-states were found to represent the micro-states, and attribution experiments revealed that micro-states with larger weights were distributed across different brain areas.

After the macro-dynamics and coarse-graining strategy were learned, the EIs at both the macro-level and micro-level could be calculated so that causal emergence could be quantified. The experiments showed that the measure of causal emergence depends on different types of noise. If noise was added to the observational data (micro-state), the degree of causal emergence increased with the level of noise. In contrast, if the noise was from the dynamical mechanism, i.e., added to macro-states, the degree of causal emergence decreased with the level of noise. This implies that coarse-graining can learn to eliminate noise in raw data so that better macro-dynamics with larger EI can emerge. However, if the noise is from the internal dynamics, it cannot be removed through coarse-graining.

\subsection{Why is Causal Emergence ``Causal''?}
In the previous sections, we discussed two frameworks that utilize machine learning and neural network techniques to address the challenge of causal emergence identification. Both frameworks place significant emphasis on maximizing EI to learn the coarse-graining strategy and macro-dynamics. Notably, the NIS+ framework extends the NIS approach to achieve real EI maximization, whereas the NIS relies on adjusting the hyperparameter $q$ to optimize EI.

Moreover, the NIS+ has demonstrated an additional capability to generalize data distributions that differ from the training data. This suggests that the NIS+ can learn an invariant causal mechanism that is independent of the input distribution shifts through EI maximization. However, before extending this method to other machine learning scenarios, an essential question needs to be addressed: Is there a genuine relationship between causal emergence, EI maximization, and causality itself?

In the previous sections, we introduced two theoretical frameworks of causal emergence: Hoel's framework and Rosas's framework.  In both frameworks, the measurement of causal emergence heavily relies on mutual information. However, it is crucial to acknowledge that mutual information primarily quantifies correlation rather than causality. So why do the authors assert that their theories pertain to ``causal'' emergence?

The key point lies in the research objectives of these frameworks, which focus on a Markov dynamical system. In this context, the dynamical mechanism is assumed to be known, and confounding variables are not considered. Under this assumption, measures related to correlation can effectively indicate causality. While mutual information itself is not a direct measure of causality, it can serve as a proxy for causality when the underlying dynamics are well understood and confounding factors are absent. Therefore, in the context of these frameworks, the use of mutual information as a measure for causal emergence is justified because it captures the relationships between variables in a way that aligns with the assumed known dynamics of the system.

When comparing the two frameworks, Hoel's framework can be considered more ``causal'' than Rosas's framework. This is because Hoel's framework introduces the do operator in the definition of effective information (EI), as expressed in Equation (\ref{eq:do_def_ei}). The do operator allows for interventions and represents a causal manipulation of variables, which strengthens the causal interpretation of the framework. Furthermore, the significance of EI lies in its deep connections with other measures of causation proposed in various fields~\cite{comolatti2022causal}. According to the information in Section \ref{sec:measure_comparisons}, these measures aim to quantify the causal impact of an intervention or treatment on an outcome. By incorporating the do operator, EI aligns with these causal effect measures and provides a valuable tool for assessing causal relationships. 

Furthermore, let us delve into the application background of the NIS and NIS+ to understand why these frameworks, based on EI maximization, are considered causal. Firstly, the input data in both frameworks, denoted as $x_t$, are generated by a Markovian dynamical system, with no other confounding factors considered, apart from the system itself. As a result, the state of two consecutive time steps, represented as $x_t$ and $x_{t+1}$, without loss of generality, can be modeled by a simple causal diagram $x_t \rightarrow x_{t+1}$ due to the Markovian property. This causal diagram naturally adheres to the exogeneity assumption in causal inference \cite{pearl2009causality}, indicating the absence of any confounders between the two states. Consequently, the ``do'' operator in EI can be replaced with conditional probability, which can be estimated from observations.

Taking all these aspects into consideration, we conclude that the theoretical frameworks of causal emergence and the machine frameworks for causal emergence identification, namely the NIS and NIS+, are indeed related to causality. Moreover, they can be extended to other scenarios within the realm of machine learning. We further explore and discuss this point in the subsequent sections.

\subsection{Causal Emergence for Machine Learning}
\label{subsec:ei_for_ml}

In this sub-section, we introduce how causal emergence can enhance machine learning in out-of-distribution scenarios. It turns out that the \textit{do} intervention introduced in EI captures the causal dependency from the data generation process and thus complements correlation-based machine learning algorithms.


\subsubsection{Out-of-Distribution Generalization}

As demonstrated in the previous subsection, the measurement of causal emergence by vanilla effective information~\cite{hoel2013quantifying} is defined for stationary dynamics systems in a self-supervised fashion since the input (cause) space and the output (effect) space share the support. However, Hoel generalized effective information to general input-output systems~\cite{chvykov2020causal} (see Section \ref{sec:continuous_EI}), where the cause and the effect space take arbitrary support, which could either be discrete or continuous. Empowered by the general notation of EI, causal emergence could be applied in supervised machine learning to evaluate the strength of causation between the feature space $\mathcal X$ (or its learned representation) and the target space $\mathcal Y$, thus enhancing the prediction from the cause (feature) to the effect (target). It is worth noting that a direct fitting from $\mathcal X$ to $\mathcal Y$ on the observations suffices for common prediction tasks with an i.i.d. assumption, implying that the training and test data are independently and identically distributed.


However, if samples are drawn from outside the training distribution, it is essential to learn a representation space that generalizes from the training to the test environment, known as out-of-distribution (OOD) generalization. Due to the common belief that causal relations generalize better than statistical correlations~\cite{arjovsky2019invariant}, causal emergence theory could serve as a criterion for the causation embedded in the representation space. The occurrence of causal emergence indicates the revelation of potential causal factors of the target, thus producing a robust representation space with regard to out-of-distribution generalization.

\subsubsection*{Introduction to OOD}

Before the detailed discussion of the connection between causal emergence and OOD generalization, we begin with a brief introduction to the latter topic. Out-of-distribution generalization toward arbitrary distributional shifts is generally considered impossible~\cite{ye2021towards}. Therefore, it is commonly assumed that there exists an underlying causal mechanism governing the data generation process, which remains invariant to distributional shifts. Formally, according to Pearl's Structural Equation Model~\cite{pearl2009causality}, which causally formulates data generation, the target $Y$ is supposed to be generated from a part of the feature $X$, i.e., $Y = f(X_{c},\epsilon)$, where $X_{c}$ represents the selected features or a representation of $X$, and $\epsilon$ is a random noise independent of $(X,Y)$. Obviously, $X_{c}$ is the true cause of the target, and the causal mechanism $f$ is presumed to be invariant across training and test environments. On the other hand, the raw feature $X$ might comprise a mixture of causal variables and non-causal variables. The latter could arise from confounding, selection bias, or anti-causal mechanisms, all of which introduce spurious correlations between the target and non-causal variables. The strength and direction of these spurious correlations vary across environments, resulting in distributional shifts that hamper the out-of-distribution performance of machine learning techniques. As a remedy, a prominent branch of OOD algorithms~\cite{arjovsky2019invariant, huh2022missing, zhao2019learning, ganin2016domain} learns an invariant representation space distilling the causal variables from the raw feature space. Typically, a metric of the representation is devised to verify the stability of the correlation between the learned representation space and the target and is adopted as a regularization for the optimization objective to encourage the representation learner to extract causal variables from features. The paradigm for invariant representation learning can be expressed as~\cite{wang2022generalizing}:
\begin{align}
\min_{f,\phi} \mathcal R(f(\phi(X)), Y) + \lambda \ell_{reg}(\phi), 
\end{align}
where $\phi(X)$ is the learned representation, $f$ is the learned predictor, $\mathcal R$ is the task-specific risk function, and $\ell_{reg}$ is the regularization.

Representation learning-based OOD algorithms are diversified in their constraints on the representation space characterized by the regularizers. Each metric of the representation fits a specific data generation pattern and does not guarantee to filter out the causal variables when the setting is violated. However, causal emergence potentially offers a united metric of representation for OOD generalization rooted in causality theory. 

Before diving into the application of causal emergence, we first discuss the representative regularizers of representation learning and their limitations. Domain-invariant representation learning proposes an aligned feature space with the constraint $g(X) \perp \mathcal E$, where $\mathcal E$ denotes the environment index for training and test environments. The regularization is implemented by the metric of Maximum Mean Discrepancy (MMD) distance of representation across environments~\cite{li2018domain} or an adversarial classifier for the environment index~\cite{ganin2016domain}. However, the alignment constraint provides a generalization guarantee only for a covariate shift~\cite{zhao2019learning}, where the distribution shift is restricted to the features $X$ without perturbing $P(Y|X)$. On the other hand, invariant risk minimization~\cite{arjovsky2019invariant,koyama2020out} pursues an invariant conditional distribution $P(Y|g(X))$ of the representation across environments, which translates to $Y \perp \mathcal E \;|\; g(X)$. Invariant risk minimization has been proven to recover the causal variable in a data generation process where the non-causal variable is anti-causally generated by the target but is reported to fail under a covariate shift~\cite{ahuja2021invariance}. Further, conditional invariant representation~\cite{li2018domain, huh2022missing} is designed specifically for image classification where the target is modeled as the cause of the feature instead of the effect. The conditional independence $g(X) \perp \mathcal E \;|\;Y$ is enforced on the representation space.

It has been theoretically proven that any statistical independence regularization is valid for a subgroup of Structural Equation Models, and it does not generalize across all data generation processes~\cite{kaur2022modeling}. As a result, the metric of representation manages to capture the causation between the representation space and the target variable only if a specific causal graph of data generation is given, hindering the general applicability of invariant representation learning. Worse still, the existence of the environment index $\mathcal E$ in the independence regularizers introduces the necessity of multiple environments for training procedures, increasing the data collection and labeling expenses. To this end, there is active demand for a unified and general metric of representation for OOD generalization, for which causal emergence might be a good candidate.

\subsubsection*{Causal Emergence and OOD}

The link between OOD generalization and causal emergence can be attributed to the do intervention of effective information. By definition~\cite{pearl2009causality}, intervening in the feature distribution with $do(X)$ modifies the structural equations of all the variables in the feature space to a constant value, whereas the equation for the target remains intact as $Y=f(X_c)$. On one hand, the assumption of the causal mechanism's invariance to distributional shift is respected since the causal correlation between causal variables and the target is reserved. On the other hand, the spurious correlation between non-causal variables and the target is removed by the intervention. For example, a non-causal variable anti-causally correlated with the target leads to a structural equation as $X_n = h(Y,\epsilon)$, and $do(X_n)$ replaces the equation with $X_n = U$, where $U\sim U_{max}$, and $U_{max}$ represents the maximum entropy distribution on the domain of $X_n$. As a result, $X_n$ is no longer caused by $Y$, and the spurious correlation vanishes. In another example, if a non-causal variable is correlated with the target by a confounder $E$ (probably in the feature space), the structural equation takes the form of $X_n = \phi(E,\epsilon)$ and $Y = f(X_c,E,\epsilon)$. Then, the do intervention would modify the equation to $X_n = U$ and confounding ceases to exist. In summary, the intervention introduced by EI captures the causal correlation between features and the target while suppressing the spurious correlation. It renders EI an ideal metric to assess the causality contained in the representation space and, thus, the generalization of the induced algorithm. 

Recall that an increment in effective information implies the occurrence of causal emergence. Therefore, we propose a conjecture that out-of-distribution generalization could be achieved simultaneously with maximizing EI (or normalized EI) for representation learning. Notably, Shannon's mutual information holds the property that any representation has no gain of mutual information with the target over the raw feature space, i.e., $I(g(X);Y) \leq I(X;Y)$, as per the data processing theorem. Thus, direct fitting inclines to absorb as much information with the target as possible into the representation space. In contrast, EI (or normalized EI) would possibly peak at a medium stage of abstraction of raw features, coinciding with the philosophy of OOD generalization that less could be more. Ideally, at the peak of EI (or normalized EI) where causal emergence occurs, all non-causal features are excluded, and the causal features are revealed, resulting in the most informative representation while remaining invariant to distributional shifts.

The removal of spurious correlations and recovery of causal mechanisms by manipulating feature distributions are principles shared by the reweighting technique widely adopted in the OOD literature. For instance, stable learning~\cite{shen2020stable,shen2020stable2,zhang2021deep} is designed for scenarios where collinearity exists in input variables, causing spurious correlations between non-causal variables and the target. It achieves this by reweighting training samples to decorrelate the features, thus reducing collinearity and eliminating spurious correlations. Similar to EI, stable learning is free of environment index labeling. Further, both sample reweighting and feature decorrelation share the philosophy of distribution intervention. In this sense, EI could be viewed as an information-theoretic abstraction of reweighting-based debiasing techniques for OOD~generalization.

\section{Discussion and Perspectives}
We have presented several quantitative theoretical frameworks on causal emergence. However, there are numerous unexplored problems and implications that warrant further discussion. In this section, we address four key topics for future research: causal emergence and causal representation learning, ontological and epidemiological causality and emergence, potential applications in complex systems, and understanding complex systems from the perspective of causal emergence.

\subsection{Causal Emergence and Causal Representation Learning}

Causal representation learning (CRL) is an emerging field in artificial intelligence (AI)~\cite{scholkopf2021toward} that combines two important fields in AI: representation learning and causal inference. Representation learning aims to extract important features~(or representations) hidden in data to make predictions automatically~\cite{bengio2013representation}. It can be regarded as a typical application of deep learning and has achieved remarkable success in various domains such as image classification~\cite{lecun2015deep}, face recognition~\cite{wang2018cosface}, language understanding~\cite{mikolov2013efficient,devlin2018bert,vaswani2017attention}, and game playing~\cite{mnih2015human,dabney2020distributional}. However, conventional representation learning suffers from a critical limitation: it can only capture the information of associations within the data but not the underlying causal relationships. Therefore, it is important to consider causality in representation learning.

{To address the problem, CRL tries to combine the advantages of the two sub-fields: representation learning and causal inference to extract the important features and the relationships with causation behind data automatically~\cite{scholkopf2021toward}.}

\begin{figure}[H]

\includegraphics[width=12cm]{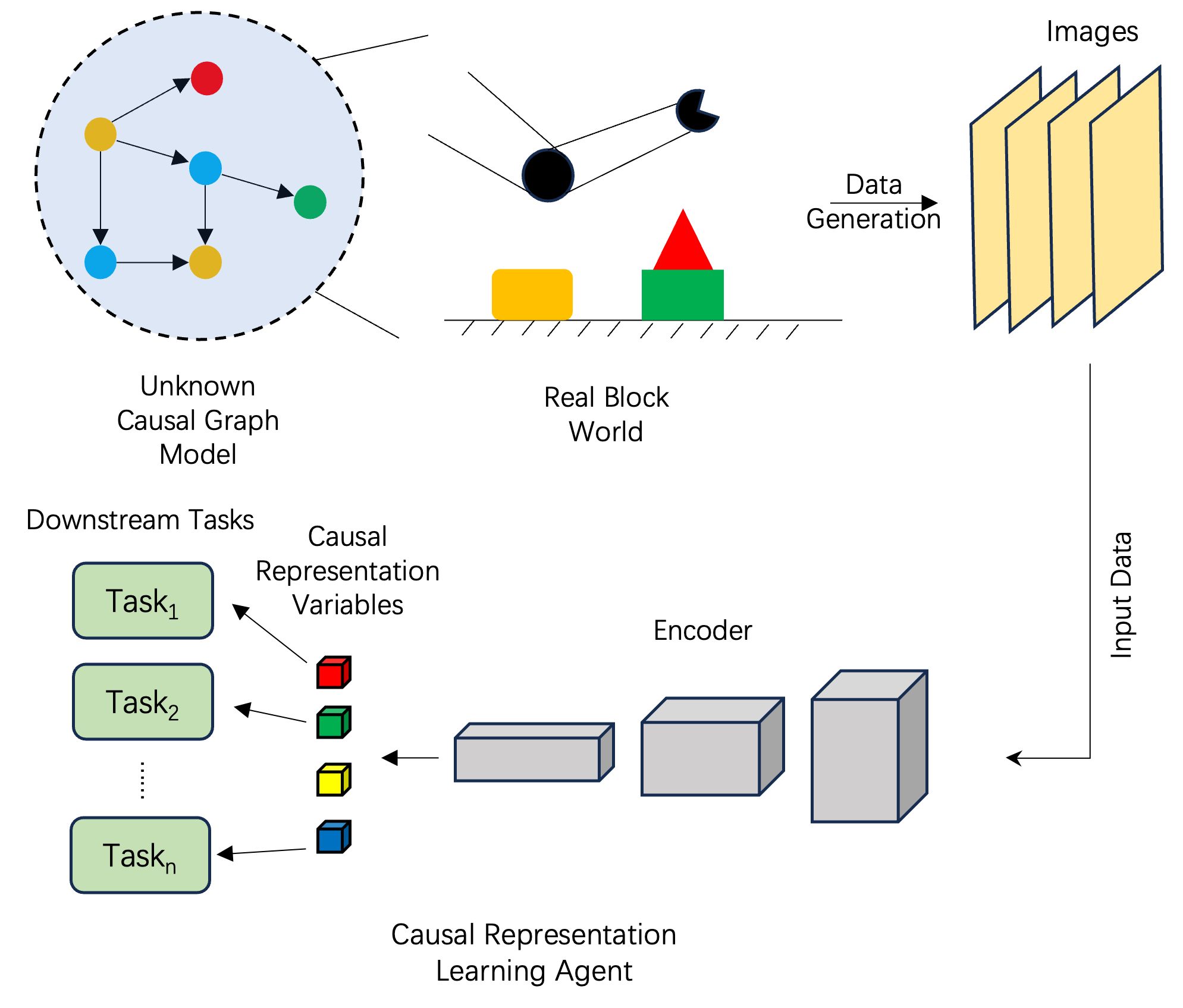}
\caption{Illustration of the workflow of a typical causal representation learning (CRL) agent. The top panel depicts a real block world with a robotic arm and the corresponding causal graph, representing relationships between variables in the real world, which is totally unknown to the agent. Additionally, the data are presented as images generated by the block world. The bottom panel depicts the structure of the causal representation learning agent. It utilizes a convolutional neural network (CNN) to extract meaningful representations from the data and perform downward tasks based on these representations. The encoder plays a crucial role in capturing relevant features and patterns from the input data, enabling the agent to learn and understand causal relationships.}
\label{fig:causal_representation_learning}
\end{figure}

A typical CRL scenario is depicted in Figure~\ref{fig:causal_representation_learning}. Suppose there is a set of variables and causal mechanisms (including causal graphs and structural equations) that describes how the world of robotic arms and colorful blocks works. However, these mechanisms are not directly observable for the agent undergoing causal representation learning. Instead, what the agent can observe is a set of images generated by robotic arms and blocks. The objective of a causal learning framework is to extract the causal variables and mechanisms from the observed images, and the variables and mechanisms can be used to implement other downstream tasks, for example, causal relationship discovery~\cite{monti2020causal} and predictions in different environments~\cite{peters2016causal} or domains~\cite{cui2022stable,kuang2018stable}.

\vspace{-2pt}

When comparing Figures \ref{fig:nis} and \ref{fig:nis+} with Figure \ref{fig:causal_representation_learning}, we found that as a framework for machine learning, Neural Information Squeezer Plus can be regarded as a kind of CRL framework. Thus, with the NIS and NIS+, we effectively employ CRL to address the problem of identifying causal emergence (CE). Both CE and CRL frameworks feature an encoder and decoder designed to represent raw data as causal variables. Latent causal mechanisms can be learned in both frameworks. In this section, we compare the similarities and differences between these two emerging fields.

\subsubsection{Similarities}

It is interesting to compare the tasks of causal emergence identification within the theoretical framework of Erik Hoel and causal representation learning. Actually, the task of causal emergence identification can be understood as a task of causal representation learning, where the macro-variables are the causal variables, the macro-dynamics are the causal mechanisms, and the coarse-graining strategy is an encoding process that transforms the original data into representations. Within this framework, EI can be understood as a measure of the strength of the causal effect on the mechanism. To identify causal emergence, we need to learn the appropriate causal variables to represent data and discover causal mechanisms at the macro-level (in the representational latent space) to ensure that the EI of the learned causal mechanism in the latent space is larger than that of the original data. Table \ref{tab1} shows a detailed comparison of causal emergence identification and causal representation learning.

\begin{table}[H]
\small

\caption{Comparison of causal emergence identification and causal representation learning.}
\label{tab1}

\setlength{\tabcolsep}{3.7mm}

\begin{tabular}{ccc}
\toprule 
\textbf{} & \textbf{\makecell[c]{Causal Emergence \\Identification}} & \textbf{\makecell[c]{Causal Representation \\Learning}}  \\
\midrule  
 
Data &   \makecell[c]{Observations (time series) \\of micro-states} & \makecell[c]{Raw data generated by some causal mechanism in real life}\\  
Latent variables&   Macro-states & Causal representations  \\ 
Causal mechanism &   Macro-dynamics & Causal mechanisms  \\ 
\makecell[c]{Mapping between data 
 \\and latent variables} &   Coarse-graining & Representation  \\ 
\makecell[c]{Optimization for \\causality}  & EI maximization & Prediction loss, disentanglement\\ 
Objective &   \makecell[c]{Finding an optimal coarse-graining strategy and a macro-dynamic that has a stronger causal effect}& \makecell[c]{Finding an optimal representation of the raw data to ensure that the independent causal mechanism can be realized by the representations}\\ 
\bottomrule  
\end{tabular}

\end{table}

With these similarities, the technologies and concepts from both fields can learn from each other. For example, the techniques of causal representation learning can be applied to discover causal emergence. 

On the other hand, the learned abstract causal representations can be viewed as macro-states, which enhances the explainability of causal representation learning. This perspective suggests that CRL essentially uncovers hidden causal emergent features within the data.

Moreover, these similarities between emergent phenomena and CRL contribute to a deeper understanding of emergence itself. By applying CRL frameworks to data generated by dynamical systems exhibiting emergent phenomena, we can extract more profound causal structures. These profound causal structures may serve as the origins of the mysteries surrounding emergence. By delving into the depths of CRL and uncovering the hidden causal relationships within complex systems, we can gain insights into the mechanisms behind emergent phenomena. These causal structures may provide a foundation for understanding the emergence of novel properties and behaviors that arise from the interactions of simpler~components. 

\subsubsection{Differences}

\begin{figure}[H]

\includegraphics[width=12cm]{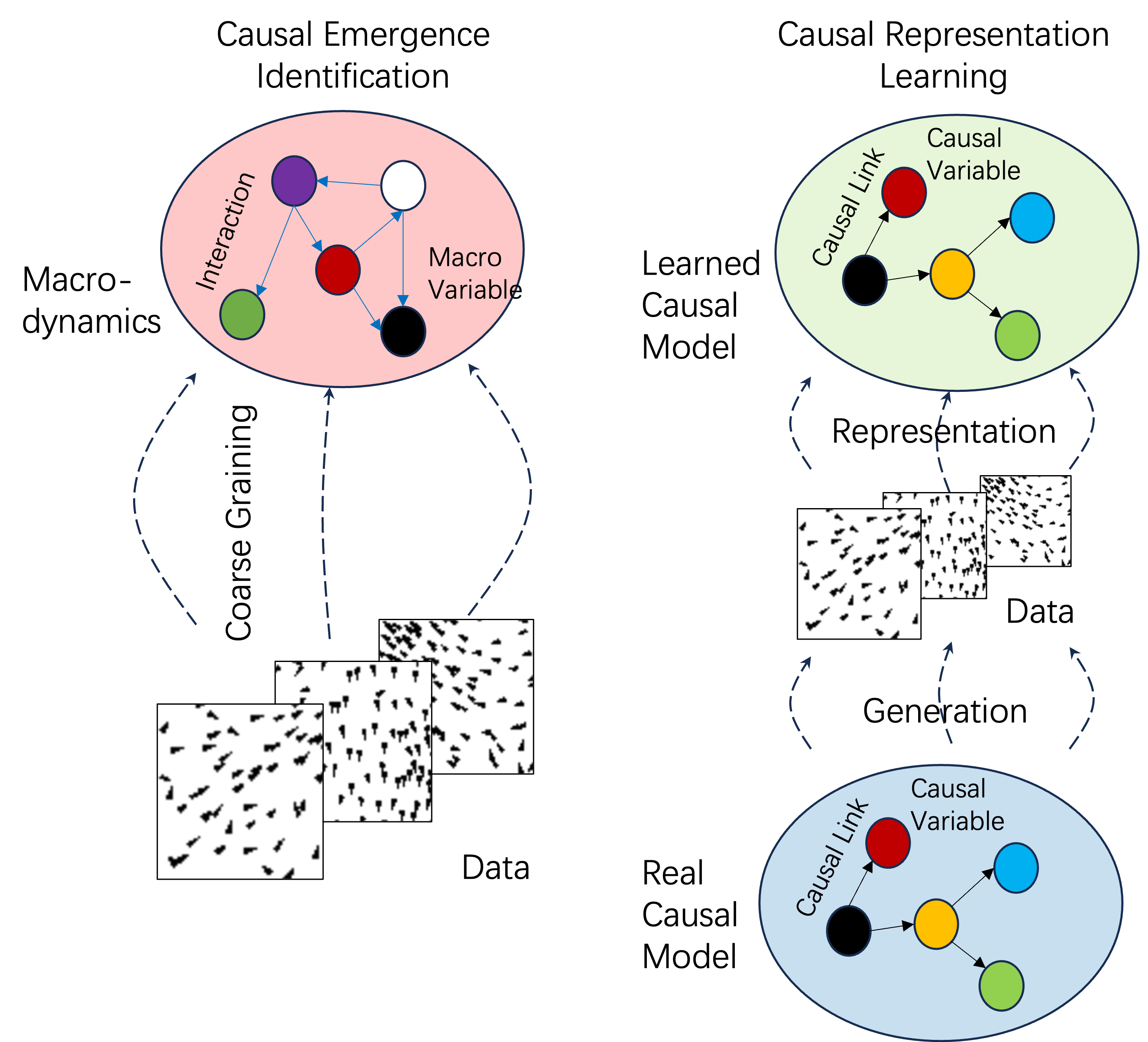}
\caption{Comparison of causal emergence identification and causal representation learning. All the potential co-founders are ignored. The black arrows represent causal relations, whereas the blue arrows represent interactions between macro-variables. The dashed lines with arrows represent the information flows for coarse-graining, representation, or data generation.}
\label{figure15}
\end{figure}

However, there are several theoretical differences between CRL and causal emergence. The biggest difference is that CRL assumes that there is a real causal mechanism behind the data, and the data are generated by this causal mechanism~\cite{scholkopf2021toward}. In contrast, for causal emergence identification, the emergent variables and mechanisms at the macro-level are just handy ways to observe and understand data, and ``real causality'' may not exist.  The differences are illustrated in Figure~\ref{figure15}.

However, if an epistemological perspective is adopted, this difference disappears because both approaches extract meaningful information from observational data to obtain representations with stronger causal effects. 

Another major difference is that the causal mechanism for macro-states in causal emergence is a dynamical system. Therefore, if the dynamics in the macro-state space are on a network, circular structures may exist because the state variables will be iterated in the dynamical system. As a result, feedback and circular indirect interactions are allowed in such a model. However, the causal mechanism is always represented by a structural causal model in CRL. Although these models are always directed acyclic graphs, loops or circular structures are not allowed. However, these differences are not important because Markovian dynamics can always be converted to causal models, as mentioned in Section \ref{sec:causal_hierarchy}.

Finally, unknown co-founders are always ignored in causal emergence, whereas they play very important roles in causal structural models.

In summary, there are deep connections between causal emergence and CRL. On one hand, the machine learning framework for causal emergence identification exhibits similar structures to the frameworks for CRL. And even the concepts can establish one-to-one correspondence, as shown in Table \ref{tab1}. Therefore, the ideas and techniques of both sides can learn from each other. 

On the other hand, studies of causal emergence can provide new perspectives and insights for causal representation learning algorithms. For example, we can interpret the learned representations and causal mechanisms as emergent causal laws. The maximization of EI may improve the efficiency of learning the causalities for both CRL and reinforcement learning agents. 

\subsection{Ontological and Epistemological Causality and Emergence}
\label{sec:ontological}
Although machine learning techniques have facilitated the learning of causal structures and models, as well as the exploration of emergent properties and causation, it is important to consider whether the results obtained through machine learning reflect ontological causality and emergence or if they are merely epistemological phenomena.

Here, ontological causality refers to the causal relationships that exist in objective reality, independent of our knowledge or understanding. Ontological causality  explores the fundamental mechanisms and interactions that give rise to causal effects in the world. Similarly, ontological emergence is concerned with the objective existence of emergent properties and their underlying mechanisms. 

Epistemological causality, on the other hand, focuses on our knowledge and understanding of causality. It deals with how we perceive, model, and explain causal relationships based on our observations and experiences, but the causality may not exist in the real world. Epistemological emergence, akin to epistemological causality, focuses on our comprehension and elucidation of emergent phenomena. In other words, epistemological causality and emergence depend on an observer, and different observers may have distinct perspectives about causality and emergence for a particular objective phenomenon.

There has been a longstanding debate regarding the ontological and epistemological aspects of causality and emergence throughout history~\cite{kim1992downward,crutchfield1989inferring,bonabeau2011detection,yurchenko2023can}. The authors of~\cite{yurchenko2023can} highlighted that the concept of ``causation'' in the literature is often vague and should be differentiated into ``cause'' and ``reason'', aligning with ontological and epistemological causality. ``Cause'' means genuine cause that sufficiently causes the effects (causal closure principle~\cite{clayton2006re} and exclusion principle~\cite{kim2007physicalism}), whereas ``reason'' serves as a mere explanation for individuals to comprehend the effects. Reason may not possess the same level of rigor as genuine cause, but it does offer a certain degree of predictability that individuals find valuable in certain circumstances. 

However, distinguishing between them, particularly when addressing specific phenomena, remains a challenging task. One such controversial concept is downward causation, which has sparked extensive discussions. The question of whether downward causation exists objectively or not remains open. In~\cite{rosas2020reconciling}, Rosas argued that downward causation not only exists independently of particular observers but also provides a characterization of the phenomenon through quantified measures using information decomposition. However, Yurchenko proposed that it is important to differentiate between two separate concepts: causality and reasoning. According to this perspective, downward causation falls under the category of ``reason'' rather than causation. In this context, causality and reasoning can be seen as representing ontological and epistemological causality,~respectively.

Similarly, debates persist regarding the nature of causal emergence. The question arises as to whether causal emergence is a genuine phenomenon that exists independently of particular observers \cite{dewhurst2021causal}. In Section \ref{sec:exampleCE}, we highlighted that different coarse-graining strategies can lead to distinct macro-dynamic mechanisms with varying measures of EI. In essence, different coarse-graining strategies can be seen as representing different observers. However, Hoel's theory proposes a criterion to differentiate between coarse-graining methods, namely EI maximization. Consequently, for a given set of Markovian dynamics, only the coarse-graining strategy and corresponding macro-dynamics that maximize the EI measure can be considered as an objective outcome. Nevertheless, a challenge arises when multiple solutions exist that maximize EI, introducing a degree of subjectivity~\cite{dewhurst2021causal}. The authors of~\cite{zhang2022neural,yang2023finding} addressed this issue by introducing constraints on prediction errors in micro-states, which partially alleviates the problem. However, the question of whether optimal results have unique solutions remains an open problem that requires further investigation.

Similarly, although the incorporation of machine learning cannot resolve debates surrounding ontological and epistemological causality and emergence, it can provide objective standards that help mitigate subjectivity. This is because machine learning algorithms strive to optimize objective functions. Thus, machine learning agents can be viewed as ``objective'' observers that make judgments on causality and emergence. This represents an additional advantage of incorporating machine learning. However, the issue of unique solutions is crucial in this approach and warrants further attention.

Are the machine learning results ontological or epistemological? The answer is that the results are epistemological and depend on machine learning algorithms. However, this does not mean that all the results of machine learning are meaningless since if the learning agents are well trained and the defined mathematical objectives are effectively optimized, the results can also be considered objective and independent of the algorithms. 

It is intriguing to pose the same question regarding human cognition. Do all cognitive results in the human brain, encompassing causality, coarse-graining strategies, and macro-dynamics, reflect objective reality, or are they purely epistemological? Firstly, answering this question is exceedingly challenging. Secondly, the cognitive results specific to a particular human brain must objectively manifest within the neural network structures. Thus, for that brain, these results are ontological. Similar reasons hold for machine learning. The cognitive results of machine learning algorithms are recorded by the structures and parameters in neural networks. Therefore, studying these structures may reveal the properties of the original systems, and the nature of the interactions between the observer and the observed objects can be objectively reflected by machines.

Furthermore, the integration of machine learning can assist in establishing a theoretical framework for modeling observers and studying the interactions between the observer and the corresponding observed complex systems. This framework allows us not only to explore the hard problems regarding causality and emergence but also to understand the limitations and boundaries of the observer. One example of this research is the emergent classicality of quantum systems from the information bottleneck formed by machine learning algorithms~\cite{zhang2023observing}. 

\subsection{Potential Applications in Complex Systems}
In Section \ref{sec:causalemergence_ai}, we discussed the techniques related to causal emergence identification, machine learning, and causal inference. In this sub-section, we discuss the potential applications of these techniques in complex systems. Complex systems can be understood as a large network, with events connected by causal links. Automatically discovering the complex causal relationships from data is a challenging problem.

The research area of causal discovery has tried to address this problem in various ways~\cite{glymour2019review,zanga2022survey,donges2009backbone,kalisch2007estimating,zheng2018dags, zhu2019causal}. However, additional challenges may arise when applying these causal discovery methods in complex systems because circular causal structures and cross-level causation may exist. 

Circular causal structures widely exist in complex dynamical systems due to the feedback effect. That is, one variable may affect itself via a direct self-loop feedback or a long chain forming a circle. This circular causal structure may pose a challenge to existing methods of causality because most of these methods study directed acyclic causal graphs. However, recent progress has seen the development of some methods for discovering these circular structures in a data-driven manner, e.g.,~\cite{richardson2013discovery,lacerda2012discovering,forre2018constraint,rantanen2020learning,zhang2019general,pamfil2020dynotears,vowels2022d,spirtes2013directed}.

Another aspect is that higher-level or cross-level causation may exist if the ingredients of scale and coarse-graining are considered. For example, downward causation describes the causal effect between macro-level and micro-level variables. Thus, the causal connections may be cross-level. If causal emergence occurs in a complex system, some strong causality may be found between macro-variables. All of these factors in complex systems present new challenges. The machine learning methods discussed in Section~\ref{sec:nis} can address the question of whether higher-level causality exists at the macro-level. However, it is important to note that this is a global property. It must be further developed and extended to find a local causal relationship between macro-variables or macro-variables and micro-variables~\cite{varley2022flickering}. Moreover, existing methods of causal discovery must be extended to consider the method of grouping a set of variables or coarse-graining a system. For example,~\cite{cheng2023cuts+} proposed a coarse-to-fine causal discovery algorithm based on Granger causality and a graph neural network. The grouping process of variables is from coarse to fine, which can improve the efficiency of the algorithm. Other multi-level causal discovery methods were proposed and discussed in~\cite{wang2023hierarchical, zhang2023wavelet,fan2023interpretable}.

Another interesting problem is emergence detection. In complex dynamical systems, various higher-level patterns, such as waves, periodic oscillations, and solitons, are ubiquitous. For instance, in the climate system, typhoons and tornadoes are emergent vortex structures. In urban areas, traffic jams also emerge as a result of interactions among a large number of cars. Social riots are another example of emergent events in higher levels of human society. Identifying these emergent patterns at an early stage is crucial and significant~\cite{haugen2023detecting, mnif2011quantitative,o2017decentralised,mnif2011quantitative,o2014towards,pazho2022survey,liu2021emergence}. Therefore, there is an urgent need for a method that can automatically detect these emergent patterns and even provide early warning signals.

In this review, we mainly focus on the emergence of causality; however, we are not limited to this specific type of emergence. More macroscopic or global-level properties could also be emergent, although causal emergence may be the most important one. For example, in~\cite{chang2020information}, the authors discussed the concept of emergent information closure, which refers to the idea that the encoded information within an agent can form a closed system that is not influenced by the external world. The authors then argued that this agent with information closure can be regarded as a kind of consciousness. Symmetry may be another interesting property within some complex systems, and it may be emergent. For example, any single trajectory of a large number of random walkers in a two-dimensional Euclidean space is random. Nevertheless, the Gaussian distribution surface describing how the number of walkers falling in each small region changes with different locations is isotropic. This kind of symmetry can only be found at the macro-level and, thus, it is emergent. It would be interesting and useful to deploy a method for automatically discovering this kind of emergent symmetry. The framework for the NIS can be extended to address the problem of emergent symmetry identification if the optimized objective, EI, is replaced with a measure of isotope or more general symmetry~\cite{liu2022machine}. Similarly, we can also find other emergent symmetries using this framework once their measurement and optimization are possible.

However, it is crucial to acknowledge the inherent limitations of applying machine learning techniques to causal discovery and emergence identification problems. For instance, in \cite{textor2016robust}, the authors highlighted the existence of 'statistically equivalent' but causally distinct DAGs, implying that different causal structures can be constructed to account for the same dataset. Consequently, when utilizing machine learning techniques to uncover causality and emergent properties, similar challenges need to be addressed. Further research in this direction warrants significant attention to overcome these obstacles.

\subsection{Understanding Complex Systems from Causal Emergence}
A profound understanding of causal emergence and emergent causality can provide insights into understanding various mysterious phenomena in complex systems, including living systems, social systems, climate systems, ecosystems, and more. Significantly, fundamental questions, such as free will~\cite{yurchenko2023can}, consciousness~\cite{tononi2015consciousness}, and life, are all intricately connected to emergent causation~\cite{walker2014top}. For instance, free will can be viewed as an emergent form of downward causation~\cite{yurchenko2023can}. Social phenomena can be comprehended through the lens of causal emergence~\cite{elder2010causal}. Interestingly, both the EI and $\phi$ID frameworks for causal emergence have connections with one of the theories of consciousness, namely the integrated information theory~\cite{tononi2016integrated}. However, understanding these abstract concepts and phenomena in the context of causal emergence is merely an initial step.

For a specific system, there are three problems that should be addressed in future studies: (1) When does causal emergence occur? (2) How does emergent causality have functional effects on the system? (3) How does emergent causality change when the system is changed to adapt to the environment? We discuss these problems one by one.

We still do not know when causal emergence will occur and how the measure of causal emergence changes with some key parameters of a system. In~\cite{yang2023finding}, the authors showed how the measure of causal emergence changes with different noises in the Boid model (see the relevant discussion in Section \ref{sec:nis+}). It is reasonable to expect that there is a phase transition of emergent causality within a complex system when some key parameter changes because the causality or strength of the causal effect is also a global property, and it may be dependent on some order parameters. 

If causal emergence occurs in the system, how does the emergent causation affect the parts and the whole of the system? For example, the emergent ``I'' can be understood as an emergent macro-variable~\cite{yurchenko2023can}. How does ``I'' influence the part, say the foot, to move? This problem is non-trivial because it relates to the problem of mind--body interaction. It is important to study the information flows at both the macro-level and micro-level together to understand this phenomenon.

Finally, what is the relationship between adaptation and emergence~\cite{holland1995hidden,holland2000emergence,holland1992adaptation}? Sometimes, when we refer to a property as emergent, we essentially mean that this property can be developed through adaptation. Therefore, adaptation or evolution serves as a causal force for certain emergent properties. This concept also applies to emergent causality. For instance, downward causation, which is commonly found in complex systems, emerges as a result of adaptation and evolution. The next crucial question is, how can we evolve an emergent property or causation? This problem resembles the issue of designing emergence~\cite{adamatzkyemergence,kreyssig2011emergent,mitchell1993revisiting}. However, our aim here is to seek an explanation rather than a design. We want to understand the specific environment and manner in which the observed emergent causation can be evolved through adaptation.

All these problems require further studies in the future. However, these problems are just a few of the more interesting problems that need to be addressed.

\vspace{6pt}
\authorcontributions{Conceptualization, J.Z.; methodology, J.Z., B.Y., P,C,; writing, B.Y., J.Z., A.L., J.W., Z.W., M.Y., K.L., M.M. All authors have read and agreed to the published version of the manuscript. 
}

\funding{This work was supported in part by National Natural Science Foundation of China (No. 62141607, U1936219). 
}

\institutionalreview{Not applicable. 
}

\dataavailability{No data available.} 

\acknowledgments{This work is inspired by the insightful discussions held in the "Causal Emergence" series reading groups organized by Swarma Club and the "Swarma-Kaifeng" Workshop. We would like to extend our sincere gratitude to Professors Yizhuang You, Jing He, Dr. Chaochao Lu, and Dr. Yanbo Zhang for their invaluable contributions and insightful discussions. We express our appreciation for the support from the Save 2050 Programme, a joint initiative of Swarma Club and X-Order. Additionally, we are grateful for the support received from Swarma Research.} 

\conflictsofinterest{The authors declare no conflicts of interest.
}

\appendixtitles{yes}
\appendixstart
\appendix

\section[\appendixname~\thesection]{Calculation of EI with Transition Probability Matrix}

\label{sec:appendix_1}
Calculating effective information ($EI$) requires knowledge of the joint probability distribution of the system. The state transition probability matrix (TPM) defines the probability of transitioning from one state of the system to another. Specifically, the $j_{th}$ element of the $i_{th}$ row in the TPM, denoted by $TPM(i, j)$, represents the probability of the system being in state $s_j$ at time $t+1$, given that it is in state $s_i$ at time $t$ (i.e., \mbox{$P(X_{t+1} = s_j|X_t = s_i))$}.
\begin{equation}
    TPM(i, j) = P(X_{t+1} = s_j | X_t = s_i)\\
= \frac{P(X_{t+1} = s_j, X_t = s_i)}{P(X_t = s_i)}
\end{equation}

Assuming we can ``force'' the system to conform to a maximum entropy distribution (i.e., uniform distribution) with $N$ states, the probability of being in any given state $i$ at time $t$, denoted by $P_u(X_t = s_i)$, is $1\over{N}$. 

As a result, the TPM can be expressed as the product of the joint probability of the system's maximum entropy distribution, denoted by $P_u(X_{t+1} = s_j, X_t = s_i)$, and the total number of states, which is $N$, as demonstrated below.
\begin{adjustwidth}{-\extralength}{0cm}
\begin{equation}
    TPM(i, j) = \frac{P_u(X_{t+1} = s_j, X_t = s_i)}{P_u(X_t = s_i)}\\
= \frac{P_u(X_{t+1} = s_j, X_t = s_i)}{1/N}\\
=N\times{P_u(X_{t+1} = s_j, X_t = s_i)}
\end{equation}
\end{adjustwidth}

Hence, the TPM of the system can be used to express both the joint probability and marginal probability at time $t+1$ of the system's maximum entropy distribution.
\begin{equation}
    \therefore P_u(X_{t+1} = s_j, X_t = s_i) = \frac{1}{N}TPM(i, j) 
\end{equation}
\begin{equation}
\therefore P_u(X_{t+1} =s_j) = \sum_iP_u(X_{t+1} = s_j, X_t = s_i) \\
= \frac{1}{N}\sum_i TPM(i, j)
\end{equation}

Using this relationship, $EI$ can be calculated directly with the TPM.
\begin{equation}
\begin{aligned}
    EI = I(I_D; E_D) &= I(X_t; X_{t+1})|_{do(X_t) \sim U} \\
&= \sum_{i,j} P_u(X_{t+1} = s_j, X_t = s_i) \log_2\Big(\frac{P_u(X_{t+1} = s_j, X_t = s_i)} { P_u(X_{t+1} =s_j)P_u(X_t = s_i)}\Big) \\
&= \sum_{i,j} \frac{1}{N}TPM(i,j)\log_2\Big(\frac{\frac{1}{N}TPM(i,j)}{\frac{1}{N} \sum_k TPM(k, j) \times \frac{1}{N}}\Big) \\
&= \frac{1}{N} \sum_{i,j} TPM(i,j)\log_2\Big(\frac{N \times TPM(i,j)}{\sum_k TPM(k, j)}\Big)
\end{aligned}
\end{equation}

\begin{adjustwidth}{-\extralength}{0cm}
\reftitle{References}
\bibliography{main_arxiv}
\PublishersNote{}
\end{adjustwidth}
\end{document}